\documentclass[a4paper,12pt]{article}
\pdfoutput=1
\usepackage{epsfig,graphicx,xcolor,amsbsy,amssymb,latexsym,amsfonts,amsmath,,tcolorbox,setspace}
\usepackage{pstricks}
\usepackage{color}
\usepackage{soul}
\usepackage[numbers,square,comma, compress]{natbib}
\usepackage{placeins}
\usepackage{wrapfig,framed,caption}

\definecolor{shadecolor}{gray}{0.925}

\usepackage{eurosym}
\usepackage[a4paper]{geometry}
\usepackage{hyperref}
\usepackage{graphicx}
\usepackage{tikz}
\usepackage{xcolor}
\usepackage{amsmath,amssymb,amsfonts,pstricks,setspace}
\usepackage[enableskew]{youngtab}

\numberwithin{equation}{section}

\newcommand{\bea}{\begin{eqnarray}\displaystyle}
\newcommand{\eea}{\end{eqnarray}}

\newcommand{\figref}[1]{Fig.~\protect\ref{#1}}

\newcommand{\fact}[2]{\mathcal{C}_{#1}^{N=2,(#2)}}
\newcommand{\factD}[2]{\mathcal{C}_{#1}^{N=3,(#2)}}
\newcommand{\factN}[2]{\mathcal{C}_{#1}^{N=4,(#2)}}
\newcommand{\factG}[2]{\mathcal{C}_{#1}^{N,(#2)}}
\newcommand{\factNp}[3]{\widetilde{\mathcal{C}}_{#1,(#3)}^{N=4,(#2)}}
\newcommand{\factE}[3]{\mathcal{C}_{#1,(#3)}^{N=2,(#2)}}
\newcommand{\factDE}[3]{\mathcal{C}_{#1,(#3)}^{N=3,(#2)}}
\newcommand{\factNE}[3]{\mathcal{C}_{#1,(#3)}^{N=4,(#2)}}
\newcommand{\factGE}[3]{\mathcal{C}_{#1,(#3)}^{N,(#2)}}
\newcommand{\buildH}[1]{H^{{(#1)}}_{N=1}}
\newcommand{\buildW}[1]{W^{(#1)}_{\text{NS}}}

\newcommand{\buildInt}[2]{\mathcal{J}^{(#1)}_{#2}}

\title{
\begin{flushright}{\vspace{-2.5cm}\small LYCEN 2020-07\\}\end{flushright}
\vspace{2.3cm}
{\bf Symmetric Orbifold Theories from\\ Little String Residues}\\[40pt]}

\author{\large \textsc{Stefan~Hohenegger\footnote{\tt s.hohenegger@ipnl.in2p3.fr}\,\,\, and\, Amer~Iqbal\footnote{\tt amer@alum.mit.edu}}}
\date{}

\begin{document}

\maketitle
\thispagestyle{empty}
\begin{center}
\renewcommand{\thefootnote}{\fnsymbol{footnote}}\vspace{-0.5cm}
${}^{\footnotemark[1]}$ Univ Lyon, Univ Claude Bernard Lyon 1, CNRS/IN2P3, IP2I Lyon,\\ UMR 5822, F-69622, Villeurbanne, France\\[0.5cm]
${}^{\footnotemark[2]}$ How I Remember It, Inc\\
Brooklyn, New York 11221\\[1.5cm]
\end{center}

\begin{abstract}
We study a class of Little String Theories (LSTs) of A type, described by $N$ parallel M5-branes spread out on a circle and which in the low energy regime engineer supersymmetric gauge theories with $U(N)$ gauge group. The BPS states in this setting correspond to M2-branes stretched between the M5-branes. Generalising an observation made in \href{https://arxiv.org/abs/1706.04425}{arXiv:1706.04425}, we provide evidence that the BPS counting functions of special subsectors of the latter exhibit a Hecke structure in the Nekrasov-Shatashvili (NS) limit, \emph{i.e.} the different orders in an instanton expansion of the supersymmetric gauge theory are related through the action of Hecke operators. We extract $N$ distinct such reduced BPS counting functions from the full free energy of the LST with the help of contour integrals with respect to the gauge parameters of the $U(N)$ gauge group. Physically, the states captured by these functions correspond to configurations where the same number of M2-branes is stretched between some of these neighbouring M5-branes, while the remaining M5-branes are collapsed on top of each other and a particular singular contribution is extracted. The Hecke structures suggest that these BPS states form the spectra of symmetric orbifold CFTs. We furthermore show that to leading instanton order (in the NS-limit) the reduced BPS counting functions factorise into simpler building blocks. These building blocks are the expansion coefficients of the free energy for $N=1$ and the expansion of a particular function, which governs the counting of BPS states of a single M5-brane with single M2-branes ending on it on either side. To higher orders in the instanton expansion, we observe new elements appearing in this decomposition, whose coefficients are related through a holomorphic anomaly equation.

\end{abstract}

\newpage

\tableofcontents

\onehalfspacing

\vskip1cm

\section{Introduction }

Little String Theories (LSTs) were first introduced in \cite{Witten:1995zh, Aspinwall:1996vc, Aspinwall:1997ye, Seiberg:1997zk, Intriligator:1997dh, Hanany:1997gh, Brunner:1997gf}. They are a type of quantum theories in 6 dimensions, which behave like ordinary quantum field theories (with point-like degrees of freedom) in the low energy regime, but whose UV completion requires the inclusion of string-like degrees of freedom. On the one hand side, LSTs serve in many aspects as toy models of string theory, with the only difference that the gravitational sector is absent: indeed, in practice, many examples of LSTs are obtained from (type II) string theory or M-theory through a particular decoupling limit, which sends the string coupling to zero, while leaving the string length finite. Thus studying properties of LSTs gives us an important window into string- or M-theory, which are intrinsically difficult to study by more direct means. On the other hand, conversely a better understanding of LSTs also provides us with more information about the (supersymmetric) gauge theories that are engineered in the low energy sector: due to their stringy origins, LSTs inherit numerous symmetries and dualities from string- and M-theory, remnants of which are still visible in the low energy gauge theories engineered by the LSTs. 

In the same spirit, there exist many (geometric and computational) tools that have been developed in the framework of full-fletched string theory (or related applications), which allow us to perform many explicit computations for LSTs. For example, geometrical methods which have been used to classify conformal field theories in 6 dimensions or less \cite{Heckman:2013pva,Heckman:2015bfa,Xie:2015rpa,Jefferson:2017ahm, Jefferson:2018irk,Caorsi:2018zsq,Bhardwaj:2018vuu, Bhardwaj:2019hhd,Apruzzi:2019opn, Bhardwaj:2019jtr, Martone:2020nsy, Argyres:2020wmq}, have recently also been deployed to attempt a classification of LSTs \cite{Bhardwaj:2015oru, Bhardwaj:2019hhd}. Indeed, while an ADE-classification of LSTs is know since some time \cite{Witten:1995zh,Aspinwall:1996vc, Aspinwall:1997ye,Seiberg:1997zk,Intriligator:1997dh,Hanany:1997gh,Brunner:1997gf}, recent efforts have focused on sharpening the list of all possible such theories. 

Furthermore, a specific class of theories which have received a lot of attention recently are LSTs of type A. Such theories, compactified on a circle, have been studied using various dual approaches in string or M-theory: on the one hand side, they can be described by configurations of $N$ parallel M5-branes that are separated along a circle $S^{1}_{\rho}$ and compactified on a circle $S^{1}_{\tau}$.\footnote{We refer the reader to \cite{Haghighat:2013gba,Hohenegger:2015btj,Hohenegger:2015cba,Hohenegger:2016eqy} for more details on the brane setup.} BPS configurations in this setting correspond to M2-branes that are stretched between neighbouring M5-branes and wrapping $S^{1}_{\tau}$. The partition function of LSTs compactified on $S^{1}_{\tau}$, $\mathcal{Z}_{N,1}$, can be calculated by analysing the theory on the intersection of the M2- and M5-branes \cite{Haghighat:2013gba,Haghighat:2013tka,Hohenegger:2013ala} using a 2-dimensional sigma-model description. In order to render $\mathcal{Z}_{N,1}$ well defined, the introduction of two regularisation parameters $\epsilon_{1,2}$ is required. From the point of view of the low energy gauge theory description, the latter correspond to the parameters of the $\Omega$-background, which are needed to regularise the Nekrasov partition function. A dual approach is given by F-theory compactified on a class toric Calabi-Yau manifolds \cite{Kanazawa:2016tnt} called $X_{N,1}$. The web diagram of the latter can directly be read off from the brane-web diagram representing the system of M2-M5-branes mentioned above \cite{Haghighat:2013gba,Haghighat:2013tka,Hohenegger:2013ala}. Furthermore, $\mathcal{Z}_{N,1}$ in this approach is captured by the topological string partition function on $X_{N,1}$, which in turn can be very efficiently calculated with the help of the topological vertex \cite{Aganagic:2003db, Iqbal:2007ii}. The regularisation parameters $\epsilon_{1,2}$ in this approach are intrinsic to the refined topological string \cite{GV1,GV2,Hollowood:2003cv} (see also \cite{Antoniadis:2010iq,Antoniadis:2013bja,Antoniadis:2013mna,Antoniadis:2015spa}).

Recent studies have exploited this efficient way to explicitly compute $\mathcal{Z}_{N,1}$ (or the corresponding free energy $\mathcal{F}_{N,1}$) to study symmetries and other properties of the corresponding LSTs. In the process, numerous very interesting and unexpected structures have been discovered, among others:
\begin{enumerate}
\item[\emph{(i)}] Dihedral and paramodular symmetries:\\
The Calabi-Yau manifold $X_{N,1}$ engineers a supersymmetric gauge theory on $\mathbb{R}^4\times S^1\times S^1$ with $U(N)$ gauge group and matter in the adjoint representation. The K\"ahler moduli space of $X_{N,1}$ can be understood as a subregion of a much larger so-called \emph{extended moduli space}. Depending on the value of $N$, there exist further regions in the latter which engineer supersymmetric gauge theories with different gauge structures and matter content. Many of these theories are dual to each other in the sense that they share the same partition function. The duality map, however, is intrinsically non-perturbative. More concretely, it was conjectured in \cite{Hohenegger:2016yuv} and proven in \cite{Bastian:2017ing,Bastian:2018dfu} that the $U(N)$ gauge theory above is dual to a circular quiver gauge theory with a gauge group made up of $M'$ factors of $U(N')$ and bifundamental matter, for any pair $(N',M')$ such that $M'N'=N$ and $\text{gcd}(M',N')=1$.\footnote{In \cite{Hohenegger:2016yuv}, the much stronger conjecture was put forth that the Calabi-Yau manifolds $X_{N,M}$ and $X_{N',M'}$ are dual to each other if $NM=N'M'$ and $\text{gcd}(N,M)=\text{gcd}(N',M')$. This implies a duality between gauge theories with gauge group $U(N)^M$ and $U(N')^{M'}$. Numerous examples have been successfully tested in \cite{Hohenegger:2016yuv,Bastian:2017ary,Bastian:2018dfu}. Furthermore, the case $\text{gcd}(N,M)=1$ has been proven in \cite{Bastian:2018dfu} for arbitrary values of $\epsilon_{1,2}$ and a proof for generic $N,M$ for $\epsilon_{1,2}\to 0$ has been presented in \cite{Haghighat:2018gqf}.}

It was shown in \cite{Bastian:2018jlf}, that this web of dualities implies additional symmetries for the partition function $\mathcal{Z}_{N,1}$ (as well as the free energy $\mathcal{F}_{N,1}$). While it is clear (due to the structure of $X_{N,1}$ as a double-elliptic fibration), that the latter are symmetric with respect to two modular groups called $SL(2,\mathbb{Z})_\rho$ and $SL(2,\mathbb{Z})_\tau$\footnote{The notation follows the K\"ahler parameters which act us modular parameters for the two groups.}, it was shown in \cite{Bastian:2018jlf} that they also enjoy a dihedral symmetry, which (from the perspective of the gauge theories) acts in an intrinsically non-perturbative fashion. Moreover, it was argued in \cite{Bastian:2019hpx}  that a particular subsector of the BPS-states (namely the sector of states which carry the same $U(1)$ charges under all the generators of the Cartan subalgebra of the $U(N)$ gauge group), is invariant under the level $N$ paramodular group $\Sigma_N\subset Sp(4,\mathbb{Q})$.

\item[\emph{(ii)}] Hecke structures:\\
In \cite{Bastian:2019hpx} evidence was presented that in the Nekrasov-Shatashvili (NS) limit \cite{Nekrasov:2009rc,Mironov:2009uv} (which in our notation essentially corresponds to the limit $\epsilon_2\to 0$), the paramodular group $\Sigma_N$ that is present in the above mentioned subsector of BPS states, is further extended to $\Sigma_N^*$. The latter is obtained from $\Sigma_N$ through the inclusion of a further generator that exchanges the modular parameters $\rho$ and $\tau$ of the two modular groups mentioned above (see appendix~\ref{App:Paramodular} for details). This result corroborates the observation made in \cite{Ahmed:2017hfr} that the states of the subsector of BPS states mentioned above (in the NS-limit) can be organised into a symmetric orbifold CFT. The latter in particular implies that the expansion coefficients of the reduced free energy (that only counts states in this particular BPS-subsector) are related through the action of Hecke operators. This relation was indeed observed in \cite{Ahmed:2017hfr} in numerous examples.

\item[\emph{(iii)}] Factorisation to leading instanton order and graph functions:\\
In \cite{Hohenegger:2019tii} non-trivial evidence was provided that the free energy $F_{N,1}$ in the so-called unrefined limit (\emph{i.e.} for $\epsilon_2=-\epsilon_1$) factorises in a very intriguing fashion: for the examples $N=2,3,4$, it was shown that the free energy to leading instanton order (from the perspective of the $U(N)$ gauge theory) can be decomposed into sums of products of the functions $\buildH{r}$ and $\buildW{r}$. The former are the expansion coefficients of the instanton expansion of the free energy $F_{N=1,1}$, while the latter are the expansion coefficients of a function that governs the counting of BPS states of an M5-branes with a single M2-brane ending on either side.\footnote{Explicit expressions for $\buildH{r}$ and $\buildW{r}$ as well as more information can be found in appendix~\ref{App:BuildingBlock}.} Furthermore, it was observed in \cite{Hohenegger:2019tii} that the coefficients appearing in this expansion of $F_{N,1}$ resemble in many respects so-called \emph{modular graph functions}, which have recently appeared in the study of scattering amplitudes in string theory \cite{Broedel:2014vla, DHoker:2015wxz, DHoker:2016mwo, DHoker:2017pvk, Zerbini:2018sox,Zerbini:2018hgs,Gerken:2018jrq,Mafra:2019ddf, Mafra:2019xms, Gerken:2019cxz}. Higher terms in the instanton expansion are more complicated: certain parts still allow to be factorised into simpler building blocks, however, on the one hand side the coefficient functions that appear in this way are more complicated and on the other hand the inclusion of Hecke transformations of $\buildH{r}$ and $\buildW{r}$ is required. While primarily dealing with the unrefined free energy, preliminary results in \cite{Hohenegger:2019tii} indicate similar decompositions (albeit more complicated) to be valid in the NS-limit.
\end{enumerate}
The current paper is a continuation of the analysis of the symmetries and structures discovered in \cite{Bastian:2019hpx,Bastian:2019wpx,Hohenegger:2019tii,Ahmed:2017hfr}: focusing on the NS-limit of the free energy, we analyse the form of the free energy that was found in \cite{Bastian:2019hpx,Bastian:2019wpx,Hohenegger:2019tii} and which is implied by the property \emph{(i)} above. We observe new subsectors of the BPS-states that show a similar Hecke structure as discussed under \emph{(ii)} above. Based on the examples of $N=2$ and $N=3$ (as well as partial results for $N=4$), we observe for given $N$ and at each order in an expansion of $\epsilon_1$, $N$ distinct subsectors of the NS-limit of the BPS-free energy $\mathcal{F}_{N,1}$ that exhibit such structures. We call the functions which count these BPS states at the $r$-th instanton order $\factGE{i}{r}{2s,0}$, where $i=1,\ldots,N$ and $s\in\mathbb{N}$ indicates the order in an expansion in powers of $\epsilon_1$. The latter can abstractly be defined for general $N$ through contour integrals of (an expansion in powers of $\epsilon_1$ of the NS-limit of) $\mathcal{F}_{N,1}$ with respect to the gauge parameters $\widehat{a}_{1,\ldots,N}$ of the $\mathfrak{a}_{N-1}$ gauge algebra (or their exponentials $Q_{\widehat{a}_i}=e^{2\pi i \widehat{a}_i}$ for $i=1,\ldots,N$). These contours extract specific coefficients in a Fourier expansion of $\mathcal{F}_{N,1}$ in $Q_{\widehat{a}_i}$ and/or particular poles in a limit where some of the $\widehat{a}_i$ vanish (see eq.~(\ref{FactGenDefinition}) for the abstract definition of the $\factGE{i}{r}{2s,0}$). From a physical perspective the functions $\factGE{i}{r}{2s,0}$ only receive contributions from M5-brane configurations where the same number of M2-branes is stretched between some of the adjacent M5-branes (see \figref{fig:BraneCollapse}) for a schematic representation. From these configurations in turn specific poles are extracted in the limit where the remaining M5-branes coincide. The remaining functional dependence of $\factGE{i}{r}{2s,0}$ is made up by two (remaining) K\"ahler moduli of $X_{N,1}$ (which we call $\rho$ and $S$). Finally, we can resum the $\factGE{i}{r}{2s,0}$ into a Laurent series expansion $\factG{i}{r}(\rho,S,\epsilon_1)$ in powers of $\epsilon_1$.

We observe that the functions $\factGE{i}{r}{2s,0}$ obtained in this fashion show numerous interesting properties. First of all, they are quasi-Jacobi forms of index $rN$ and weight $2s-2i$. Moreover, the functions for $r>1$ can be obtained through the action of the $r$-th Hecke operator $\mathcal{H}_r$ (see (\ref{DefHecke}) in appendix~\ref{App:ModularStuff} for a definition) on $\factGE{i}{r=1}{2s,0}(\rho,S)$
\begin{align}
&\factGE{i}{r}{2s,0}(\rho,S)=\mathcal{H}_r\left[\factGE{i}{1}{2s,0}(\rho,S)\right]\,,&&\begin{array}{l}\forall r\geq 1\,,\\ \forall i=1,\ldots,N\,.\end{array}
\end{align}
Following the logic of \cite{Ahmed:2017hfr}, this suggests that the corresponding BPS states can be organised into a symmetric torus orbifold CFT. However, since the seed function (\emph{i.e.} the initial function $\factGE{i}{r=1}{2s,0}$) is different in each case, the corresponding target spaces are different for all $i=1,\ldots,N$. Indeed, the functions $\factG{i}{r=1}$ can be factorised in terms of $\buildH{r=1}$ and $\buildW{r=1}$ in a very simple fashion (see eq.~(\ref{FactorisationO1})). For $r>1$, the $\factG{i}{r}$ still can mostly be decomposed into $\buildH{r=1}(\rho,S)$ and $\buildW{r=1}(\rho,S)$, up to remainder functions (see eq.~(\ref{FactorisationOr})). The latter, however, are not arbitrary, but are connected by equations that strongly resemble holomorphic anomaly equations \cite{Bershadsky:1993cx}. These results generalise the properties of the free energy discussed under \emph{(iii)} above.  Since the results in this paper are obtained by studying the examples of $N=2$ and $N=3$ (as well as partially $N=4$) their generalisations to higher $N$ have to be considered conjectures. However, the large number of examples that all follow the same pattern, provides rather strong evidence in their favour.

This paper is organised as follows: In section~\ref{Sect:LSTFree} we review the LST partition function $\mathcal{Z}_{N,1}$ and the associated free energy $\mathcal{F}_{N,1}$, as well as some of their properties discovered in recent publications. Due to the technical nature of some of the subsequent discussions, we provide a summary of the results of this paper in section~\ref{Sect:Summary}. The sections~\ref{Sect:CaseN2}, \ref{Sect:CaseN3} and \ref{Sect:CaseN4} provide a detailed discussion of the LST free energies for $N=2$, $N=3$ and $N=4$ respectively.  Finally, section~\ref{Sect:Conclusion} contains our conclusions. Additional details on modular objects, explicit discussions of properties of the free energy, the discussion (and explicit expressions for some of their expansion coefficients) of the fundamental building blocks $\buildH{r}$ and $\buildW{r}$ as well as the definition of paramodular groups, which have been deemed too long for the main body of this paper have been relegated to four appendices.

\section{Little String Free Energies and Their Properties}\label{Sect:LSTFree}
The Little String Theories (LSTs) of A-type that we are interested in can be studied by ex-

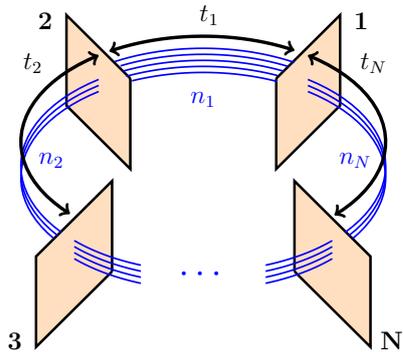
\begin{wrapfigure}{r}{0.35\textwidth}
\begin{center}
${}$\\[-1cm]
\scalebox{0.8}{\parbox{6.7cm}{\begin{tikzpicture}[scale = 1]
\draw[very thick,fill=orange!25!white] (1.5,-1.25) -- (2.75,-2.5) -- (2.75,-1) -- (1.5,0.25) -- (1.5,-1.25);
\draw[very thick,fill=orange!25!white] (-1.5,-1.25) -- (-2.75,-2.5) -- (-2.75,-1) -- (-1.5,0.25) -- (-1.5,-1.25);
\draw[very thick,fill=orange!25!white] (1.2,1.95) -- (2.25,3) -- (2.25,1.5) -- (1.2,0.45) -- (1.2,1.95);
\draw[very thick,fill=orange!25!white] (-1.2,1.95) -- (-2.25,3) -- (-2.25,1.5) -- (-1.2,0.45) -- (-1.2,1.95);
\draw [thick,blue,domain=-35:55,yshift=0.3cm] plot ({3*cos(\x)}, {1.75*sin(\x)});
\draw [thick,blue,domain=-37:55,yshift=0.4cm] plot ({3*cos(\x)}, {1.75*sin(\x)});
\draw [thick,blue,domain=-39:55,yshift=0.5cm] plot ({3*cos(\x)}, {1.75*sin(\x)});
\node[blue] at (2.5,0.6) {$n_N$}; 
\draw [thick,blue,domain=-135:-110,yshift=0.2cm] plot ({3*cos(\x)}, {1.75*sin(\x)});
\draw [thick,blue,domain=-135:-110,yshift=0.3cm] plot ({3*cos(\x)}, {1.75*sin(\x)});
\draw [thick,blue,domain=-135:-110,yshift=0.4cm] plot ({3*cos(\x)}, {1.75*sin(\x)});
\draw [thick,blue,domain=-135:-110,yshift=0.5cm] plot ({3*cos(\x)}, {1.75*sin(\x)});
\draw [thick,blue,domain=-70:-45,yshift=0.2cm] plot ({3*cos(\x)}, {1.75*sin(\x)});
\draw [thick,blue,domain=-70:-45,yshift=0.3cm] plot ({3*cos(\x)}, {1.75*sin(\x)});
\draw [thick,blue,domain=-70:-45,yshift=0.4cm] plot ({3*cos(\x)}, {1.75*sin(\x)});
\draw [thick,blue,domain=-70:-45,yshift=0.5cm] plot ({3*cos(\x)}, {1.75*sin(\x)});
\node[blue] at (0,-1.3) {\Large $\cdots$}; 
%
\draw [thick,blue,domain=125:215,yshift=0.3cm] plot ({3*cos(\x)}, {1.75*sin(\x)});
\draw [thick,blue,domain=125:217,yshift=0.4cm] plot ({3*cos(\x)}, {1.75*sin(\x)});
\draw [thick,blue,domain=125:219,yshift=0.5cm] plot ({3*cos(\x)}, {1.75*sin(\x)});
\node[blue] at (-2.5,0.6) {$n_2$}; 
\draw [thick,blue,domain=62:117,yshift=0.6cm] plot ({3*cos(\x)}, {1.75*sin(\x)});
\draw [thick,blue,domain=60:119,yshift=0.7cm] plot ({3*cos(\x)}, {1.75*sin(\x)});
\draw [thick,blue,domain=66:113,yshift=0.3cm] plot ({3*cos(\x)}, {1.75*sin(\x)});
\draw [thick,blue,domain=66:115,yshift=0.4cm] plot ({3*cos(\x)}, {1.75*sin(\x)});
\draw [thick,blue,domain=64:116,yshift=0.5cm] plot ({3*cos(\x)}, {1.75*sin(\x)});
\node[blue] at (0,1.6) {$n_1$}; 
\node at (2.6,2.9) {\bf 1};
\node at (-2.6,2.9) {\bf 2};
\node at (-3.1,-2.4) {\bf 3};
\node at (3.1,-2.4) {\bf N};
\draw [ultra thick,<->,domain=-44:55,yshift=0.9cm] plot ({3*cos(\x)}, {1.75*sin(\x)});
\node at (2.8,2.2) {$t_N$}; 
\draw [ultra thick,<->,domain=60:121,yshift=0.9cm] plot ({3*cos(\x)}, {1.75*sin(\x)});
\node at (0.1,3) {$t_1$}; 
\draw [ultra thick,<->,domain=125:223,yshift=0.9cm] plot ({3*cos(\x)}, {1.75*sin(\x)});
\node at (-2.8,2.2) {$t_2$}; 
\end{tikzpicture}}}
\end{center} 
\caption{{\it $N$ parallel M5-branes (orange) with $(n_1,\ldots,n_N)$ M2-branes (blue) stretched between them. The distances between the M5-branes are $t_{1,\ldots,N}$.}}
\label{fig:M5M2branes}
${}$\\[-1.5cm]
\end{wrapfigure}

\noindent
ploiting various dual descriptions in string or M-theory. On the one hand side, they can be described through configurations of parallel M5-branes compactified on a circle of circumference $\tau$ and spread out on a circle with circumference $\rho$, where the distances between the neighbouring M5-branes are denoted $(t_1,\ldots,t_N)$ such that,
\bea
\rho=t_{1}+t_{2}+\ldots+t_{N}\,.
\eea
BPS states in this setting are given by M2-branes. The latter are stretched between the neighbouring M5-branes and appear as strings in their worldvolums, wrapping the circle $S^{1}_{\tau}$ on which the M5-branes are wrapped \cite{Haghighat:2013gba}.  In this context, arbitrarily many M2-branes can be stretched between any of the neighbouring M5-branes (a schematic example is shown in \figref{fig:M5M2branes}). The space transverse to the M2-branes inside the M5-brane worldvolume is $\mathbb{R}^{4}_{\parallel}$ and the M2-branes appear as point particles in this space. The worldvolume theory of M2-branes has $\mathcal{N}=(4,4)$ supersymmetry which is broken down to $\mathcal{N}=(0,2)$ by a $U(1)_{\epsilon_1}\times U(1)_{\epsilon_2}\times U(1)_{m}$ action on $R^{4}_{\parallel}\times \mathbb{R}^{4}_{\perp}$~\cite{Haghighat:2013gba}~\footnote{We define $q=e^{i \epsilon_1}$ and $t=e^{- i \epsilon_2}$ so that the unrefined limit is $q=t$.},
\bea
(z_{1},z_{2},w_{1},w_{2})\in \mathbb{C}^{2}_{\parallel}\times \mathbb{C}^{2}_{\perp}\mapsto (z_{1}\,e^{i \epsilon_{1}}, z_{2}\,e^{ i \epsilon_{2}}\,, w_{1}\,e^{ i (m+\epsilon_{+})},w_{2}\,e^{ i (m-\epsilon_{+})})\,.
\eea
The BPS degeneracies of the M2-branes is captured by the elliptic genus of the worldvolume theory 
which depends on the parameters $(\tau,t_{1,\cdots,N},m,\epsilon_{1,2})$. This can be calculated by 
studying the gauge and matter content of the $(0,2)$ worldvolume theory and using the techniques 
developed in \cite{EG3, EG1, EG2}. A different approach is to calculate the $(0,2)$ elliptic genus of the sigma model to 
which the worldvolume theory flows in the infrared. The target space of the sigma model in this case is 
the product of Hilbert schemes of points on $\mathbb{C}^{2}_{\parallel}$ and the equivariant elliptic genus 
can be calculated using the details of the $U(1)_{\epsilon_{1}}\times U(1)_{\epsilon_{2}}$ action on the 
target space \cite{HNBook, Nakajima:2003pg, Haghighat:2013gba}. The theory on the worldvolume of the compactified M5-branes is 
the five dimensional ${\cal N}=1^{\star}$ quiver gauge theory. The partition function of this 
gauge theory captures the M2-brane BPS states as well and is given by the generating function of 
equivariant elliptic genera of the rank $N$ charge $K$ instanton moduli spaces $M(N,k)$ \cite{Haghighat:2013gba, Hohenegger:2013ala}. 

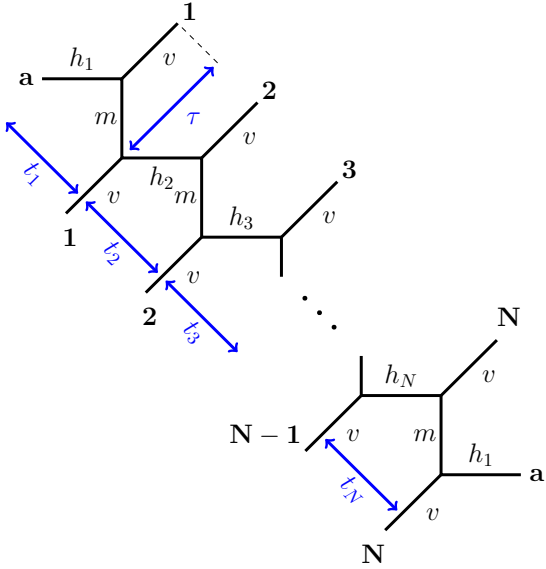
\begin{wrapfigure}{l}{0.47\textwidth}
\begin{center}
\vspace{0cm}
\scalebox{0.7}{\parbox{10.3cm}{\begin{tikzpicture}[scale = 1.50]
\draw[ultra thick] (-1,0) -- (0,0) -- (0,-1) -- (1,-1) -- (1,-2) -- (2,-2) -- (2,-2.5);
\node[rotate=315] at (2.5,-3) {\Huge $\cdots$};
\draw[ultra thick] (3,-3.5) -- (3,-4) -- (4,-4) -- (4,-5) -- (5,-5);
\draw[ultra thick] (0,0) -- (0.7,0.7);
\draw[ultra thick] (1,-1) -- (1.7,-0.3);
\draw[ultra thick] (2,-2) -- (2.7,-1.3);
\draw[ultra thick] (4,-4) -- (4.7,-3.3);
\draw[ultra thick] (0,-1) -- (-0.7,-1.7);
\draw[ultra thick] (1,-2) -- (0.3,-2.7);
\draw[ultra thick] (3,-4) -- (2.3,-4.7);
\draw[ultra thick] (4,-5) -- (3.3,-5.7);
\node at (-1.2,0) {\large {\bf $\mathbf a$}};
\node at (5.2,-5) {\large {\bf $\mathbf a$}};
\node at (0.85,0.85) {\large {$\mathbf 1$}};
\node at (1.85,-0.15) {\large {$\mathbf 2$}};
\node at (2.85,-1.15) {\large {$\mathbf 3$}};
\node at (4.85,-3) {\large {$\mathbf{N}$}};
\node at (-0.65,-2) {\large {$\mathbf{1}$}};
\node at (0.35,-3) {\large {$\mathbf{2}$}};
\node at (1.8,-4.5) {\large {$\mathbf{N-1}$}};
\node at (3.15,-6) {\large {$\mathbf{N}$}};
\node at (-0.5,0.25) {\large  {\bf $h_1$}};
\node at (0.5,-1.25) {\large  {\bf $h_2$}};
\node at (1.5,-1.75) {\large  {\bf $h_3$}};
\node at (3.5,-3.75) {\large  {\bf $h_N$}};
\node at (4.5,-4.75) {\large  {\bf $h_1$}};
\node at (-0.2,-0.5) {\large  {\bf $m$}};
\node at (0.8,-1.5) {\large  {\bf $m$}};
\node at (3.8,-4.5) {\large  {\bf $m$}};
\node at (0.6,0.25) {\large  {\bf $v$}};
\node at (1.6,-0.75) {\large  {\bf $v$}};
\node at (2.6,-1.75) {\large  {\bf $v$}};
\node at (4.6,-3.75) {\large  {\bf $v$}};
\node at (-0.1,-1.5) {\large  {\bf $v$}};
\node at (0.9,-2.5) {\large  {\bf $v$}};
\node at (2.9,-4.5) {\large  {\bf $v$}};
\node at (3.9,-5.5) {\large  {\bf $v$}};
%
%
\draw[ultra thick,blue,<->] (-1.45,-0.55) -- (-0.55,-1.45);
\node[blue,rotate=315] at (-1.1,-1.2) {{\large {\bf {$t_1$}}}};
\draw[ultra thick,blue,<->] (-0.45,-1.55) -- (0.45,-2.45);
\node[blue,rotate=315] at (-0.1,-2.2) {{\large {\bf {$t_2$}}}};
\draw[ultra thick,blue,<->] (0.55,-2.55) -- (1.45,-3.45);
\node[blue,rotate=315] at (0.9,-3.2) {{\large {\bf {$t_3$}}}};
\draw[ultra thick,blue,<->] (2.55,-4.55) -- (3.45,-5.45);
\node[blue,rotate=315] at (2.9,-5.2) {{\large {\bf {$t_N$}}}};
\draw[dashed] (1.2,0.2) -- (0.7,0.7);
\draw[ultra thick,blue,<->] (0.1,-0.9) -- (1.15,0.15);
\node[blue] at (0.9,-0.5) {{\large {\bf {$\tau$}}}};
\end{tikzpicture}}}
\caption{\sl Web diagram of $X_{N,1}$.}
\label{Fig:N1web}
\end{center}
${}$\\[-3cm]
\end{wrapfigure}

A dual approach to describe the same LST is through F-theory compactified on a toric Calabi-Yau threefold \cite{Kanazawa:2016tnt} which in \cite{Hohenegger:2015btj} was called $X_{N,1}$. The BPS string states are given by D3-branes wrapping various rational curves in the base of the Calabi-Yau threefold with K\"ahler parameters $t_{1},\cdots, t_{N}$. The web diagram of the latter can be directly read off from the brane web configuration discussed before and is shown in \figref{Fig:N1web}. The latter figure also includes a definition (shown in blue) of a basis of the K\"ahler parameters of $X_{N,1}$: these include besides
$t_{1},\ldots,t_N$ also $\tau$ and $m$
which can  can be expressed in terms of the basis $(h_1,\ldots,h_N,m,v)$. From the perspective of F-theory compactified on $X_{N,1}$, the little string partition function $\mathcal{Z}_{N,1}$ is captured by the topological string partition function on $X_{N,1}$ \cite{Bhardwaj:2015oru, Hohenegger:2015btj}. The latter can be computed in an efficient manner using the refined topological vertex formalism \cite{Aganagic:2003db,Iqbal:2007ii}.\\[-36pt]

\noindent

\noindent
In \cite{Haghighat:2013gba, Haghighat:2013tka, Hohenegger:2013ala, Hohenegger:2015btj,Bastian:2018fba,Bastian:2017jje} two different expansion of ${\cal Z}_{N,1}$ and their interpretations were studied,
\bea\label{TwoExpansions}
{\cal Z}_{N,1}&=&\sum_{k}Q_{\tau}^{k}\,Z_{k}(t_{1,\cdots,N},m,\epsilon_{1,2})\,,\\\nonumber
&=&\sum_{k_{1},\cdots,k_{N}}Q_{t_{1}}^{k_1}\cdots Q_{t_{N}}^{k_N}\,Z_{k_{1}\cdots k_{N}}(\tau,m,\epsilon_{1,2})\,,
\eea
where $Q_{\tau}=e^{2\pi i \tau}$ and $Q_{t_{i}}=e^{2\pi i t_{i}}$.

As discussed in Appendix~\ref{App:N1PartitionFunction},  $Z_{k}(t_{1,\cdots,N},m,\epsilon_{1,2})$ is the equivariant $(2,2)$ elliptic genus of $M(N,k)$. This expansion of the partition function is natural when considering the theory on the M5-branes. The expansion in the second line in Eq.(\ref{TwoExpansions}) gives the functions $Z_{k_{1}\cdots k_N}$ which capture the degeneracy of configurations of M2-branes in which $k_{i}$ M2-branes are stretched between the $i$-th and $(i+1)$-th M5-brane. The $Z_{k_{1}\cdots k_{N}}$ is the equivariant $(0,2)$ elliptic genus with target space $\otimes_{i=1}^{N}\text{Hilb}^{k_{i}}[\mathbb{C}^2]$ with right moving fermions coupled to a bundle $V_{k_{1}\cdots k_{N}}$, the details of which are given in  \cite{Haghighat:2013gba, Hohenegger:2013ala}.

In \cite{Hohenegger:2013ala, Hohenegger:2015btj} the following little string free energy ${\cal F}^{\text{plet}}_{N,1}$ was discussed
\begin{align}
{\cal F}^{\text{plet}}_{N,1}(t_{1,\ldots,N},m,\tau,\epsilon_{1,2})=\mbox{Plog} \,\mathcal{Z}_{N,1}(t_{1,\ldots,N},m,\tau,\epsilon_{1,2})\,,\label{FreeEnergy}
\end{align}
where, Plog denotes the plethystic logarithm\footnote{The plethystic logarithm of a function $g(x_{1},x_{2},\cdots,x_{K})$ is given by $\text{Plog}\,g(x_{1},x_{2},\cdots,x_{K})=\sum_{n=1}^{\infty}\frac{\mu(n)}{n}\,\text{log}\,g(nx_{1},nx_{2},\cdots,nx_{K})$ where $\mu(n)$ is the M\"obius function.} of $\mathcal{Z}_{N,1}$. The exact form of ${\cal Z}_{N,1}$ is given in Appendix~\ref{App:N1PartitionFunction}. From a physical perspective, ${\cal F}^{\text{plet}}_{N,1}$ only counts single-particle BPS excitations of the LST, projecting out multi-particle states. Similar to the two equivalent expansions of the partition function in eq. (\ref{TwoExpansions}), one can similarly consider  two different ways of expanding~${\cal F}^{\text{plet}}_{N,1}$
\begin{align}
{\cal F}^{\text{plet}}_{N,1}(t_{1,\ldots,N},m,\tau,\epsilon_{1,2})=\sum_{k}Q_{\tau}^{k}\,F^{\text{plet}}_{k}(t_{1,\ldots,N},m,\epsilon_{1,2})=\sum_{k_{1}\cdots k_{N}}Q_{t_{1}}^{k_1}\cdots Q_{t_{N}}^{k_N}\,F_{\text{plet}}^{(k_{1},\ldots, k_{N})}(\tau,m,\epsilon_{1,2})\,.\label{ExpandPlet}
\end{align}

\noindent
In previous work numerous properties of the free energy ${\cal F}^{\text{plet}}_{N,1}$ (or some of the coefficients appearing in these two expansions) have been discovered. In the following we shall discuss some of them, which will turn out important for our current work:
\begin{itemize}
\item Recursion relation:\\
In \cite{Hohenegger:2015cba,Hohenegger:2016eqy} the counting functions of a particular class of single BPS states has been discussed: these states correspond to M-brane configurations of the type schematically shown in \figref{fig:M5M2branes}, however, they are special in the sense that they have one (or several neighbouring) M5-brane(s) with only a single M2-brane ending on them on either side. In the notation of \figref{fig:M5M2branes}, these are characterised by the fact that several adjacent $n_i$ are identical to 1, \emph{i.e.}
\begin{align}
(n_1,\ldots,n_k,\underbrace{1,\ldots,1}_{m\text{-times}},n_{k+m},\ldots,n_N)\,.
\end{align}

The BPS degeneracy of such states is captured by $F_{\text{plet}}^{(k_1,\ldots, k_N)}$ (defined in Eq.(\ref{ExpandPlet})) with $(k_{1},\cdots,k_N)=(n_1,\ldots,n_k,\underbrace{1,\ldots,1}_{m\text{-times}},n_{k+m},\ldots,n_N)$. It was observed that in this case,
\bea
F_{\text{plet}}^{(n_1,\ldots,n_k,1,\ldots,1,n_{k+m},\ldots,n_N)}=\,F_{\text{plet}}^{(n_1,\ldots,n_k,1,n_{k+m},\ldots,n_N)}\,W(\tau,m,\epsilon_{1,2})^{m-1}\,.
\eea
The relative factor $W$ appearing in this relation is a quasi-modular form and is given by,
\begin{align}
&W(\tau,m,\epsilon_{1,2})=\frac{\theta_1(\tau,m+\epsilon_+)\theta_1(\tau,m-\epsilon_+)-\theta_1(\tau,m+\epsilon_-)\theta_1(\tau,m-\epsilon_-)}{\theta_1(\tau,\epsilon_1)\theta_2(\tau,\epsilon_2)}\,,\label{ExpansionFunctionW}
\end{align}
with $\epsilon_\pm =\tfrac{\epsilon_1\pm \epsilon_2}{2}$. Further information on this function and particular expansions that will be useful in the remainder of this paper can be found in Appendix~\ref{App:ExpansionW}.

\item {\it Self-similarity:}
In \cite{Hohenegger:2016eqy} it has been observed that in the NS-limit and in a certain region of the K\"ahler moduli space of $X_{N,1}$, the part of the free energy that counts only single particle states, becomes directly related to the BPS counting function for the LST with $N=1$ (and thus proportional to the free energy ${\cal F}^{\text{plet}}_{N=1,1}$). With the notation introduced above, the particular region in the moduli space is defined as
\bea 
t_1=t_2=\ldots=t_N=\frac{\rho}{N}
\eea
so that the M5-branes are all at equal distance from each other on the circle. In this region of the moduli space the free energy is a function of $(\tau,\rho,m)$ only and
\bea
{\cal F}^{\text{plet,NS}}_{N,1}\Big(\tfrac{\rho}{N},\cdots, \tfrac{\rho}{N},\tau,m,\epsilon_{1}\Big)={\cal F}^{\text{plet,NS}}_{N=1,1}\Big(\tfrac{\rho}{N},\tau,m,\epsilon_{1}\Big)\,,\label{SelfSimilarity}
\eea
where the Nekrasov-Shatashvili (NS) limit \cite{Nekrasov:2009rc,Mironov:2009uv} is defined as,
\bea
{\cal F}^{\text{plet,NS}}_{N,1}(t_{1},\cdots,t_{N},\tau,m,\epsilon_{1})=
\lim_{\epsilon_{2}\mapsto 0}\,\frac{\epsilon_{2}}{\epsilon_1}\,{\cal F}^{\text{plet}}_{N,1}(t_{1},\cdots t_{N},\tau,m,\epsilon_{1})\,.
\eea
\item {\it Hecke structures and torus orbifold:}
In \cite{Ahmed:2017hfr} contributions to the free energy coming from BPS states which carry the same charges under all the generators of the Cartan subalgebra of the gauge algebra $\mathfrak{a}_{N-1}$ were studied. In the language of the M-brane description, these correspond to configurations in which an equal number of M2-branes is stretched between any of the adjacent M5-branes. The degeneracy of such states are captured by,
\bea
{\cal F}^{\text{orb}}_{N,1}(\rho,\tau,m,\epsilon_{1,2})=\sum_{k}Q_{\rho}^{k}\,F_{\text{plet}}^{(k,\ldots, k)}(\tau,m,\epsilon_{1,2})\,.\label{DefOrbFreeEnerg}
\eea
 Based on the study of numerous examples (and supported by modular arguments), it has been conjectured in \cite{Ahmed:2017hfr} that\footnote{Here ${\cal F}^{\text{orb,NS}}_{N,1}$ is the NS-limit of ${\cal F}^{\text{orb}}_{N,1}$ in (\ref{DefOrbFreeEnerg}), \emph{i.e.}  ${\cal F}^{\text{orb,NS}}_{N,1}(\rho,\tau,m,\epsilon_1)=\lim_{\epsilon_{2}\mapsto 0}\,\frac{\epsilon_{2}}{\epsilon_1}\,{\cal F}^{\text{orb}}_{N,1}(\rho,\tau,m,\epsilon_{1,2})$.} $\mbox{exp}\Big({\cal F}^{\text{orb,NS}}_{N,1}(\rho,\tau,m,\epsilon_{1})\Big)$ is the partition function of a two dimensional torus orbifold theory whose target space is the symmetric product of the moduli space ${\cal M}_{1\cdots1}$ of monopole strings with charge $(1,\ldots,1)$
 \begin{align}
\mbox{exp}\Big({\cal F}^{\text{orb,NS}}_{N,1}(\rho,\tau,m,\epsilon_{1})\Big)&=\sum_{k}Q_{\rho}^{k}\,\chi_{\text{ell}}(\mbox{Sym}^{k}{\cal M}_{1\cdots1})=\prod_{k,n,\ell,r}\Big(1-Q_{\rho}^{k}Q_{\tau}^{n}Q_{m}^{\ell}\,q^{r}\Big)^{-c(kn,\ell,r,s)}\,.
\end{align}
Here $c(k,\ell,r)$ are the coefficients in the Fourier expansion of $\chi_{\text{ell}}({\cal M}_{1\cdots1})$
\bea
\chi_{\text{ell}}({\cal M}_{1\cdots1})&=&\sum_{k,\ell,r}c(k,\ell,r)\,Q_{\tau}^{k}\,Q_{m}^{\ell}\,q^{r}\,.
 \eea
 \end{itemize}
In this paper we shall discuss novel properties of the free energy, which (in a certain sense) 

\noindent
\begin{wrapfigure}{l}{0.52\textwidth}
\begin{center}
\vspace{0.1cm}
\scalebox{0.7}{\parbox{11.2cm}{\begin{tikzpicture}[scale = 1.50]
\draw[ultra thick] (-1,0) -- (0,0) -- (0,-1) -- (1,-1) -- (1,-2) -- (2,-2) -- (2,-2.5);
\node[rotate=315] at (2.5,-3) {\Huge $\cdots$};
\draw[ultra thick] (3,-3.5) -- (3,-4) -- (4,-4) -- (4,-5) -- (5,-5);
\draw[ultra thick] (0,0) -- (0.7,0.7);
\draw[ultra thick] (1,-1) -- (1.7,-0.3);
\draw[ultra thick] (2,-2) -- (2.7,-1.3);
\draw[ultra thick] (4,-4) -- (4.7,-3.3);
\draw[ultra thick] (0,-1) -- (-0.7,-1.7);
\draw[ultra thick] (1,-2) -- (0.3,-2.7);
\draw[ultra thick] (3,-4) -- (2.3,-4.7);
\draw[ultra thick] (4,-5) -- (3.3,-5.7);
\node at (-1.2,0) {\large {\bf $\mathbf a$}};
\node at (5.2,-5) {\large {\bf $\mathbf a$}};
\node at (0.85,0.85) {\large {$\mathbf 1$}};
\node at (1.85,-0.15) {\large {$\mathbf 2$}};
\node at (2.85,-1.15) {\large {$\mathbf 3$}};
\node at (4.85,-3) {\large {$\mathbf{N}$}};
\node at (-0.65,-2) {\large {$\mathbf{1}$}};
\node at (0.35,-3) {\large {$\mathbf{2}$}};
\node at (1.8,-4.5) {\large {$\mathbf{N-1}$}};
\node at (3.15,-6) {\large {$\mathbf{N}$}};
\node at (-0.5,0.25) {\large  {\bf $h_1$}};
\node at (0.5,-1.25) {\large  {\bf $h_2$}};
\node at (1.5,-1.75) {\large  {\bf $h_3$}};
\node at (3.5,-3.75) {\large  {\bf $h_N$}};
\node at (4.5,-4.75) {\large  {\bf $h_1$}};
\node at (-0.2,-0.5) {\large  {\bf $m$}};
\node at (0.8,-1.5) {\large  {\bf $m$}};
\node at (3.8,-4.5) {\large  {\bf $m$}};
\node at (0.6,0.25) {\large  {\bf $v$}};
\node at (1.6,-0.75) {\large  {\bf $v$}};
\node at (2.6,-1.75) {\large  {\bf $v$}};
\node at (4.6,-3.75) {\large  {\bf $v$}};
\node at (-0.1,-1.5) {\large  {\bf $v$}};
\node at (0.9,-2.5) {\large  {\bf $v$}};
\node at (2.9,-4.5) {\large  {\bf $v$}};
\node at (3.9,-5.5) {\large  {\bf $v$}};
\draw[ultra thick,red,<->] (1.05,0.95) -- (1.95,0.05);
\node[red,rotate=315] at (1.75,0.65) {{\large {\bf {$\widehat{a}_1$}}}};
\draw[ultra thick,red,<->] (2.05,-0.05) -- (2.95,-0.95);
\node[red,rotate=315] at (2.75,-0.35) {{\large {\bf {$\widehat{a}_2$}}}};
\draw[ultra thick,red,<->] (5.05,-3.05) -- (5.95,-3.95);
\node[red,rotate=315] at (5.75,-3.35) {{\large {\bf {$\widehat{a}_N$}}}};
%
%
%
\draw[dashed] (-0.7,-1.7) -- (-1.2,-2.2);
\draw[dashed] (5,-5) -- (3.25,-6.75);
\draw[ultra thick,red,<->] (3.2,-6.7) -- (-1.1,-2.2);
\node[red] at (0.9,-4.7) {{\large {\bf {$S$}}}};
\draw[dashed] (4,-5) -- (-1.5,-5);
\draw[dashed] (-0.7,-1.7) -- (-1.5,-1.7);
\draw[ultra thick,red,<->] (-1.5,-4.95) -- (-1.5,-1.75);
\node[red,rotate=90] at (-1.75,-3.3) {{\large {\bf{$R-NS$}}}};
\end{tikzpicture}}}
\caption{\sl Web diagram of $X_{N,1}$ labelled by the parameters $(\widehat{a}_{1},\ldots,\widehat{a}_{N}, S, R)$.}
\label{Fig:N2web}
\end{center}
${}$\\[-1.7cm]
\end{wrapfigure}
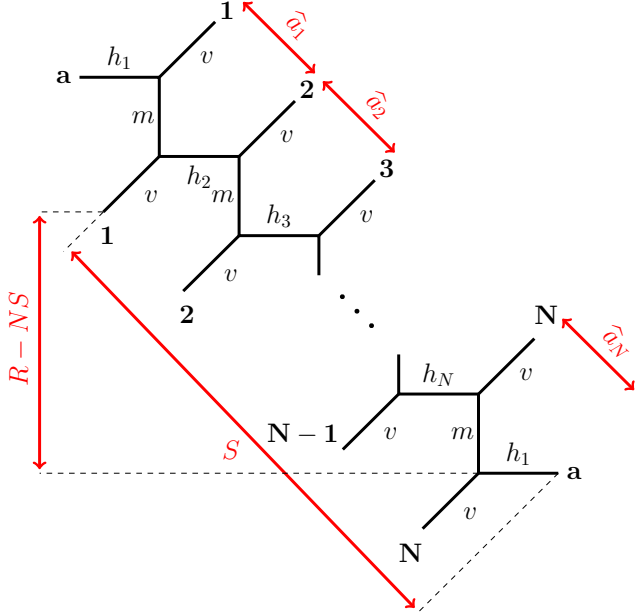 

\noindent
generalise some of the points mentioned above. However, to render some of these properties more clearly visible, we shall choose to slightly modify two important points:
\begin{itemize}
\item Instead of the basis $(t_1,\ldots,t_N,m,\tau)$, which is defined in \figref{Fig:N1web} as certain K\"ahler parameters of $X_{N,1}$, we prefer to work in a different basis given by the parameters $(\widehat{a}_1,\ldots,\widehat{a}_N,S,R) $: this basis was first introduced in \cite{Bastian:2017ing,Bastian:2017ary,Bastian:2018jlf} and allows for a more streamlined definition of some of the symmetries of the free energy. With respect to the web diagram of $X_{N,1}$, this basis is shown in \figref{Fig:N2web}. Furthermore, as was discussed in \cite{Bastian:2018jlf,Bastian:2019hpx}, it can be obtained from the basis $(t_{1},\cdots,t_{N},m,\tau)$, through the following linear transformation\footnote{We implicitly use $t_{N+1}=t_{1}$.} (with $v=\tau-m$).
\begin{align}
&R=\tau-2N\,m+N\,\rho\,,&&S=-m+\rho\,,&&\widehat{a}_{i}=t_{i+1}\,,&&\forall\hspace{0.25cm}i=1,\cdots,N\,.
\end{align}
This transformation is part of a symmetry group $\mathbb{G}_N\times \text{Dih}_N$, where $\text{Dih}_N$ is a subgroup of the Weyl group of the $U(N)$ gauge group and $\mathbb{G}_N$ is a (dihedral) symmetry group that is implied by a web of dualities of the little string theory (see \cite{Bastian:2018jlf} for more details). Since $\mathbb{G}_N\times \text{Dih}_N$ leaves the free energy invariant, the results discussed above also hold when formulated in the new basis $(\widehat{a}_1,\ldots,\widehat{a}_N,S,R)$. For further convenience we also introduce
\begin{align}
&Q_{R}=e^{2\pi i R}\,,&&Q_{S}=e^{2\pi i S}\,,&&Q_{\widehat{a}_{j}}=e^{2\pi i \widehat{a}_{j}}\hspace{0.5cm}\forall j=1,\ldots,N\,.
\end{align}
\end{itemize}

\begin{itemize}
\item Instead of $\mathcal{F}_{N,1}^{\text{plet}}$ in (\ref{FreeEnergy}) (which only counts single particle BPS states), we work with the full free energy
\begin{align}
{\cal F}_{N,1}(a_{1,\ldots,N},S,R,\epsilon_{1,2})=\ln \,\mathcal{Z}_{N,1}(\widehat{a}_{1,\ldots,N},S,R,\epsilon_{1,2})\,,\label{FullFreeEnergy}
\end{align}
${\cal F}_{N,1}$ can be expanded in powers of $Q_R$,
\begin{align}
{\cal F}_{N,1}(\widehat{a}_{1,\ldots,N},S,R;\epsilon_{1,2})&=\sum_{r} Q_R^r P_N^{(r)}(\widehat{a}_{1,\ldots,N},S;\epsilon_{1,2})\,,
\label{FreeEnergyExpansion}
\end{align}
where we can also expand the coefficients $P^{(r)}(\widehat{a}_{1,\ldots,N},S;\epsilon_{1,2})$ in powers of $\epsilon_{1,2}$
\begin{align}
P_N^{(r)}(\widehat{a}_{1,\ldots,N},S;\epsilon_{1,2})=\sum_{s_1,s_2}\epsilon_1^{s_1-1}\epsilon_2^{s_2-1}\,P^{(r)}_{N,(s_1,s_2)}(\widehat{a}_{1,\ldots,N},S)\,.\label{FreeEnergGen}
\end{align}
We will mostly be interested in the NS-limit, \emph{i.e.} $s_2=0$ and $s_1\in\mathbb{N}_{\text{even}}$. Finally, we can expand the $P^{(r)}_{N,(s_1,s_2)}$ in powers of $Q_{\widehat{a}_i}=e^{2\pi i \widehat{a}_i}$
\begin{align}
P^{(r)}_{N,(s_1,s_2)}(\widehat{a}_{1,\ldots,N},S)=\sum_{n_1,\ldots,n_N}Q_{\widehat{a}_1}^{n_1}\ldots Q_{\widehat{a}_N}^{n_N}\,P^{(r),\{n_1,\ldots,n_N\}}_{(s_1,s_2)}(S)\,.\label{PpartFree}
\end{align}
For later convenience, we will also use the notation $\underline{n}=\{n_1,\ldots,n_N\}$. From the $P^{(r),\underline{n}}_{N,(s_1,s_2)}$ we construct the following (a priori formal) object
\begin{align}
H^{(r),\{n_1,\ldots,n_N\}}_{(s_1,s_2)}(\rho,S)=\sum_{k=0}^\infty Q_\rho^k\, P^{(r),\{n_1+k,n_2+k,\ldots,n_N+k\}}_{N,(s_1,s_2)}(S)\,,\label{DefAbstractH}
\end{align}
where $Q_\rho=e^{2\pi i\sum_{j=1}^N\widehat{a}_j}$. The $P^{(r)}_{N,(s_1,s_2)}(\widehat{a}_{1,\ldots,N2},S)$ in (\ref{PpartFree}) are resummed as
\begin{align}
P^{(r)}_{N,(s_1,s_2)}(\widehat{a}_{1,\ldots,N},S)=H^{(r),\{0,\ldots,0\}}_{(s_1,s_2)}(\rho,S)+\sum'_{\underline{n}} H^{(r),\underline{n}}_{(s_1,s_2)}(\rho,S)\,Q_{\widehat{a}_1}^{n_1}\ldots Q_{\widehat{a}_N}^{n_N} \,.\label{ResumP}
\end{align}
Here the summation is over all $\underline{n}=\{n_1,\ldots,n_N\}\in(\mathbb{N}\cup\{0\})^N$ such that at least one of the $n_i=0$. Furthermore, we implicitly assume that $\widehat{a}_N=\rho-\sum_{i=1}^{N-1}\widehat{a}_i$.

\end{itemize}
\noindent
In the remainder of this paper we identify a limit in which the NS-limit of the free energy diverges but the residue of the second order pole counts BPS stats of a symmetric orbifold theory: For example, the partition function for the case $N=2$ is discussed in Appendix~\ref{App:N1PartitionFunction}. In this case the partition function has a pole at $\widehat{a}_{1}=\pm\, 2\epsilon_{+}$, while in the NS limit the free energy ${\cal F}_{N=2,1}$ has a pole at $\widehat{a}_{1}=\pm\,\epsilon_{1}$. Terms of different order in $Q_R$ have different order poles at $\widehat{a}_{1}=\pm\, \epsilon_{1}$ with different residues. If we expand the NS 
free energy in powers of $\epsilon_{1}$ then the coefficients have different order poles at $\widehat{a}_{1}=0$ with residues now shifted because of the $\epsilon_1$ 
expansion. The lowest order pole is of order two. 

On a technical level, just as in previous work, we rely on studying series expansions of examples for small values of $N$, which reveal certain patterns. However, since the corresponding computations are rather technical, we will summarise our observations in the following section, before presenting the computations for $N=2$, $N=3$ and $N=4$ respectively.


\section{Summary of Results}\label{Sect:Summary}

Due to the technical nature of some of the results of this paper, we provide a short overview of our main observations. For given $N$, we start by extracting the following $N$ functions from the (expansion coefficients of the) free energy  $P_{N,(2s,0)}^{(r)}(\widehat{a}_{1,\ldots,N},S)$ in (\ref{FreeEnergGen})
\begin{align}
&\factGE{i}{r}{2s,0}(\rho,S)=\frac{1}{(2\pi i)^{N}r^{i-1}}\sum_{\ell=0}^\infty Q_\rho^\ell \oint_0 d\widehat{a}_1\,\widehat{a}_1\oint_{-\widehat{a}_1}d\widehat{a}_2\,(\widehat{a}_1+\widehat{a}_2)\ldots\oint_{-\widehat{a}_1-\ldots-\widehat{a}_{i-2}} d\widehat{a}_{i-1}(\widehat{a}_1+\ldots+\widehat{a}_{i-1})\nonumber\\
&\hspace{2.5cm}\times \oint_0\frac{dQ_{\widehat{a}_{i}}}{Q_{\widehat{a}_{i}}^{1+\ell}}\ldots \oint_0\frac{dQ_{\widehat{a}_{N}}}{Q_{\widehat{a}_{N}}^{1+\ell}}\,P_{N,(2s,0)}^{(r)}(\widehat{a}_1,\ldots,\widehat{a}_N,S)\,,\hspace{2cm}\forall i=1,\ldots,N\,.\label{FactGenDefinition}
\end{align}
The latter can be resummed into a (formal) series expansion in $\epsilon_1$ \footnote{Although a priori it is a formal expansion but the $i=1$ case given in eq. (\ref{Ci1example}) and the $i=2$ example discussed in Appendix B for $N=2$ (see eq. (\ref{AppendixBEq})) shows that it is a Jacobi form involving $\theta_{1}(\rho,z)$ and its derivatives.}
\begin{align}
\factG{i}{r}(\rho,S,\epsilon_1)=\sum_{s=0}^\infty \epsilon_1^{2s-2i}\,\factGE{i}{r}{2s,0}(\rho,S)\,,\label{FactGenDefinitionExpand}
\end{align}
which defines a Jacobi form of weight zero and index $r\,N$.
\noindent
The contour integrals in (\ref{FactGenDefinition}) are

\begin{wrapfigure}{l}{0.41\textwidth}
\vspace{-0.5cm}
\begin{center}
\scalebox{0.8}{\parbox{8.4cm}{\begin{tikzpicture}[scale = 1]
\draw[very thick,fill=orange!25!white] (1.5,-1.25) -- (2.75,-2.5) -- (2.75,-1) -- (1.5,0.25) -- (1.5,-1.25);
\draw[very thick,fill=orange!25!white] (-1.5,-1.25) -- (-2.75,-2.5) -- (-2.75,-1) -- (-1.5,0.25) -- (-1.5,-1.25);
\draw[very thick,fill=orange!25!white] (1.2,1.95) -- (2.25,3) -- (2.25,1.5) -- (1.2,0.45) -- (1.2,1.95);
\draw[very thick,fill=orange!25!white] (-1.2,1.95) -- (-2.25,3) -- (-2.25,1.5) -- (-1.2,0.45) -- (-1.2,1.95);
\draw[very thick,fill=orange!25!white,yshift=0.3cm] (-3.8,-0.8) -- (-2.4,-0.8) -- (-2.4,0.9) -- (-3.8,0.9) -- (-3.8,-0.8);
\draw[very thick,fill=orange!25!white,yshift=0.3cm] (3.8,-0.8) -- (2.4,-0.8) -- (2.4,0.9) -- (3.8,0.9) -- (3.8,-0.8);
\draw [thick,blue,domain=-35:0,yshift=0.3cm] plot ({3*cos(\x)}, {1.75*sin(\x)});
\draw [thick,blue,domain=-37:0,yshift=0.4cm] plot ({3*cos(\x)}, {1.75*sin(\x)});
\draw [thick,blue,domain=-39:0,yshift=0.5cm] plot ({3*cos(\x)}, {1.75*sin(\x)});
\draw [thick,blue,domain=31:55,yshift=0.3cm] plot ({3*cos(\x)}, {1.75*sin(\x)});
\draw [thick,blue,domain=28:55,yshift=0.4cm] plot ({3*cos(\x)}, {1.75*sin(\x)});
\draw [thick,blue,domain=24:55,yshift=0.5cm] plot ({3*cos(\x)}, {1.75*sin(\x)});
\draw [thick,blue,domain=-135:-110,yshift=0.2cm] plot ({3*cos(\x)}, {1.75*sin(\x)});
\draw [thick,blue,domain=-135:-110,yshift=0.3cm] plot ({3*cos(\x)}, {1.75*sin(\x)});
\draw [thick,blue,domain=-135:-110,yshift=0.4cm] plot ({3*cos(\x)}, {1.75*sin(\x)});
%
\draw [thick,blue,domain=-70:-45,yshift=0.2cm] plot ({3*cos(\x)}, {1.75*sin(\x)});
\draw [thick,blue,domain=-70:-45,yshift=0.3cm] plot ({3*cos(\x)}, {1.75*sin(\x)});
\draw [thick,blue,domain=-70:-45,yshift=0.4cm] plot ({3*cos(\x)}, {1.75*sin(\x)});
\node[blue] at (0,-1.3) {\Large $\cdots$}; 
\draw [thick,red,domain=180:213,yshift=0.2cm] plot ({3*cos(\x)}, {1.75*sin(\x)});
\draw [thick,red,domain=180:215,yshift=0.3cm] plot ({3*cos(\x)}, {1.75*sin(\x)});
\draw [thick,red,domain=181:217,yshift=0.4cm] plot ({3*cos(\x)}, {1.75*sin(\x)});
\draw [thick,red,domain=182:219,yshift=0.5cm] plot ({3*cos(\x)}, {1.75*sin(\x)});
\draw [thick,red,domain=125:149,yshift=0.3cm] plot ({3*cos(\x)}, {1.75*sin(\x)});
\draw [thick,red,domain=125:153,yshift=0.4cm] plot ({3*cos(\x)}, {1.75*sin(\x)});
\draw [thick,red,domain=125:157,yshift=0.5cm] plot ({3*cos(\x)}, {1.75*sin(\x)});

\draw [thick,red,domain=62:75,yshift=0.6cm] plot ({3*cos(\x)}, {1.75*sin(\x)});
\draw [thick,red,domain=60:75,yshift=0.7cm] plot ({3*cos(\x)}, {1.75*sin(\x)});
\draw [thick,red,domain=66:75,yshift=0.3cm] plot ({3*cos(\x)}, {1.75*sin(\x)});
\draw [thick,red,domain=66:75,yshift=0.4cm] plot ({3*cos(\x)}, {1.75*sin(\x)});
\draw [thick,red,domain=64:75,yshift=0.5cm] plot ({3*cos(\x)}, {1.75*sin(\x)});
\draw [thick,red,domain=105:117,yshift=0.6cm] plot ({3*cos(\x)}, {1.75*sin(\x)});
\draw [thick,red,domain=105:113,yshift=0.3cm] plot ({3*cos(\x)}, {1.75*sin(\x)});
\draw [thick,red,domain=105:115,yshift=0.4cm] plot ({3*cos(\x)}, {1.75*sin(\x)});
\draw [thick,red,domain=105:116,yshift=0.5cm] plot ({3*cos(\x)}, {1.75*sin(\x)});
\node[red] at (0,2.2) {\Large $\cdots$}; 
\node at (2.6,2.9) {\bf $\mathbf{1}$};
\node at (-2.8,2.9) {\bf $\mathbf{i-1}$};
\node at (-4,1) {\bf $\mathbf{i}$};
\node at (-3.3,-2.4) {\bf $\mathbf{i+1}$};
\node at (4.1,1) {\bf $\mathbf{N}$};
\node at (3.45,-2.4) {\bf $\mathbf{N-1}$};
\node[blue] at (2.65,1.65) {$\ell$}; 
\node[blue] at (2.8,-0.7) {$\ell$}; 
\node[blue] at (1.4,-1.6) {$\ell$}; 
\node[blue] at (-1.4,-1.6) {$\ell$}; 
\node[red] at (-2.8,-0.7) {$n_i$}; 
\node[red] at (-1.6,1.4) {$n_{i-1}$}; 
\node[red] at (1.2,2.5) {$n_1$}; 
\node[red] at (-1.1,2.5) {$n_{i-2}$}; 
\draw [ultra thick,green!50!black,<-,domain=70:111,yshift=-0.1cm] plot ({3*cos(\x)}, {1.75*sin(\x)});
\draw [ultra thick,green!50!black,<-,domain=70:140,yshift=-0.4cm] plot ({3*cos(\x)}, {1.75*sin(\x)});
\draw [ultra thick,green!50!black,<-,domain=70:90,yshift=-0.7cm] plot ({3*cos(\x)}, {1.75*sin(\x)});
\draw[ultra thick,green!50!black] (0,1.05) to [out=180,in=45] (-2.,0.5) to [out=225,in=150] (-1.9,0.1);
\draw [ultra thick,<-,domain=55:70,yshift=1.2cm] plot ({3*cos(\x)}, {1.75*sin(\x)});
\draw [ultra thick,dashed,domain=70:80,yshift=1.2cm] plot ({3*cos(\x)}, {1.75*sin(\x)});
\node at (1.1,3.2) {$\widehat{a}_1$};
\draw [ultra thick,<-,domain=125:110,yshift=1.2cm] plot ({3*cos(\x)}, {1.75*sin(\x)});
\draw [ultra thick,dashed,domain=110:100,yshift=1.2cm] plot ({3*cos(\x)}, {1.75*sin(\x)});
\node at (-1.1,3.2) {$\widehat{a}_{i-2}$};
\draw [ultra thick,<->,domain=130:177,yshift=1.2cm] plot ({3*cos(\x)}, {1.75*sin(\x)});
\node at (-3.3,2.1) {$\widehat{a}_{i-1}$};
\draw [ultra thick,<->,domain=185:228,yshift=1.2cm] plot ({3*cos(\x)}, {1.75*sin(\x)});
\node at (-3.2,0.6) {$\widehat{a}_{i}$};
\draw [ultra thick,<-,domain=235:250,yshift=1.2cm] plot ({3*cos(\x)}, {1.75*sin(\x)});
\draw [ultra thick,dashed,domain=250:260,yshift=1.2cm] plot ({3*cos(\x)}, {1.75*sin(\x)});
\node at (-0.8,-0.2) {$\widehat{a}_{i+1}$};
\draw [ultra thick,->,domain=285:305,yshift=1.2cm] plot ({3*cos(\x)}, {1.75*sin(\x)});
\draw [ultra thick,dashed,domain=275:285,yshift=1.2cm] plot ({3*cos(\x)}, {1.75*sin(\x)});
\node at (0.8,-0.1) {$\widehat{a}_{N-2}$};
\draw [ultra thick,<->,domain=311:344,yshift=1.2cm] plot ({3*cos(\x)}, {1.75*sin(\x)});
\node at (1.9,0.4) {$\widehat{a}_{N-1}$};
\draw [ultra thick,<->,domain=48:0.3,yshift=1.2cm] plot ({3*cos(\x)}, {1.75*sin(\x)});
\node at (3,2.2) {$\widehat{a}_{N}$};
\end{tikzpicture}}}
\end{center} 
\caption{{\it Brane configuration for extracting $\factG{i}{r}$}}
\label{fig:BraneCollapse}
\vspace{-0.5cm}
\end{wrapfigure}
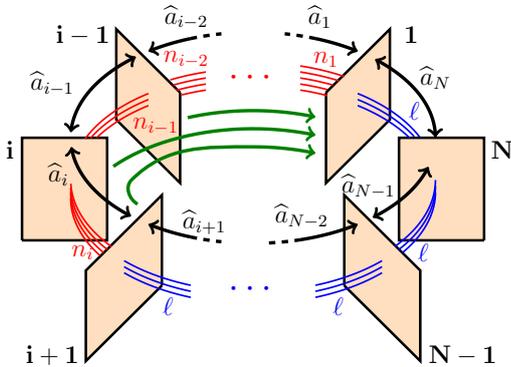

\noindent
understood to be small circles centred around the specified points.\footnote{We implicitly assume here that we are in a generic point in the moduli space, such that the free energy has isolated poles with respect to the various variables. In this case, the precise form of the contour is not crucial, and the prescription is simply designed to extract their residues.} From a mathematical point of view, this extracts and combines specific coefficients in a mixed Fourier-Taylor expansion of $P_N^{(r)}$: with respect to the variables $\widehat{a}_j$ for $1\leq j\leq i-1$ the prescription computes successively the coefficient of the second order pole at $\widehat{a}_j=-\widehat{a}_1-\ldots-a_{j-1}$. For the variables $\widehat{a}_j$ for $i\leq j\leq N$ the prescription sums (weighted by $Q_\rho^\ell$) the coefficients of the term $Q_{\widehat{a}_i}^{\ell} Q_{\widehat{a}_{i+1}}^{\ell}\ldots Q_{\widehat{a}_N}^{\ell}$ in a Fourier series expansion in terms of $Q_{\widehat{a}_j}=e^{2\pi i \widehat{a}_j}$. From a physical perspective, the latter prescription combines contributions from M2-brane configurations (weighted by $Q_\rho^\ell$), in which an equal number of $\ell$ M2-branes is stretched between the M5-branes $j$ and $j+1$ for $i\leq j\leq N$,\footnote{In this notation it is understood that the M5 brane $N+1$ is in fact the M5-brane $1$.} (see \figref{fig:BraneCollapse} for a schematic picture of a generic configuration, as well as the figures~\figref{fig:PartM5braneN2Config}, \figref{fig:M5braneConfigN3} and \figref{fig:M5braneConfigN4} for examples of the cases $N=2,3,4$). Concerning the remaining M5-branes, we consider a region in the moduli space, in which they all form a  stack on top of each other: from the resulting divergent expression, the contour integrals for $\widehat{a}_j$ for $1\leq j\leq i-1$ in (\ref{FactGenDefinition}) extract successively the poles of the form $(\widehat{a}_1+\widehat{a}_2+\ldots+\widehat{a}_{i-1})^{-2}$, $(\widehat{a}_1+\widehat{a}_2+\ldots+\widehat{a}_{i-2})^{-2}$, $\ldots$, $\widehat{a}_1^{-2}$. The total order of the divergence selected in this fashion is $2i-2$: from all the examples we have explicitly calculated, this seems to be the highest singularity that appears in the prepotential at leading instanton order. This matches the analysis of the singularities of the partition function in appendix~\ref{App:N1PartitionFunction}.

The brane configuration of parallel M5-branes labelled $1, \cdots ,i-1$ stacked on top of each other (which is used to define the ${\cal C}_{i}^{N,(r)}$) can also be understood geometrically: we recall that the Calabi-Yau threefold $X_{N,1}$ dual to the generic M5-brane configuration has a resolved $A_{N-1}$ singularity. The case in which $i-1$ M5-branes are on top of each other corresponds to a partial resolution of the $A_{N-1}$ singularity so that it becomes a $A_{i-1}$ singularity,
\bea
A_{N-1}\mapsto A_{i-2}\times \underbrace{A_{N-i+1}}_{\text{resolved}}.
\eea
The functions ${\cal C}^{N,(r)}_{i=1}(\rho,S,\epsilon_1)$ were already studied before in \cite{Ahmed:2017hfr}. As already briefly discussed in the previous section, there (based on the study of numerous explicit examples) the following relation was observed (we are using the notation introduced above)
\begin{align}
&{\cal C}^{N,(r)}_{i=1}(\rho,S,\epsilon_1)=\mathcal{H}_r\left[{\cal C}^{N,(1)}_{i=1}(\rho,S,\epsilon_1)\right]\,,&&\forall r\geq 1\label{HeckeNPaperRev}
\end{align}
where $\mathcal{H}_r$ is the $r$-th Hecke operator (see eq.~(\ref{DefHecke}) in appendix~\ref{App:ModularStuff} for the definition) and 
\begin{align}
{\cal C}^{N,(1)}_{i=1}(\rho,S,\epsilon_1)=
\frac{\theta_{+}\,\theta_{-}(\theta_{+}\theta_{-}'-\theta_{-}\theta_{+}')^{N-1}}{\eta(\rho)^{3N}\,\theta_{1}(\rho,\epsilon_{1})^{N}}
\label{Ci1example}
\end{align}
with $\theta_{\pm}=\theta_{1}(\rho,-S+\rho\pm \frac{\epsilon_{1}}{2})$. 
In this paper, based on a detailed study of $\factG{i}{r}(\rho,S,\epsilon_1)$ for $N=2$ and $N=3$ (and partially also for $N=4$), we provide evidence that (\ref{HeckeNPaperRev}) can be generalised to all the functions defined in (\ref{FactGenDefinition}):
\begin{align}
&{\cal C}^{N,(r)}_{i,(2s,0)}(\rho,S)=\mathcal{H}_r\left[{\cal C}^{N,(1)}_{i,(2s,0)}(\rho,S)\right]\,,&&\begin{array}{l}\forall r\geq 1\,,\\ \forall i=1,\ldots,N\,, s\geq 0.\end{array}\label{HeckeNCon}
\end{align}
The ${\cal C}^{N,(1)}_{i,(2s,0)}$ are index $N$ Jacobi forms of weight $2s-2i$ and are coefficients in the $\epsilon_1$ expansion of ${\cal C}^{N,(1)}_{i}(\rho,S,\epsilon_1)$ which is a Jacobi form in two complex variables $(S,\epsilon_1)$ \cite{krieg}. The action of the Hecke operator ${\cal H}_{r}$ extends to the case of multiple complex variables as given in eq.~(\ref{DefHeckeMultiple}) in appendix~\ref{App:ModularStuff}. From eq. \ref{HeckeNCon} it then follows that,
\begin{align}
&{\cal C}^{N,(r)}_{i}(\rho,S,\epsilon_1)=\mathcal{H}_r\left[{\cal C}^{N,(1)}_{i}(\rho,S,\epsilon_1)\right]\,,&&\begin{array}{l}\forall r\geq 1\,,\\ \forall i=1,\ldots,N\,.\end{array}\label{HeckeNCon2}
\end{align}
Assuming that this relation indeed holds for generic $r$, a generating function can be formed capturing the degeneracies BPS states in this subsector
\begin{align}
Z^{(N)}_{i}(R, \rho,S,\epsilon_1)&=\text{exp}\Big(\sum_{r=1}^\infty Q_{R}^{Nr}\,{\cal C}^{N,(r)}_{i}(\rho,S,\epsilon_{1})\Big)\,.\label{DefSumZn}
\end{align}
The eq.(\ref{HeckeNCon2}) together with the fact that the ``seed function" ${\cal C}^{N,(1)}_{i}(\rho,S,\epsilon_1)$ is a weight zero (index $N$) Jacobi form implies that $Z^{(N)}_{i}(R,\rho,S,\epsilon)$
can be interpreted as the partition functions of symmetric orbifold conformal field theories.\footnote{The power $Nr$ of $Q_R$ in the summation in (\ref{DefSumZn}) is chosen such that $Z^{(N)}_{i}$ can be recognised more readily as a paramodular form with respect to the group $\Sigma_N^*$ (see appendix~\ref{App:Paramodular}).} These symmetric orbifold theories arise from a special subsector of the full theory and are extracted using the contour integrals involving $a_{1,\cdots,i-1}$ in the NS limit. The fact that in this special subsector  the moduli space of charge $r$ instantons can be realised as symmetric product of $r$ charge $1$ instantons suggests that this special subsector is getting contributions from well separated instantons only \cite{Khoze:1998gy,Dorey:2002ik}.

Furthermore, the study of the above mentioned examples has brought to light numerous interesting patterns, which suggest additional interesting properties of the functions $\factG{i}{r}(\rho,S,\epsilon_1)$: it was already remarked in \cite{Hohenegger:2019tii} that to order $\mathcal{O}(Q_R)$ (\emph{i.e.} for $r=1$), the free energy can be decomposed into simpler building blocks, which are given by the expansion coefficients of the free energy for $N=1$ (see appendix~\ref{App:N1Expansion} for the definition) as well as the expansion of the function $W$ (see (\ref{ExpansionFunctionW}) in appendix~\ref{App:ExpansionW}), which governs the counting of BPS states of a single M5-brane with single M2-branes ending on it on either side (see sections~\ref{Sect:N2FactOrd1} and \ref{FactorisationR1N3} for the details in the cases $N=2$ and $N=3$ respectively). This decomposition is also reflected at the level of the functions $\factG{i}{r=1}(\rho,S,\epsilon_1)$ (and accordingly also for their expansion coefficients $\factGE{i}{r=1}{2s,0}(\rho,S)$), for which we find order by order in $\epsilon_1$
\begin{align}
&\factG{i}{r=1}(\rho,S,\epsilon_1)=\varkappa^{(1)}_{i,N}\,(\buildH{1}(\rho,S,\epsilon_1))^i\,(\buildW{1}(\rho,S,\epsilon_1))^{N-i}\,,&&\forall i=1,\ldots,N\,,\label{FactorisationO1}
\end{align}
where $\varkappa^{(1)}_{i,N}$ is a numerical factor: from the study of the examples $N=2,3,4$ examples we conjecture that the factor $\varkappa^{(r)}_{i,N}$ only depends on $i$ and $N$. Modulo the factor $\varkappa^{(1)}_{i,N}$ the functions ${\cal C}_{i}^{N,(r=1)}(\rho,S,\epsilon_1)$ satisfy the recursive relation:
\bea
{\cal C}_{i}^{N+1,(r=1)}(\rho,S,\epsilon_1)&\sim&{\cal C}_{i}^{N,(r=1)}(\rho,S,\epsilon_1)\,W^{(1)}_{NS}(\rho,S,\epsilon_1)\,,\\\nonumber
{\cal C}_{i+1}^{N+1,(r=1)}(\rho,S,\epsilon_1)&\sim&{\cal C}_{i}^{N,(r=1)}(\rho,S,\epsilon_1)\,H^{(1)}_{N=1}(\rho,S,\epsilon_1)
\eea
Starting from a configuration of $(N+1)$ M5-branes with $i$ of them collapsed to form a stack, the first recursion relation suggests that the BPS states that contribute to the poles in ${\cal C}_{i}^{N+1,(r=1)}$ can be counted from a similar configuration, where we remove one of the M5-branes that is not part of the stack and it is related to ${\cal C}_{i}^{N,(r=1)}$ through multiplication with $W^{(1)}_{NS}(\rho,S,\epsilon_1)$. Similarly, the second recursion relation suggests that the effect of removing an M5-brane from the stack of collapsed branes on the counting function ${\cal C}_{i}^{N+1,(r=1)}$ is described by multiplying  ${\cal C}_{i}^{N,(r=1)}$ with the function $H^{(1)}_{N=1}(\rho,S,\epsilon_1)$.

To higher orders in $Q_R$ (\emph{i.e.} for $r>1$), the decomposition is more complicated. While we did not manage to identify all coefficients uniquely\footnote{They are, however, implicitly given through the relation (\ref{HeckeNCon}).}, the examples we have studied suggest
\begin{align}
\factG{i}{r}(\rho,S,\epsilon_1)=\varkappa^{(r)}_{i,N}\,(\buildH{r}(\rho,S,\epsilon_1))^i\,(\buildW{r}(\rho,S,\epsilon_1))^{N-i}+\mathfrak{R}_i^{N,(r)}(\rho,S,\epsilon_1)\,,\label{FactorisationOr}
\end{align}
where the rest-term $\mathfrak{R}_i^{N,(r)}$ itself can be decomposed into combinations of $\buildH{r}(\rho,S,\epsilon_1)$ where the coefficients are quasi-modular forms that only depend on $\rho$ and which can be written as harmonic polynomials in the Eisenstein series $(E_2(\rho),E_4(\rho),E_6(\rho))$ (see appendix~\ref{App:ModularStuff} for the definitions). Moreover, different such polynomials are related through derivatives with respect to the Eisenstein series $E_2$, in the style of holomorphic anomaly equations. We refer the reader to sections~\ref{Sect:FactorisationN2} and \ref{Sect:FactorisationN3} for the details in the cases $N=2$ and $N=3$ respectively. 

In the following sections we shall present detailed computations for $N=2$ and $N=3$ (and partially also $N=4$), which support the observations just outlined. After that we shall conclude in section~\ref{Sect:Conclusion}.

\section{Little String Theory with $N=2$}\label{Sect:CaseN2}
The simplest non-trivial example is to consider a model of Little Strings, which is engineered by two M5-branes on a circle that probe a flat $\mathbb{R}^4$ transverse space\footnote{The partition function function of the case when the transverse space is $\mathbb{C}^{2}/\mathbb{Z}_{M}$ is given in \cite{Haghighat:2013tka, Hohenegger:2013ala}. It would be interesting to see if the Hecke structure we study in this paper is also present for $M>1$.}. 
\subsection{Decomposition of the Free Energy}
As explained above, the partition function and free energy of this particular LST is captured by the topological string on the toric Calabi-Yau threefold, $X_{2,1}$, whose web-diagram is shown 

\begin{wrapfigure}{l}{0.38\textwidth}
\begin{center}
\vspace{-0.5cm}
\scalebox{0.7}{\parbox{6.6cm}{\begin{tikzpicture}[scale = 1.45]
\draw[ultra thick] (-1,0) -- (0,0) -- (0,-1) -- (1,-1) -- (1,-2) -- (2,-2);
\draw[ultra thick] (0,0) -- (0.7,0.7);
\draw[ultra thick] (1,-1) -- (1.7,-0.3);
\draw[ultra thick] (0,-1) -- (-0.7,-1.7);
\draw[ultra thick] (1,-2) -- (0.3,-2.7);
\node at (-1.2,0) {\large {\bf $\mathbf a$}};
\node at (2.2,-2) {\large {\bf $\mathbf a$}};
\node at (0.85,0.85) {\large {$\mathbf 1$}};
\node at (1.85,-0.15) {\large {$\mathbf 2$}};
\node at (-0.65,-2) {\large {$\mathbf{1}$}};
\node at (0.35,-3) {\large {$\mathbf{2}$}};
\node at (-0.5,0.25) {\large  {\bf $h_1$}};
\node at (0.5,-1.25) {\large  {\bf $h_2$}};
\node at (-0.2,-0.5) {\large  {\bf $m$}};
\node at (0.8,-1.5) {\large  {\bf $m$}};
\node at (0.6,0.25) {\large  {\bf $v$}};
\node at (1.6,-0.75) {\large  {\bf $v$}};
\node at (-0.1,-1.5) {\large  {\bf $v$}};
\node at (0.9,-2.5) {\large  {\bf $v$}};
\draw[ultra thick,red,<->] (1.05,0.95) -- (1.95,0.05);
\node[red,rotate=315] at (1.75,0.65) {{\large {\bf {$\widehat{a}_1$}}}};
\draw[ultra thick,red,<->] (2.05,-0.05) -- (2.95,-0.95);
\node[red,rotate=315] at (2.75,-0.35) {{\large {\bf {$\widehat{a}_2$}}}};
\draw[dashed] (-0.7,-1.7) -- (-1.2,-2.2);
\draw[dashed] (2,-2) -- (0.25,-3.75);
\draw[ultra thick,red,<->] (0.3,-3.6) -- (-1.1,-2.2);
\node[red] at (-0.7,-3.2) {{\large {\bf {$S$}}}};
\end{tikzpicture}}}
\caption{\sl Web diagram of $X_{2,1}$.}
\label{Fig:21web}
\vspace{-0.4cm}
\end{center}
\end{wrapfigure}
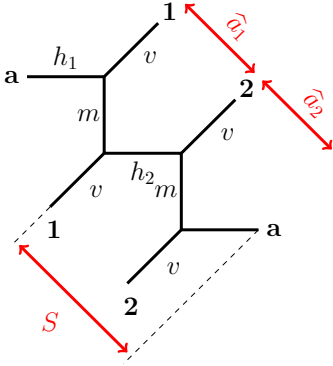 

\noindent
in \figref{Fig:21web}. Here, we use a basis of  K\"ahler parameters $(R,S,\rho,\widehat{a}_1)$ where in addition to the parameters given in the figure, we have
\begin{align}
&\rho=\widehat{a}_1+\widehat{a}_2\,,&&R-2S=v-m\,.
\end{align}
Starting from the partition function $\mathcal{Z}_{2,1}$, we define the free energy
\begin{align}
\mathcal{F}_{2,1}(\widehat{a}_{1,2},S,R,\epsilon_{1,2})=\log \mathcal{Z}_{2,1}(\widehat{a}_{1,2},S,R,\epsilon_{1,2})\,.
\end{align}
We decompose the latter in terms of $H^{(r),\{n,0\}}_{(s_1,s_2)}(\rho,S)$ (for $n\in\mathbb{N}\cup\{0\}$) as described in Section~\ref{Sect:LSTFree}. Upon using the symmetries of the former, the summation in (\ref{ResumP}) becomes
\begin{align}
P^{(r)}_{2,(s_1,s_2)}(\widehat{a}_{1,2},S)=H^{(r),\{0,0\}}_{(s_1,s_2)}(\rho,S)+\sum_{n=1}^\infty H^{(r),\{n,0\}}_{(s_1,s_2)}(\rho,S)\,\left(Q_{\widehat{a}_1}^n+\frac{Q_\rho^n}{Q_{\widehat{a}_1}^n}\right)\,.\label{DefPN2Ord}
\end{align}
In the following we shall only discuss the so-called NS-limit \cite{Nekrasov:2009rc,Mironov:2009uv}, \emph{i.e.} we consider $s_2=0$. Only for $n=0$ (which corresponds to the part of the free energy discussed in \cite{Ahmed:2017hfr}), the $H^{(r),\{n,0\}}_{(s,0)}$ are (quasi) Jacobi forms. For $n>0$, the $H^{(r),\{n,0\}}_{(s,0)}$ are no longer modular objects. However, following \cite{Bastian:2019wpx,Hohenegger:2019tii}, based on studying series expansions in $Q_\rho$ (and exploiting certain patterns arising in the expansion coefficients) we can conjecture the following generic form\footnote{We have verified that these expressions agree with an expansion of $P^{(r)}_{2,(s_1,s_2)}$ in (\ref{DefPN2Ord}) following from the general definition (\ref{FullFreeEnergy}) and (\ref{FreeEnergyExpansion}) in terms of the partition function defined in (\ref{TopStringAppend}) and (\ref{Zlambdas}), up to order $\mathcal{O}(Q_\rho^{30})$ for $r=1$, $\mathcal{O}(Q_\rho^{16})$ for $r=2$ and $\mathcal{O}(Q_\rho^{12})$ for $r=3$ and up to $2s=8$.}
\begin{align}
H^{(r),\{n,0\}}_{(2s,0)}(\rho,S)=\left\{\begin{array}{lcl} \mathfrak{h}_{0,(2s)}^{(r)}(\rho,S)& \text{for} & n=0\\ \frac{1}{1-Q_\rho^n}\,\sum_{k=1}^{rs+1}\,n^{2k-1}\,\mathfrak{h}^{(r)}_{k,(2s)}(\rho,S) & \text{for} & n>0\end{array}\right.\,,\label{HfunctionsN2}
\end{align}
where $\mathfrak{h}^{(r)}_{k,(2s)}$ is a (quasi-)Jacobi form of index $2r$ and weight $2s-2-2k$. Using the standard Jacobi-forms $\phi_{-2,1}$ and $\phi_{0,1}$ (see appendix~\ref{App:ModularStuff} for the definition), they can be cast into the form
\begin{align}
\mathfrak{h}_{k,(2s)}^{(r)}(\rho,S)=\sum_{a=0}^{2r}h_{a,k,(2s)}^{(r)}\,\left(\phi_{-2,1}(\rho,S)\right)^a\,\left(\phi_{0,1}(\rho,S)\right)^{2r-a}\,,
\end{align}
where $h_{a,k,(2s)}^{(r)}$ are (quasi-)modular forms of weight $2s-2-2k+2a$ and depth $sr+\delta_{k,0}$ which can be expressed as homogeneous polynomials of the Eisenstein series $\{E_{2,4,6}\}$. For later convenience, the coefficients $h_{a,k,(2s)}^{(r)}$ for $r=1$, $r=2$ and $r=3$ are tabulated in Tables~\ref{Tab:CoeffsExpr1}, \ref{Tab:CoeffsExpr2} and \ref{Tab:CoeffsExpr3}, respectively.

\begin{table}[htbp]
\begin{center}
\begin{tabular}{|c||c||c|c|c|}\hline
&&&&\\[-12pt]
$s$ & $k$ & $h^{(r=1)}_{0,k,(2s)}$ & $h^{(r=1)}_{1,k,(2s)}$ & $h^{(r=1)}_{2,k,(2s)}$\\[4pt]\hline\hline
&&&&\\[-12pt]
$0$ & $0$ & $0$ & $-\frac{1}{12}$ & $-\frac{E_2}{6}$\\[4pt]\hline
&&&&\\[-12pt]
& $1$ & $0$ & $0$ & $-2$\\[4pt]\hline\hline
&&&&\\[-12pt]
$1$ & $0$ & $\frac{1}{1152}$ & $0$ & $\frac{E_4-2 E_2^2}{288} $\\[4pt]\hline
&&&&\\[-12pt]
& $1$ & $0$ & $\frac{1}{24}$ & $-\frac{E_2 }{12}$\\[4pt]\hline
&&&&\\[-12pt]
 & $2$ & $0$ & $0$ & $\frac{1}{3}$\\[4pt]\hline\hline
 &&&&\\[-12pt]
$2$ & $0$ & $\frac{E_2 }{55296}$ & $\frac{5 E_2 ^2-7 E_4 }{69120}$ & $\frac{-10 E_2 ^3-3 E_2  E_4 +4 E_6 }{69120}$\\[4pt]\hline
 &&&&\\[-12pt]
& $1$ & $-\frac{1}{4608}$ & $\frac{E_2 }{576}$ & $-\frac{10 E_2 ^2+13 E_4 }{5760}$\\[4pt]\hline
 &&&&\\[-12pt]
 & $2$ & $0$ & $-\frac{1}{144}$ & $\frac{E_2 }{72}$\\[4pt]\hline
  &&&&\\[-12pt]
 & $3$ & $0$ & $0$ & $-\frac{1}{60}$\\[4pt]\hline\hline
  &&&&\\[-12pt]
$3$ & $0$ & $\frac{E_4 }{2211840}$ & $\frac{35 E_2 ^3-21 E_2  E_4 -29 E_6 }{17418240}$ & $\frac{-70 E_2 ^4-168 E_2 ^2 E_4 -8 E_2  E_6 +123 E_4 ^2}{34836480}$\\[4pt]\hline
& $1$ & $-\frac{E_2 }{110592}$ & $\frac{E_2 ^2+E_4 }{27648}$ & $\frac{-70 E_2 ^3-273 E_2  E_4 -92 E_6 }{2903040}$\\[4pt]\hline
  &&&&\\[-12pt]
 & $2$ & $\frac{1}{27648}$ & $-\frac{E_2 }{3456}$ & $\frac{10 E_2 ^2+13 E_4 }{34560}$\\[4pt]\hline
   &&&&\\[-12pt]
 & $3$ & $0$ & $\frac{1}{2880}$ & $-\frac{E_2 }{1440}$\\[4pt]\hline
    &&&&\\[-12pt]
 & $4$ & $0$ & $0$ & $\frac{1}{2520}$\\[4pt]\hline\hline
 &&&&\\[-12pt]
 $4$ & $0$ & $\frac{118 E_6+105 E_2  E_4-70 E_2 ^3  }{13377208320}$ & $\frac{175 E_2 ^4+210 E_2 ^2 E_4 -130 E_2  E_6 -381 E_4 ^2}{5573836800}$ & $\frac{1682 E_{10}-350 E_2 ^5-2030 E_2 ^3 E_4 -1000 E_2 ^2 E_6 +177 E_2  E_4 ^2 }{16721510400}$\\[4pt]\hline
     &&&&\\[-12pt]
  & $1$ & $\frac{-10 E_2 ^2-7 E_4 }{53084160}$ & $\frac{10 E_2 ^3+30 E_2  E_4 +11 E_6 }{19906560}$ & $\frac{-350 E_2 ^4-2730 E_2 ^2 E_4 -1840 E_2  E_6 -2283 E_4 ^2}{1393459200}$\\[4pt]\hline
&&&&\\[-12pt]
  & $2$ & $\frac{E_2 }{663552}$ & $\frac{-E_2 ^2-E_4 }{165888}$ & $\frac{70 E_2 ^3+273 E_2  E_4 +92 E_6 }{17418240}$\\[4pt]\hline
  &&&&\\[-12pt]
  & $3$ & $-\frac{1}{552960}$ & $\frac{E_2 }{69120}$ & $\frac{-10 E_2 ^2-13 E_4 }{691200}$\\[4pt]\hline
 &&&&\\[-12pt]
  & $4$ & $0$ & $-\frac{1}{120960}$ & $\frac{E_2 }{60480}$\\[4pt]\hline
   &&&&\\[-12pt]
  & $5$ & $0$ & $0$ & $-\frac{1}{181440}$\\[4pt]\hline
\end{tabular}
\end{center}
\caption{Expansion coefficients $h_{a,k,(2s)}^{(r=1)}$.}
\label{Tab:CoeffsExpr1}
\end{table}

\begin{table}[htbp]
\begin{center}
\begin{tabular}{|c||c||c|c|c|c|c|}\hline
&&&&\\[-12pt]
$s$ & $k$ & $h^{(r=2)}_{0,k,(2s)}$ & $h^{(r=2)}_{1,k,(2s)}$ & $h^{(r=2)}_{2,k,(2s)}$ & $h^{(r=2)}_{3,k,(2s)}$ & $h^{(r=2)}_{4,k,(2s)}$ \\[4pt]\hline\hline
&&&&&&\\[-12pt]
$0$ & $0$ & $0$ & $-\frac{1}{4608}$ & $-\frac{E_2 }{1152}$ & $-\frac{E_4 }{1152}$ & $\frac{E_6 -E_2  E_4 }{144}$\\[4pt]\hline
&&&&&&\\[-12pt]
 & $1$ & $0$ & $0$ & $-\frac{1}{96}$ & $0$ & $-\frac{E_4 }{12}$\\[4pt]\hline
&&&&&&\\[-12pt]
& $2$ & $0$ & $0$ & $0$ & $-\frac{1}{12}$ & $0$\\[4pt]\hline
&&&&&&\\[-12pt]
& $3$ & $0$ & $0$ & $0$ & $0$ & $-\frac{1}{24}$\\[4pt]\hline\hline
&&&&&&\\[-12pt]
$1$ & $0$ & $\frac{1}{442368}$ & $0$ & $\frac{5 E_4 -4 E_2 ^2}{55296}$ & $\frac{4 E_2  E_4 -3 E_6 }{6912}$ & $\frac{-16 E_2 ^2 E_4 -16 E_2  E_6 +37 E_4 ^2}{27648}$\\[4pt]\hline
&&&&&&\\[-12pt]
 & $1$ & $0$ & $\frac{1}{4608}$ & $-\frac{E_2 }{1152}$ & $\frac{E_4 }{128}$ & $-\frac{E_2  E_4 +2 E_6 }{144}$\\[4pt]\hline
&&&&&&\\[-12pt]
& $2$ & $0$ & $0$ & $\frac{1}{128}$ & $-\frac{E_2 }{144}$ & $\frac{53 E_4 }{1440}$\\[4pt]\hline
&&&&&&\\[-12pt]
& $3$ & $0$ & $0$ & $0$ & $\frac{13}{576}$ & $-\frac{E_2 }{288}$\\[4pt]\hline
&&&&&&\\[-12pt]
& $4$ & $0$ & $0$ & $0$ & $0$ & $\frac{1}{120}$\\[4pt]\hline\hline
&&&&&&\\[-12pt]
$2$ & $0$ & $\frac{E_2 }{32\cdot 24^4}$ & $\frac{10 E_2 ^2-21 E_4 }{13271040}$ & $\frac{87 E_6-20 E_2 ^3-59 E_2  E_4  }{6635520}$ & $\frac{170 E_2 ^2 E_4 +140 E_2  E_6 -251 E_4 ^2}{3317760}$ & $\frac{E_4(746 E_6-80 E_2 ^3) -240 E_2 ^2 E_6 -351 E_2  E_8 }{3317760}$\\[4pt]\hline
&&&&&&\\[-12pt]
& $1$ & $\frac{-1}{884736}$ & $\frac{E_2 }{55296}$ & $\frac{-20 E_2 ^2-109 E_4 }{552960}$ & $\frac{9 E_2  E_4 +13 E_6 }{13824}$ & $\frac{-80 E_2 ^2 E_4 -320 E_2  E_6 -721 E_4 ^2}{276480}$\\[4pt]\hline
&&&&&&\\[-12pt]
& $2$ & $0$ & $-\frac{5}{36864}$ & $\frac{E_2 }{1536}$ & $\frac{-40 E_2 ^2-593 E_4 }{138240}$ & $\frac{371 E_2  E_4 +535 E_6 }{120960}$\\[4pt]\hline
&&&&&&\\[-12pt]
& $3$ & $0$ & $0$ & $-\frac{301}{184320}$ & $\frac{13 E_2 }{6912}$ & $\frac{-20 E_2 ^2-901 E_4 }{138240}$\\[4pt]\hline
&&&&&&\\[-12pt]
& $4$ & $0$ & $0$ & $0$ & $-\frac{7}{2880}$ & $\frac{E_2 }{1440}$\\[4pt]\hline
&&&&&&\\[-12pt]
& $5$ & $0$ & $0$ & $0$ & $0$ & $-\frac{73}{120960}$\\[4pt]\hline
%
\end{tabular}
\end{center}
\caption{Expansion coefficients $h_{a,k,(2s)}^{(r=2)}$.}
\label{Tab:CoeffsExpr2}
\end{table}

\begin{table}[htbp]
\begin{center}
\rotatebox{90}{\parbox{23.5cm}{\begin{tabular}{|c||c||c|c|c|c|c|c|c|}\hline
&&&&&&&&\\[-12pt]
$s$ & $k$ & $h^{(r=3)}_{0,k,(2s)}$ & $h^{(r=3)}_{1,k,(2s)}$ & $h^{(r=3)}_{2,k,(2s)}$ & $h^{(r=3)}_{3,k,(2s)}$ & $h^{(r=3)}_{4,k,(2s)}$ & $h^{(r=3)}_{5,k,(2s)}$ & $h^{(r=3)}_{6,k,(2s)}$ \\[4pt]\hline\hline
&&&&&&&&\\[-12pt]
$0$ & $0$ & $0$ & $\frac{-1}{2985984}$ & $-\frac{E_2 }{497664}$ & $-\frac{E_4 }{124416}$ & $\frac{22 E_6 -27 E_2  E_4 }{186624}$ & $\frac{8 E_2  E_6 -9 E_4 ^2}{20736}$ & $\frac{E_4  (20 E_6 -21 E_2  E_4 )}{31104}$\\[4pt]\hline
&&&&&&&&\\[-12pt]
 & $1$ & $0$ & $0$ & $-\frac{1}{41472}$ & $0$ & $-\frac{E_4 }{576}$ & $\frac{E_6 }{216}$ & $-\frac{7 E_4 ^2}{864}$\\[4pt]\hline
&&&&&&&&\\[-12pt]
& $2$ & $0$ & $0$ & $0$ & $-\frac{1}{1296}$ & $0$ & $-\frac{E_4 }{90}$ & $\frac{11 E_6 }{1134}$\\[4pt]\hline
&&&&&&&&\\[-12pt]
& $3$ & $0$ & $0$ & $0$ & $0$ & $-\frac{1}{432}$ & $0$ & $-\frac{7 E_4 }{1080}$\\[4pt]\hline
&&&&&&&&\\[-12pt]
& $4$ & $0$ & $0$ & $0$ & $0$ & $0$ & $-\frac{1}{810}$ & $0$\\[4pt]\hline
&&&&&&&&\\[-12pt]
& $5$ & $0$ & $0$ & $0$ & $0$ & $0$ & $0$ & $-\frac{1}{7560}$\\[4pt]\hline\hline
&&&&&&&&\\[-12pt]
$1$ & $0$ & $\frac{1}{36\cdot 24^5}$ & $0$ & $\frac{13 E_4 -6 E_2 ^2}{23887872}$ & $\frac{9 E_2  E_4 -8 E_6 }{1119744}$ & $\frac{-36 E_2 ^2 E_4 -96 E_2  E_6 +127 E_4 ^2}{1990656}$ & $\frac{9 E_2 ^2 E_6 +42 E_2  E_4 ^2-53 E_4  E_6 }{186624}$ & $\frac{-378 E_2 ^2 E_4 ^2-1440 E_2  E_4  E_6 +1233 E_4 ^3+544 E_6 ^2}{4478976}$\\[4pt]\hline
&&&&&&&&\\[-12pt]
 & $1$ & $0$ & $\frac{1}{1990656}$ & $-\frac{E_2 }{331776}$ & $\frac{E_4 }{9216}$ & $\frac{-27 E_2  E_4 -94 E_6 }{124416}$ & $\frac{24 E_2  E_6 +139 E_4 ^2}{41472}$ & $-\frac{E_4  (21 E_2  E_4 +100 E_6 )}{20736}$\\[4pt]\hline
&&&&&&&&\\[-12pt]
& $2$ & $0$ & $0$ & $\frac{13}{248832}$ & $-\frac{E_2 }{10368}$ & $\frac{59 E_4 }{25920}$ & $-\frac{E_2  E_4 }{720}-\frac{97 E_6 }{18144}$ & $\frac{11 E_2  E_6 }{9072}+\frac{79 E_4 ^2}{8640}$\\[4pt]\hline
&&&&&&&&\\[-12pt]
& $3$ & $0$ & $0$ & $0$ & $\frac{5}{10368}$ & $-\frac{E_2 }{3456}$ & $\frac{287 E_4 }{51840}$ & $-\frac{7 E_2  E_4 }{8640}-\frac{55 E_6 }{13608}$\\[4pt]\hline
&&&&&&&&\\[-12pt]
& $4$ & $0$ & $0$ & $0$ & $0$ & $\frac{7}{8640}$ & $-\frac{E_2 }{6480}$ & $\frac{13 E_4 }{6480}$\\[4pt]\hline
&&&&&&&&\\[-12pt]
& $5$ & $0$ & $0$ & $0$ & $0$ & $0$ & $\frac{353}{1088640}$ & $-\frac{E_2 }{60480}$\\[4pt]\hline
&&&&&&&&\\[-12pt]
& $6$ & $0$ & $0$ & $0$ & $0$ & $0$ & $0$ & $\frac{1}{34020}$\\[4pt]\hline
\end{tabular}}}
\end{center}
\caption{Expansion coefficients $h_{a,k,(2s)}^{(r=3)}$.}
\label{Tab:CoeffsExpr3}
\end{table}

\subsection{Factorisation at Order $\mathcal{O}(Q_R)$}\label{Sect:N2FactOrd1}
As was already conjectured in \cite{Hohenegger:2019tii}, the coefficients $H^{(r=1),\{n,0\}}_{(2s,0)}(\rho,S)$ can be factorised in terms of $H^{(1),\{0\}}_{(s,0)}$ and $W_{(s,0)}(\rho,S)$, \emph{i.e.} the coefficients that appear in the expansion of the free energy for $N=1$ and the function $W^{(1)}_{\text{NS}}$ defined in Eq.~(\ref{DefBuildW}) (as reviewed in Appendix~\ref{App:N1Expansion}), concretely
\begin{align}
H^{(r=1),\{0,0\}}_{(2s,0)}(\rho,S)&=2\sum_{a+b=s}H^{(1),\{0\}}_{(2a,0)}(\rho,S)\,W_{(2b,0)}(\rho,S)\,,\nonumber\\
H^{(r=1),\{n,0\}}_{(2s,0)}(\rho,S)&=\frac{1}{1-Q_\rho^n}\,\sum_{a,b=0}^{s} H^{(1),\{0\}}_{(2a,0)}(\rho,S) H^{(1),\{0\}}_{(2b,0)}(\rho,S)\,\mathcal{M}_{ab}^{(s)}(n)\,,&&\forall \hspace{0.25cm}n\geq 1\,.\label{DecompN2MatrixFirst}
\end{align}
Here $\mathcal{M}^{(s)}$ is a symmetric $(s+1)\times (s+1)$ dimensional matrix, that only depends on $n$, which in \cite{Hohenegger:2019tii} was conjectured to take the form 
\begin{align}
&\mathcal{M}_{ab}^{(s)}=-2\,\frac{(-1)^{s+a+b}\,n^{2s+1-2(a+b)}}{\Gamma(2s-2(a+b-1))}\,,&&a,b\in\{0,\ldots,s\}\,,
\end{align}
where it is understood that $1/\Gamma(-m)=0$ for $m\in\mathbb{N}\cup\{0\}$. The first few instances of $\mathcal{M}^{(s)}$ are
\begin{align}
&\mathcal{M}^{(0)}=(-2n)\,,&&\mathcal{M}^{(1)}=\left(\begin{array}{cc}\frac{n^3}{3} & -2n \\ -2n & 0\end{array}\right)\,,&&\mathcal{M}^{(2)}=\left(\begin{array}{ccc} -\frac{n^5}{60} & \frac{n^3}{3} & -2 n \\ \frac{n^3}{3} & -2 n & 0 \\ -2 n & 0 & 0 \\ \end{array}\right)\,.
\end{align}
In \cite{Hohenegger:2016eqy} it was shown that the NS limit of the partition functions ${\cal Z}_{N,1}$ have a self-similar behavior\footnote{The precise relation that was shown in \cite{Hohenegger:2016eqy} is eq.~(\ref{SelfSimilarity}).} in a certain region of the K\"ahler moduli space i.e., for $\widehat{a}_{1}=\widehat{a}_{2}=\cdots=\widehat{a}_{N}=\frac{\rho}{N}$. Relations such as (\ref{SelfSimilarity}) allow to infer non-trivial information about the free energy for generic $N$, just based on the knowledge of the (much simpler) free energy for the configuration $N=1$, albeit only at a specific point in the moduli space. From this perspective, (\ref{DecompN2MatrixFirst}) is similar in spirit to this self-similarity: they allow to obtain non-trivial information about the $N=2$ free energy at leading instanton order just from the configuration $N=1$. We shall see that relations of this type also exist for $N>2$ and (to some extent) also generalise to higher orders in $Q_R$.

\subsection{Hecke Structures}\label{Sect:HeckeOrder2}
The coefficients $H^{(r),\{n,0\}}_{(2s,0)}(\rho,S)$ for $r>1$ do not seem to exhibit simple factorisations of the type (\ref{DecompN2MatrixFirst}). We shall, however, in the following identify particular subsectors of the free energy (as introduced in (\ref{FactGenDefinition}) for generic $N$) that in fact do again factorise.

To this end, we define the following contour integrals
\begin{align}
\factE{1}{r}{2s,0}(\rho,S)&:=\frac{1}{(2\pi i)^2}\,\sum_{\ell=0}^\infty Q_\rho^\ell\oint_0\frac{dQ_{\widehat{a}_1}}{Q_{\widehat{a}_1}^{1+\ell}}\oint_0 \frac{dQ_{\widehat{a}_2}}{Q_{\widehat{a}_2}^{1+\ell}}\,P^{(r)}_{2,(2s,0)}(\widehat{a}_{1},\widehat{a}_2,S)\,,\label{FactDefinition1}\\
\factE{2}{r}{2s,0}(\rho,S)&:=\frac{1}{(2\pi i)}\,\frac{1}{r}\,\oint_0 d\widehat{a}_1\,\widehat{a}_1\,P^{(r)}_{2,(2s,0)}(\rho,\widehat{a}_{1},S)\,,\label{FactDefinition2}
\end{align}
where all contours are small circles around the origin\footnote{The integrals are in fact designed to precisely extract the residues in a Laurent series expansion.} and in (\ref{FactDefinition2}) we have implicitly used $\widehat{a}_2=\rho-\widehat{a}_1$. With these coefficient functions, we define the (a priori formal) series in~$\epsilon_1$
\begin{align}
\fact{a}{r}(\rho,S,\epsilon_1)=\sum_{s=0}^\infty \epsilon_1^{2s-2a}\,\factE{a}{r}{2s,0}(\rho,S)\,,&&\forall a=1,2\,.
\end{align}
From the perspective of the M-brane-web, the functions (\ref{FactDefinition1}) and (\ref{FactDefinition2}) count certain BPS configurations of M2-branes stretched between two M5-branes on a circle. Due to the contour prescriptions, however, only certain configurations contribute, which are visualised in \figref{fig:PartM5braneN2Config}:

\begin{figure}[h]
\begin{center}
\scalebox{0.8}{\parbox{6.1cm}{\begin{tikzpicture}[scale = 1]
\draw[very thick,fill=orange!25!white] (-3,-1.5) -- (-3,1.5) -- (-1,1.5) -- (-1,-1.5) -- (-3,-1.5);
\draw[very thick,fill=orange!25!white] (3,-1.5) -- (3,1.5) -- (1,1.5) -- (1,-1.5) -- (3,-1.5);
\draw [thick,blue,domain=-185:5,yshift=0.1cm] plot ({2*cos(\x)}, {0.59*sin(\x)}); 
\draw [thick,blue,domain=-180:-0] plot ({2*cos(\x)}, {0.6*sin(\x)});
\draw [thick,blue,domain=-175:-5,yshift=-0.1cm] plot ({2*cos(\x)}, {0.61*sin(\x)}); 
\draw [thick,blue,domain=119:61,yshift=0.1cm] plot ({2*cos(\x)}, {0.59*sin(\x)}); 
%
\draw [thick,blue,domain=119:61] plot ({2*cos(\x)}, {0.6*sin(\x)});
%
\draw [thick,blue,domain=119:61,yshift=-0.1cm] plot ({2*cos(\x)}, {0.61*sin(\x)}); 
%
\draw [ultra thick,domain=119:61,yshift=0.35cm] plot ({2*cos(\x)}, {0.59*sin(\x)}); 
\draw [ultra thick,<-,dashed,domain=175:119,yshift=0.35cm] plot ({2*cos(\x)}, {0.59*sin(\x)}); 
\draw [ultra thick,->,dashed,domain=61:5,yshift=0.35cm] plot ({2*cos(\x)}, {0.59*sin(\x)}); 
\draw [ultra thick,<->,domain=-175:-5,yshift=-0.4cm] plot ({2*cos(\x)}, {0.61*sin(\x)}); 
\node[blue] at (0.6,0.15) {$\ell$}; 
\node[blue] at (-0.6,-0.15) {$\ell$}; 
\node at (0,-1.4) {$\widehat{a}_2$}; 
\node at (0,1.3) {$\widehat{a}_1$}; 
\node at (0,-2.4) {\large\text{\bf{(a)}}};
\end{tikzpicture}}}
\hspace{3cm}
\scalebox{0.8}{\parbox{6.1cm}{\begin{tikzpicture}[scale = 1]
\draw[very thick,fill=orange!25!white] (-3,-1.5) -- (-3,1.5) -- (-1,1.5) -- (-1,-1.5) -- (-3,-1.5);
\draw[very thick,fill=orange!25!white] (3,-1.5) -- (3,1.5) -- (1,1.5) -- (1,-1.5) -- (3,-1.5);
\draw [thick,blue,domain=-185:5,yshift=0.1cm] plot ({2*cos(\x)}, {0.59*sin(\x)}); 
\draw [thick,blue,domain=-180:-0] plot ({2*cos(\x)}, {0.6*sin(\x)});
\draw [thick,blue,domain=-175:-5,yshift=-0.1cm] plot ({2*cos(\x)}, {0.61*sin(\x)}); 
\draw [thick,red,domain=119:61,yshift=0.3cm] plot ({2*cos(\x)}, {0.59*sin(\x)}); 
\draw [thick,red,domain=119:61,yshift=0.2cm] plot ({2*cos(\x)}, {0.59*sin(\x)}); 
\draw [thick,red,domain=119:61,yshift=0.1cm] plot ({2*cos(\x)}, {0.59*sin(\x)}); 
%
\draw [thick,red,domain=119:61] plot ({2*cos(\x)}, {0.6*sin(\x)});
%
\draw [thick,red,domain=119:61,yshift=-0.1cm] plot ({2*cos(\x)}, {0.61*sin(\x)}); 
%
\node[red] at (0.6,0.15) {$n_1$}; 
\node[blue] at (-0.6,-0.15) {$n_2$}; 
\draw [ultra thick,domain=119:61,yshift=0.6cm] plot ({2*cos(\x)}, {0.59*sin(\x)}); 
\draw [ultra thick,<-,dashed,domain=175:119,yshift=0.6cm] plot ({2*cos(\x)}, {0.59*sin(\x)}); 
\draw [ultra thick,->,dashed,domain=61:5,yshift=0.6cm] plot ({2*cos(\x)}, {0.59*sin(\x)}); 
\draw [ultra thick,<->,domain=-175:-5,yshift=-0.4cm] plot ({2*cos(\x)}, {0.61*sin(\x)}); 
\node at (0,-1.35) {$\rho-\widehat{a}_1$}; 
\node at (0,1.5) {$\widehat{a}_1$};
\node at (0,-2.4) {\large\text{\bf{(b)}}};
\end{tikzpicture}}}

\end{center} 
\caption{{\it Brane web configurations made up from $N=2$ M5-branes (drawn in orange) spaced out on a circle, with various M2-branes (drawn in red and blue) stretched between them. (a) An equal number $\ell$ of M2-branes is stretched between the M5-branes on either side of the circle. Configurations of this type are relevant for the computation of $\fact{1}{r}$. (b) $n_1$ M2-branes are stretched on one side of the circle and $n_2$($\neq n_1$) on the other side of the circle. Configurations of this type are relevant for the contributions $\fact{2}{r}$.}}
\label{fig:PartM5braneN2Config}
\end{figure}
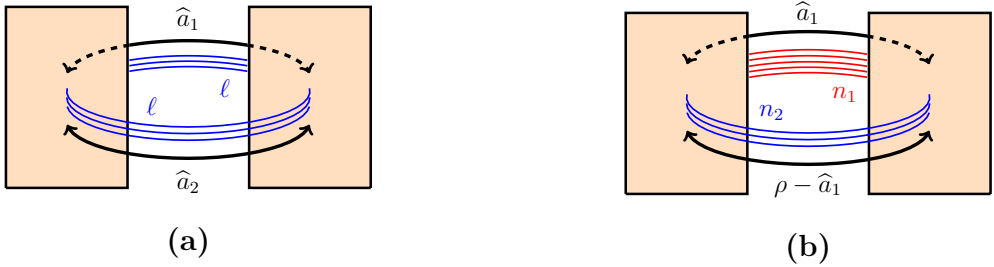

\begin{itemize}
\item Combination $\factE{1}{r}{2s,0}$:\\
Upon writing $P^{(r)}_{2,(2s,0)}$ as a Fourier expansion in $Q_{\widehat{a}_{1,2}}$ (similar to $H^{(r),\{n_1,\ldots,n_N\}}_{(2s,0)}$ in eq.~(\ref{PpartFree}))
\begin{align}
P^{(r)}_{2,(2,0)}(\widehat{a}_{1,2},S,\epsilon_{1,2})=\sum_{n_1,n_2=0}^\infty Q_{\widehat{a}_1}^{n_1} Q_{\widehat{a}_2}^{n_2}\,P^{(r),\{n_1,n_2\}}_{(2s,0)}(S)\,,
\end{align}
the contour prescriptions in (\ref{FactDefinition1}) extract all terms with $n_1=n_2$. $\factE{1}{r}{2s,0)}$ thus receives contributions only from those brane configurations, where an equal number of M2-branes is stretched between the two M5-branes on either side of the circle, as shown in \figref{fig:PartM5braneN2Config}~(a). In fact, $\factE{1}{r}{2s,0}$ can equivalently be written as
\begin{align}
\factE{1}{r}{2s,0}(\rho,S)=H_{(2s,0)}^{(r),\{0,0\}}(\rho,S)\,,\label{Fact1intoH}
\end{align}
and $\fact{1}{r}(\rho,S,\epsilon_1)$ is in fact exactly the reduced free energy studied in \cite{Ahmed:2017hfr}. Explicit expansions of $\factE{1}{r}{2s,0}$ for $r=1$, $r=2$ and $r=3$ can be recovered from Tables~\ref{Tab:CoeffsExpr1}, \ref{Tab:CoeffsExpr2} and \ref{Tab:CoeffsExpr3}, respectively from the coefficients with $k=0$.
\item Combination $\factE{2}{r}{2s,0}$:\\
The function $\factE{2}{r}{2s,0}$ in (\ref{FactDefinition2}) receives contributions from configurations in which $n_1$ M2-branes are stretched between the M5-branes on one side of the circle and $n_2$ (with $n_2\neq n_1$) on the other side, as schematically shown in \figref{fig:PartM5braneN2Config}~(b). Furthermore, from each of these contributions, the contour integral extracts the pole of the type $\widehat{a}_1^{-2}$ (where it is important to write $\widehat{a}_2=\rho-\widehat{a}_1$).

In terms of the functions $H_{(2s,0)}^{(r),\{n,0\}}$ in (\ref{HfunctionsN2}), the contour prescription in fact extracts the contributions of $\mathfrak{h}_{k=1,(2s)}^{(r)}$
\begin{align}
\factE{2}{r}{2s,0}(\rho,S)=\frac{1}{r}\,\mathfrak{h}_{k=1,(2s)}^{(r)}\,.\label{ExtractCh2}
\end{align}
To intuitively understand this result, we introduce \cite{Bastian:2019wpx}
\begin{align}
&\mathcal{I}_\alpha(\rho,\widehat{a}_1)=D^{2\alpha}_{\widehat{a}_1}\mathcal{I}_0
=D^{2\alpha}_{\widehat{a}_1}\sum_{n=1}^\infty\frac{n}{1-Q_\rho^n}\left(Q^n_{\widehat{a}_1}+\frac{Q_\rho^n}{Q^n_{\widehat{a}_1}}\right)\,,&&\text{with} &&D_{\widehat{a}_1}=Q_{\widehat{a}_1}\,\frac{\partial}{\partial Q_{\widehat{a}_1}}\,.\label{DefIalpha}
\end{align} 
As argued in \cite{Bastian:2019wpx}, $\mathcal{I}_0$ can be written in terms of Weierstrass's elliptic function $\wp$ and the second Eisenstein series (see appendix~\ref{App:ModularStuff} for the definitions)
\begin{align}
\mathcal{I}_0(\rho,\widehat{a}_1)=\frac{1}{(2\pi i)^2}\left[2\zeta(2)E_2(\rho)+\wp(\widehat{a}_1;\rho)\right]\,.
\end{align}
Since Weierstrass's elliptic function affords the following series expansion
\begin{align}
&\wp(z;\rho)=\frac{1}{z^2}+\sum_{k=1}^\infty 2\zeta(2k+2)(2k+1)\,E_{2k+2}(\rho)\,z^{2k}\,,
\end{align}
we have for the contour integral 
\begin{align}
\oint d\widehat{a}_1\,\widehat{a}_1\,\mathcal{I}_\alpha(\rho,\widehat{a}_1)=2\pi i\,\delta_{\alpha 0}\,,\label{ContourIalpha}
\end{align} 
such that with (\ref{DefPN2Ord}) and (\ref{HfunctionsN2}) we have (\ref{ExtractCh2}). The factor $1/r$ in the latter relation is simply a convenient normalisation factor as will become apparent later on.

A more direct way to arrive at (\ref{ExtractCh2}) is to start from the decomposition (\ref{DefPN2Ord}) and exchange\footnote{This is possible since the sum over $k$ is finite.} the summations over $k$ and $n$
\bea\label{Divergence}
P^{(r)}_{(2s,0)}(\widehat{a}_{1,2},S)=H^{(r),\{0,0\}}_{(2s,0)}(\rho,S)+\sum_{k=1}^{rs}\mathfrak{h}_{k,(2s)}^{(r)}(\rho,S)\,X_{k}(\widehat{a}_{1,2})\,,
\eea
where we introduced the shorthand notation
\begin{align}
X_{k}(\rho,\widehat{a}_{1,2})=\sum_{n=1}^{\infty}\frac{n^{2k-1}}{1-Q_{\rho}^n}\Big(Q_{\widehat{a}_{1}}^{n}+Q_{\widehat{a}_{2}}^n\Big)=\sum_{n=1}^{\infty}\sum_{b=0}^{\infty}n^{2k-1}\,Q_{\rho}^{n\,b}\Big(Q_{\widehat{a}_{1}}^{n}+Q_{\widehat{a}_{2}}^n\Big)\,.\label{DefXkSums}
\end{align}
We can express $X_{k}$ in terms of the q-Polygamma function $\psi^{(m)}_{q}(z)$,
\begin{align}
&\psi_{q}(z)=\frac{d\,\mbox{ln}\Gamma_{q}(z)}{dz}=-\mbox{ln}(1-q)+\mbox{ln}(q)\sum_{n=0}^{\infty}\frac{q^{n+z}}{1-q^{n+z}}\,,&&\psi^{(m)}_{q}(z)=\frac{d^{m}\,\psi_{q}(z)}{dz^m}\,,\label{DefqPolyGamma}
\end{align}
where $\Gamma_q$ is the q-Gamma function
\begin{align}
\Gamma_{q}(z)=(1-q)^{1-z}\prod_{n=0}^{\infty}\frac{1-q^{n+1}}{1-q^{n+z}}\,.
\end{align}
To this end, we interchange\footnote{This is possible for $|Q_\rho|<1$ and $|Q_{\widehat{a}_{1,2}}|<1$. To see this, we consider for example
\begin{align}
\sum_{n=1}^\infty\sum_{b=0}^\infty\,n^{2k-1} |Q_\rho|^{nb}|Q_{\widehat{a}_1}|^n\leq \sum_{n=1}^\infty\sum_{b=0}^\infty\,n^{2k-1} |Q_\rho|^{b}|Q_{\widehat{a}_1}|^n =\left(\sum_{n=1}^\infty n^{2k-1}|Q_{\widehat{a}_1}|^n\right)\left(\sum_{b=0}^\infty |Q_{\rho}|^b\right)=\frac{\text{Li}_{1-2k}(|Q_{\widehat{a}_1}|)}{1-|Q_\rho|}\,,\nonumber
\end{align}
where $\text{Li}_{1-2k}$ denotes the polylogarithm. Thus (\ref{DefXkSums}) is absolutely convergent, and therefore the summations can be interchanged.} the sums in the last expression in (\ref{DefXkSums}) and find with~(\ref{DefqPolyGamma})
\begin{align}
&X_{k}(\rho,\widehat{a}_{1,2})= \frac{1}{\mbox{ln}(Q_{\rho})^{k+1}}\Big(\psi^{(2k-1)}_{Q_{\rho}}(\widehat{a}_{1}/\rho)+\psi^{(2k-1)}_{Q_{\rho}}(\widehat{a}_{2}/\rho)\Big) && \text{for} && k\geq 1\,.
\end{align}
The $q$-Gamma function $\Gamma_{q}(z)$ satisfies the identity $\Gamma_{q}(z+1)=\frac{1-q^{z}}{1-q}\,\Gamma_{q}(z)$ and, therefore, for $z\mapsto 0$ we obtain
\begin{align}
\Gamma_q(z)=-\frac{1-q}{z\,\ln(q)}+\mathcal{O}(z^{0})\,.
\end{align}
Thus the function $X_{k}(\rho,\widehat{a}_{1,2})$ diverges for $\widehat{a}_{1}\mapsto 0$ and in fact has a pole of order $k+1$
\bea
X_{k}(\rho,\widehat{a}_{1,2})\sim - \frac{(2k-1)!}{\widehat{a}_{1}^{2k}}+O(\widehat{a}_{1}^0)
\eea
Therefore the only contribution to the contour integral in (\ref{FactDefinition2}) (which extracts the pole of order 2) stems from $X_1(\rho,\widehat{a}_{1,2})$, thus yielding (\ref{ExtractCh2}).
\end{itemize}

\noindent
By comparing the explicit expressions for the contributions $\factE{1}{r}{2s,0}$ and $\factE{2}{r}{2s,0}$ to the free energy, we find that they satisfy the following recursion relation
\begin{align}
&\factE{1}{r}{2s,0}(\rho,S)=\mathcal{H}_r\left[\factE{1}{1}{2s,0}(\rho,S)\right]\,,&&\factE{2}{r}{2s,0}(\rho,S)=\mathcal{H}_r\left[\factE{2}{1}{2s,0}(\rho,S)\right]\,.\label{HeckeN2}
\end{align}
The normalisation factor $1/r$ appearing in the definition (\ref{FactDefinition2}) was chosen to normalise the right hand side of the second equation above.

\subsection{Decomposition of $\factE{1}{r}{2s,0}$ and $\factE{2}{r}{2s,0}$} \label{Sect:FactorisationN2}
In section~\ref{Sect:N2FactOrd1} we have seen that the free energy in the NS limit factorises to order $\mathcal{O}(Q_R)$ as in eq.~(\ref{DecompN2MatrixFirst}) with the basic building blocks given by the expansion coefficients of the free energy in the case $N=1$. While the complete free energy at higher orders $\mathcal{O}(Q_R^r)$ (for $r>1$) does not exhibit such a behaviour, the particular contributions $\factE{1}{r}{2s,0}$ and $\factE{2}{r}{2s,0}$ defined in (\ref{FactDefinition2}) lend themselves to a generalisation of (\ref{DecompN2MatrixFirst}).

\subsubsection{Factorisation at Order $Q_R^1$}
The first step is to establish the  factorisation of $\factE{1}{r=1}{2s,0}$ and $\factE{2}{r=1}{2s,0}$, which are in fact induced by (\ref{DecompN2MatrixFirst}). Indeed, using (\ref{Fact1intoH}) as well as (\ref{ExtractCh2}), we have immediately
\begin{align}
\factE{1}{r=1}{2s,0}(\rho,S)=2\sum_{i,j=0}^s\delta_{s,i+j}H^{(1),\{0\}}_{(2i,0)}(\rho,S)\,W_{(2j,0)}(\rho,S)\,,\label{DecomposeN2HeckeSeed}\\
\factE{2}{r=1}{2s,0}(\rho,S)=-2\sum_{i,j=0}^s\delta_{s,i+j}H^{(1),\{0\}}_{(2i,0)}(\rho,S)\,H^{(1),\{0\}}_{(2j,0)}(\rho,S)\,.\label{DecomposeN2HeckeSeed2}
\end{align}
Combining these expansion coefficients (in a series of $\epsilon_1$), we can equivalently write the following relations for the (a priori formal) series expansions
\begin{align}
\fact{1}{r=1}(\rho,S,\epsilon_1)&=2\,\buildH{1}(\rho,S,\epsilon_1)\,\buildW{1}(\rho,S,\epsilon_1)\,,\nonumber\\
\fact{2}{r=1}(\rho,S,\epsilon_1)&=-2\,\left[\buildH{1}(\rho,S,\epsilon_1)\right]^2\,,\label{R1FactorisationN2}
\end{align}
where the coefficients $\buildH{1}$ and $\buildW{1}$ are defined in (\ref{DefBuildH}) and (\ref{DefBuildW}) respectively. 
\subsubsection{Factorisation at Order $Q_R^2$}\label{Sect:FactN2OrdR2}
Based on eq.(\ref{DecomposeN2HeckeSeed}) and  (\ref{DecomposeN2HeckeSeed2}), the first attempt to factorise the function $\fact{1,2}{r=2}$ to order $\mathcal{O}(Q_R^2)$ would be to use a similar decomposition, except to replace $\buildH{1}$ and $\buildW{1}$ by their order $\mathcal{O}(Q_R^2)$ counterparts $\buildH{2}$ and $\buildW{2}$ respectively. However, this does not fully reproduce the correct answer, instead we have\footnote{The relation (\ref{N2FactorisationOrd2}) as well as the remaining equations in this subsection are understood to hold order by order in an expansion in powers of $\epsilon_1$ and we have checked it up to order $\epsilon_1^6$. To safe writing, however, in the following we state our results in terms of the formal series expansions.} 
{\allowdisplaybreaks
\begin{align}
\fact{1}{r=2}(\rho,S,\epsilon_1)&=\frac{4}{3}\,\buildH{2}(\rho,S,\epsilon_1)\,\buildW{2}(\rho,S,\epsilon_1)+\mathfrak{R}^{(2)}_1(\rho,S,\epsilon_1)\,,\nonumber\\
\fact{2}{r=2}(\rho,S,\epsilon_1)&=-\frac{4}{3}\,\left[\buildH{2}(\rho,S,\epsilon_1)\right]^2+\mathfrak{R}^{(2)}_2(\rho,S,\epsilon_1)\,,\label{N2FactorisationOrd2}
\end{align}}
The additional contributions $\mathfrak{R}^{(2)}_{1,2}$ are formal expansions in powers of $\epsilon_1$
\begin{align}
&\mathfrak{R}^{(2)}_{a}(\rho,S,\epsilon_1)=\sum_{s=0}^\infty\epsilon_1^{2s-2a}\,\mathfrak{R}^{(2)}_{a,(2s,0)}(\rho,S)\,,&&\forall a=1,2\,,
\end{align}
where the $\mathfrak{R}^{(2)}_{a,(2s,0)}(\rho,S)$ in turn can be decomposed as
\begin{align}
\mathfrak{R}^{(2)}_{a,(2s,0)}(\rho,S)=\sum_{i=0}^4\mathfrak{r}^{(2)}_{a,i,(2,0)}(\rho)\,(\phi_{-2,1}(\rho,S))^i\,\phi_{0,1}(\rho,S))^{4-i}\,,
\end{align}
and the $\mathfrak{r}^{(2)}_{a,i,(2,0)}(\rho)$ are (quasi-)modular forms of weight $2s-2+2i-2a$ and the first few expressions are tabulated for $a=1$ in Table~\ref{Tab:CoefsCorrR21} and for $a=2$ in Table~\ref{Tab:CoefsCorrR22}.

\begin{table}[htbp]
\begin{center}
\begin{tabular}{|c||c|c|c|c|c|}\hline
&&&&&\\[-12pt]
$s$ & $\mathfrak{r}^{(2)}_{1,0,(2s,0)}$ & $\mathfrak{r}^{(2)}_{1,1,(2s,0)}$ & $\mathfrak{r}^{(2)}_{1,2,(2s,0)}$ & $\mathfrak{r}^{(2)}_{1,3,(2s,0)}$ & $\mathfrak{r}^{(2)}_{1,4,(2s,0)}$\\[4pt]\hline\hline
&&&&&\\[-12pt]
$0$ & $0$ & $0$ & $0$ & $0$ & $\frac{E_6 -E_2  E_4 }{144}$\\[4pt]\hline
&&&&&\\[-12pt]
$1$ & $0$ & $0$ & $0$ & $\frac{E_2  E_4 -E_6 }{3456}$ & $\frac{E_2 ^2 (-E_4 )-E_2  E_6 +2 E_4 ^2}{1728}$\\[4pt]\hline
&&&&&\\[-12pt]
$2$ & $0$ & $0$ & $\frac{E_6 -E_2  E_4 }{221184}$ & $\frac{E_2 ^2 E_4 +E_2  E_6 -2 E_4 ^2}{41472}$ & $\frac{-20 E_2 ^3 E_4 -60 E_2 ^2 E_6 -69 E_2  E_4 ^2+149 E_4  E_6 }{829440}$\\[4pt]\hline
\end{tabular}
\end{center}
\caption{Coefficients appearing in the expansion of the correction term $\mathfrak{R}_1^{(2)}$.}
\label{Tab:CoefsCorrR21}
\end{table}

\begin{table}[htbp]
\begin{center}
\begin{tabular}{|c||c|c|c|c|c|}\hline
&&&&&\\[-12pt]
$s$ & $\mathfrak{r}^{(2)}_{2,0,(2s,0)}$ & $\mathfrak{r}^{(2)}_{2,1,(2s,0)}$ & $\mathfrak{r}^{(2)}_{1,2,(2s,0)}$ & $\mathfrak{r}^{(2)}_{2,3,(2s,0)}$ & $\mathfrak{r}^{(2)}_{2,4,(2s,0)}$\\[4pt]\hline\hline
&&&&&\\[-12pt]
$0$ & $0$ & $0$ & $0$ & $0$ & $-\frac{E_4 }{24}$\\[4pt]\hline
&&&&&\\[-12pt]
$1$ & $0$ & $0$ & $0$ & $\frac{E_4 }{576}$ & $-\frac{E_2  E_4 +2 E_6 }{288}$\\[4pt]\hline
&&&&&\\[-12pt]
$2$ & $0$ & $0$ & $-\frac{E_4 }{36864}$ & $\frac{E_2  E_4 +2 E_6 }{6912}$ & $-\frac{20 E_2 ^2 E_4 +80 E_2  E_6 +149 E_4 ^2}{138240}$\\[4pt]\hline
&&&&&\\[-12pt]
$3$ & $0$ & $\frac{E_4 }{5308416}$ & $\frac{-E_2  E_4 -2 E_6 }{442368}$ & $\frac{8 E_2 ^2 E_4 +32 E_2  E_6 +59 E_4 ^2}{1327104}$ & $\frac{-140 E_2 ^3 E_4 -840 E_2 ^2 E_6 -3129 E_2  E_4 ^2-5056 E_4  E_6 }{34836480}$\\[4pt]\hline
\end{tabular}
\end{center}
\caption{Coefficients appearing in the expansion of the correction term $\mathfrak{R}_2^{(2)}$.}
\label{Tab:CoefsCorrR22}
\end{table}

\noindent
The functions $\mathfrak{R}^{(2)}_{a}$ can themselves again be factorised where the basic building blocks are $\buildH{1}$
\begin{align}
\mathfrak{R}_a^{(2)}(\rho,S,\epsilon_1)=\mathfrak{S}^{(2)}_{a,4}(\rho,\epsilon_1)\left[\buildH{1}(\rho,S,\epsilon_1)\right]^4\,.
\end{align}
The only novel feature is the appearance of the functions $\mathfrak{S}^{(2)}_{a,4}$, which are $S$-independent (quasi)Jacobi forms that are characterised through
\begin{align}
&\frac{d \mathfrak{S}^{(2)}_{1,4}}{dE_2}(\rho,\epsilon_1)=\frac{\epsilon_1^2}{6}\,\mathfrak{S}^{(2)}_{2,4}(\rho,\epsilon_1)\,,&&\mathfrak{S}_{2,4}(\rho,\epsilon_1)=\sum_{s=1}^\infty\epsilon_1^{2s-6}\frac{(4^{s+1}-1)(2s+1)}{3\cdot 4^s\pi^{2(s+1)}}\,\zeta(2s+2)\,E_{2s+2}(\rho)\,.\label{InitialHoloAnomN2}
\end{align}
While we cannot write a closed form expression for  the holomorphic anomaly in $\mathfrak{R}_1^{(2)}$, we have
\begin{align}
\mathfrak{S}^{(2)}_{2,4}=\int dE_2\,\mathfrak{S}^{(2)}_{1,4}+\sum_{s=1}^\infty\epsilon_1^{2s-6}\frac{(4^{s+1}-1)(2s+1)}{18\cdot 4^s\pi^{2(s+1)}}\,\zeta(2s+2)\,\mathfrak{e}_{2s+4}(\rho)\,,
\end{align}
where $\mathfrak{e}_{2s+4}$ is a polynomial in $E_{4,6}$ of weight $2s+4$, normalised such that $\mathfrak{e}_{2s+4}(\rho)=1+\mathcal{O}(Q_\rho)$.\footnote{Implicitly $\mathfrak{e}_{2s+4}(\rho)$ is of course fixed uniquely through the relation (\ref{HeckeN2}).}

\subsubsection{Factorisation at Order $Q_R^r$ for $r>2$}\label{Sect:N2HigherFactorisation}
Following the decomposition (\ref{N2FactorisationOrd2}) at order $Q_R^2$, we can consider similar expressions to higher orders. From the explicit examples we find to order $Q_R^3$ 
{\allowdisplaybreaks
\begin{align}
\fact{1}{r=3}(\rho,S,\epsilon_1)&=\frac{3}{2}\,H_{N=1}^{(3)}\,W_{\text{NS}}^{(3)}+(H_{N=1}^{(1)})^6\,\mathfrak{S}^{(r=3)}_{1,(6,0)}(\rho,\epsilon_1)\nonumber\\
&\hspace{0.2cm}+(H_{N=1}^{(1)})^4\,H_{N=1}^{(2)}\,\mathfrak{S}^{(r=3)}_{1,(4,1)}(\rho,\epsilon_1)+(H_{N=1}^{(1)})^2\,(H_{N=1}^{(2)})^2\,\mathfrak{S}^{(r=3)}_{1,(2,2)}(\rho,\epsilon_1)\,,\nonumber\\
\fact{2}{r=3}(\rho,S,\epsilon_1)&=-\frac{3}{2}\,H_{N=1}^{(3)}\,H_{N=1}^{(3)}+(H_{N=1}^{(1)})^6\,\mathfrak{S}^{(r=3)}_{2,(6,0)}(\rho,\epsilon_1)\nonumber\\
&\hspace{0.2cm}+(H_{N=1}^{(1)})^4\,H_{N=1}^{(2)}\,\mathfrak{S}^{(r=3)}_{2,(4,1)}(\rho,\epsilon_1)+(H_{N=1}^{(1)})^2\,(H_{N=1}^{(2)})^2\,\mathfrak{S}^{(r=3)}_{2,(2,2)}(\rho,\epsilon_1)\,,
\end{align}}
and to order $Q_R^4$ 
\begin{align}
\fact{1}{r=4}(\rho,S,\epsilon_1)&=\frac{8}{7}\,H_{N=1}^{(4)}\,W_{\text{NS}}^{(4)}+(H_{N=1}^{(1)})^8\,\mathfrak{S}^{(r=4)}_{1,(8,0,0)}(\rho,\epsilon_1)+(H_{N=1}^{(1)})^6\,H_{N=1}^{(2)}\,\mathfrak{S}^{(r=4)}_{1,(6,1,0)}(\rho,\epsilon_1)\nonumber\\
&\hspace{0.2cm}+(H_{N=1}^{(1)})^4\,(H_{N=1}^{(2)})^2\,\mathfrak{S}^{(r=3)}_{1,(4,2,0)}(\rho,\epsilon_1)+(H_{N=1}^{(1)})^2\,(H_{N=1}^{(2)})^3\,\mathfrak{S}^{(r=3)}_{1,(2,3,0)}(\rho,\epsilon_1)\nonumber\\
&\hspace{0.2cm}+(H_{N=1}^{(1)})^2\,(H_{N=1}^{(3)})^2\,\mathfrak{S}^{(r=3)}_{1,(,0,2)2}(\rho,\epsilon_1)\,,\nonumber\\
\fact{2}{r=4}(\rho,S,\epsilon_1)&=-\frac{8}{7}\,H_{N=1}^{(4)}\,H_{N=1}^{(4)}+(H_{N=1}^{(1)})^8\,\mathfrak{S}^{(r=4)}_{2,(8,0,0)}(\rho,\epsilon_1)+(H_{N=1}^{(1)})^6\,H_{N=1}^{(2)}\,\mathfrak{S}^{(r=4)}_{2,(6,1,0)}(\rho,\epsilon_1)\nonumber\\
&\hspace{0.2cm}+(H_{N=1}^{(1)})^4\,(H_{N=1}^{(2)})^2\,\mathfrak{S}^{(r=3)}_{2,(4,2,0)}(\rho,\epsilon_1)+(H_{N=1}^{(1)})^2\,(H_{N=1}^{(2)})^3\,\mathfrak{S}^{(r=3)}_{2,(2,3,0)}(\rho,\epsilon_1)\nonumber\\
&\hspace{0.2cm}+(H_{N=1}^{(1)})^2\,(H_{N=1}^{(3)})^2\,\mathfrak{S}^{(r=3)}_{2,(2,0,2)}(\rho,\epsilon_1)\,.
\end{align}
Here $\mathfrak{S}^{(r),\ell}_{i,k,}(\rho,\epsilon_1)$ are independent of $S$ and we find the following $\epsilon_1$-expansions for $r=3$
{\allowdisplaybreaks
\begin{align}
\frac{1}{\epsilon_1^{10}}\mathfrak{S}^{(3)}_{1,(6,0)}&=\frac{E_4  (E_6 -E_2  E_4 )}{2592}+\frac{\epsilon_1  ^2 \left(E_6 ^2-3 E_2  E_4  E_6 +2
   E_4 ^3\right)}{15552}\nonumber\\
   &\hspace{1cm}+\frac{\epsilon_1  ^4 \left(43 E_4 ^2
   E_6-28 E_2  E_4 ^3-15 E_2  E_6 ^2 \right)}{622080}+O\left(\epsilon_1  ^6\right)\,,\nonumber\\
\frac{1}{\epsilon_1^{8}}\mathfrak{S}^{(3)}_{1,(4,1)}&= \frac{E_4 ^2-E_2  E_6 }{162   }-\frac{5 \epsilon_1^2   E_4  (E_2 
   E_4 -E_6 )}{1944}+\frac{\epsilon_1  ^4 \left(-196 E_2  E_4  E_6 +123 E_4 ^3+73
   E_6 ^2\right)}{233280}+O\left(\epsilon_1  ^6\right)\,,\nonumber\\  
\frac{1}{\epsilon_1^{6}}\mathfrak{S}^{(3)}_{1,(2,2)}&=\frac{2 (E_6 -E_2  E_4 )}{81 }+\frac{2\epsilon_1^2}{243} \left(E_4 ^2-E_2  E_6 \right)+\frac{\epsilon_1 ^4 E_4}{405}
   (E_6 -E_2  E_4 )+O\left(\epsilon_1  ^6\right)\,,\nonumber\\[8pt]
\frac{1}{\epsilon_1^{8}}\mathfrak{S}^{(3)}_{2,(6,0)}&=-\frac{E_4 ^2}{648}-\frac{E_4  E_6  \epsilon_1  ^2}{1296}-\frac{\epsilon_1  ^4 \left(28 E_4 ^3+15
   E_6 ^2\right)}{155520}+O\left(\epsilon_1  ^6\right)\,,\nonumber\\
\frac{1}{\epsilon_1^{6}}\mathfrak{S}^{(3)}_{2,(4,1)}&=\frac{-2 E_6 }{81}-\frac{5 E_4 ^2 \epsilon_1^2  }{486}-\frac{49 E_4  E_6  \epsilon_1 
   ^4}{14580}+O\left(\epsilon_1  ^6\right)\,,\hspace{0.2cm}
\frac{1}{\epsilon_1^{4}}\mathfrak{S}^{(3)}_{2,(2,2)}=\frac{-8 E_4 }{81}-\frac{8  \epsilon_1  ^2 E_6 }{243}-\frac{4 E_4 ^2 \epsilon_1  ^4}{405}+O\left(\epsilon_1  ^6\right)\,.\nonumber
\end{align}}
and for $r=4$
{\allowdisplaybreaks
\begin{align}
\frac{1}{\epsilon_1^{14}}\,\mathfrak{S}^{(4)}_{1,(8,0,0)}&=\frac{ \left(-21 E_2  E_4 ^3-31 E_2  E_6 ^2+52 E_4 ^2 E_6 \right)}{1741824}-\frac{13
   \epsilon_1  ^2 (E_4  E_6  (E_2  E_4 -E_6 ))}{497664}\nonumber\\
 &\hspace{1cm}+\frac{\epsilon_1  ^4 \left(-2 E_2  \left(654
   E_4 ^4+4129 E_4  E_6 ^2\right)+6179 E_4 ^3 E_6 +3387 E_6 ^3\right)}{627056640}+O\left(\epsilon_1 
   ^6\right)\,,\nonumber\\
\frac{1}{\epsilon_1^{12}}\,\mathfrak{S}^{(4)}_{1,(6,1,0)}&=\frac{\left(-181 E_2  E_4  E_6 +129 E_4 ^3+52 E_6 ^2\right)}{108864}+\frac{\epsilon_1  ^2
   \left(-651 E_2  E_4 ^3-313 E_2  E_6 ^2+964 E_4 ^2 E_6 \right)}{653184}\nonumber\\
   &\hspace{1cm}+\frac{ \epsilon_1^4 E_4 
   \left(-12075 E_2  E_4  E_6 +5269 E_4 ^3+6806 E_6 ^2\right)}{13063680}+O\left(\epsilon_1  ^6\right)\,,\nonumber\\
\frac{1}{\epsilon_1^{10}}\,\mathfrak{S}^{(4)}_{1,(4,2,0)}&=-\frac{5}{756} (E_4  (E_2  E_4 -E_6 ))+\frac{\epsilon_1  ^2 \left(-919 E_2  E_4  E_6 +540
   E_4 ^3+379 E_6 ^2\right)}{163296}\nonumber\\
   &\hspace{1cm}+\frac{\epsilon_1  ^4 \left(-2118 E_2  E_4 ^3-1303 E_2  E_6 ^2+3421
   E_4 ^2 E_6 \right)}{979776}+O\left(\epsilon_1  ^6\right)\,,\nonumber\\
\frac{1}{\epsilon_1^{8}}\,\mathfrak{S}^{(4)}_{1,(2,3,0)}&=\frac{7 \left(E_4 ^2-E_2  E_6 \right)}{243 }-\frac{103 \epsilon_1^2   (E_4  (E_2 
   E_4 -E_6 ))}{5103}\nonumber\\
   &\hspace{1cm}+\frac{\epsilon_1  ^4 \left(5386
   E_6 ^2-13887 E_2  E_4  E_6 +8501 E_4 ^3\right)}{1224720}+O\left(\epsilon_1  ^6\right)\,,\nonumber\\
\frac{1}{\epsilon_1^{6}}\,\mathfrak{S}^{(4)}_{1,(2,0,2)}&=-\frac{15 (E_2  E_4 -E_6 )}{128 }+\frac{17\epsilon_1^2}{256} \left(E_4 ^2-E_2  E_6 \right)-\frac{1511
   \epsilon_1  ^4 (E_4  (E_2  E_4 -E_6 ))}{43008}+O\left(\epsilon_1  ^6\right)\,,\nonumber\\[8pt]
\frac{1}{\epsilon_1^{12}}\,\mathfrak{S}^{(4)}_{2,(8,0,0)}&=-\frac{ \left(21 E_4 ^3+31 E_6 ^2\right)}{580608}-\frac{13 E_4 ^2 E_6  \epsilon_1^2}{165888}-\frac{\epsilon_1  ^4 \left(654 E_4 ^4+4129 E_4  E_6 ^2\right)}{104509440}+O\left(\epsilon_1  ^6\right)\,,\nonumber\\
\frac{1}{\epsilon_1^{10}}\,\mathfrak{S}^{(4)}_{2,(6,1,0)}&=-\frac{181 E_4  E_6}{36288}-\frac{\epsilon_1  ^2 \left(651 E_4 ^3+313 E_6 ^2\right)}{217728}-\frac{115
   E_4 ^2 E_6  \epsilon_1  ^4}{41472}+O\left(\epsilon_1  ^6\right)\,,\nonumber\\
\frac{1}{\epsilon_1^{8}}\,\mathfrak{S}^{(4)}_{2,(4,2,0)}&=-\frac{5 E_4 ^2}{252}-\frac{919 E_4  E_6  \epsilon_1  ^2}{54432}-\frac{\epsilon_1  ^4 \left(2118 E_4 ^3+1303
   E_6 ^2\right)}{326592}+O\left(\epsilon_1  ^6\right)\,,\nonumber\\
\frac{1}{\epsilon_1^{6}}\,\mathfrak{S}^{(4)}_{2,(2,3,0)}&=-\frac{7 E_6 }{81  }-\frac{103 E_4 ^2 \epsilon_1^2  }{1701}-\frac{1543 E_4  E_6  \epsilon_1 
   ^4}{45360}+O\left(\epsilon_1  ^6\right)\,,\nonumber\\
\frac{1}{\epsilon_1^{4}}\, \mathfrak{S}^{(4)}_{2,(2,0,2)}&=-\frac{45 E_4 }{128 }-\frac{51 \epsilon_1  ^2 E_6 }{256}-\frac{1511 E_4 ^2 \epsilon_1  ^4}{14336}+O\left(\epsilon_1  ^6\right)\,.\nonumber
\end{align}}
Comparing these expressions suggests the following form
\begin{align}
\left.\begin{array}{r}\fact{1}{r}\\[6pt] \fact{2}{r} \end{array}\right\}=\sum'_{i_1,\ldots,i_{r}}\mathfrak{S}^{(r)}_{a,(i_1,\ldots,i_{r})}\,(H_{N=1}^{(1)})^{i_1}\ldots (H_{N=1}^{(1)})^{i_{r}}+ \frac{2r}{\sigma_1(r)}\left\{\begin{array}{lcl}H_{N=1}^{(r)}\,W_{\text{NS}}^{(r)} & \text{for} & a=1 \\[6pt] (-1)H_{N=1}^{(r)}\,H_{N=1}^{(r)} & \text{for} & a=2\end{array}\right.\label{GenFormDecomposeN2R3}
\end{align}
Here the prime on the summation denotes the following conditions on $(i_1,\ldots,i_{r-1})$
\begin{align}
&\sum_{j=1}^{r}j i_j=2r\,,&&\text{and}&&i_1\in\mathbb{N}_{\text{even}} \,,&&\text{and}&&i_1>0\,,
\end{align}
and $\mathfrak{S}^{(r)}_{a,(i_1,\ldots,i_{r-1})}$ are quasi-modular forms depending on $\rho$ and $\epsilon_1$ which in particular satisfy 
\begin{align}
&\frac{\partial \mathfrak{S}^{(r)}_{2,(i_1,\ldots,i_r)}}{\partial E_2(\rho)} (\rho,\epsilon_1)=0\,,&&\mathfrak{S}^{(r)}_{2,(i_1,\ldots,i_{r})}(\rho,\epsilon_1)=\frac{12}{r\,\epsilon_1^2}\,\frac{\partial \mathfrak{S}^{(r)}_{1,(i_1,\ldots,i_{r})}}{\partial E_2(\rho)}(\rho,\epsilon_1)\,,&&\forall r>1\,.
\end{align}
This generalises the first relation in (\ref{InitialHoloAnomN2}) and also implies that $\mathfrak{S}^{(r)}_{2,(i_1,\ldots,i_r)}$ is a holomorphic Jacobi form. Notice also, for all examples we have computed so far $\mathfrak{S}^{(r)}_{2,(i_1,\ldots,i_{r})}=0$ for $i_r>0$.
\section{Hecke Structure for $N=3$}\label{Sect:CaseN3}
After discussing the free energy of the $N=2$ LST, we continue with $N=3$.

\subsection{Decomposition of the Free Energy}\label{Sect:FreeEnergN3}
The starting point is to compute the decomposition of the free energy. The web diagram re-

\begin{wrapfigure}{l}{0.38\textwidth}
\begin{center}
\vspace{-0.8cm}
\scalebox{0.7}{\parbox{8cm}{\begin{tikzpicture}[scale = 1.45]
\draw[ultra thick] (-1,0) -- (0,0) -- (0,-1) -- (1,-1) -- (1,-2) -- (2,-2) -- (2,-3) -- (3,-3);
\draw[ultra thick] (0,0) -- (0.7,0.7);
\draw[ultra thick] (1,-1) -- (1.7,-0.3);
\draw[ultra thick] (2,-2) -- (2.7,-1.3);
\draw[ultra thick] (0,-1) -- (-0.7,-1.7);
\draw[ultra thick] (1,-2) -- (0.3,-2.7);
\draw[ultra thick] (2,-3) -- (1.3,-3.7);
\node at (-1.2,0) {\large {\bf $\mathbf a$}};
\node at (3.2,-3) {\large {\bf $\mathbf a$}};
\node at (0.85,0.85) {\large {$\mathbf 1$}};
\node at (1.85,-0.15) {\large {$\mathbf 2$}};
\node at (2.85,-1.15) {\large {$\mathbf 3$}};
\node at (-0.65,-2) {\large {$\mathbf{1}$}};
\node at (0.15,-2.9) {\large {$\mathbf{2}$}};
\node at (1.15,-3.9) {\large {$\mathbf{3}$}};
\node at (-0.5,0.25) {\large  {\bf $h_1$}};
\node at (0.5,-1.25) {\large  {\bf $h_2$}};
\node at (1.5,-2.25) {\large  {\bf $h_3$}};
\node at (-0.2,-0.5) {\large  {\bf $v$}};
\node at (0.8,-1.5) {\large  {\bf $v$}};
\node at (1.8,-2.5) {\large  {\bf $v$}};
\node at (0.6,0.25) {\large  {\bf $m$}};
\node at (1.6,-0.75) {\large  {\bf $m$}};
\node at (2.6,-1.75) {\large  {\bf $m$}};
\node at (-0.1,-1.5) {\large  {\bf $m$}};
\node at (0.9,-2.5) {\large  {\bf $m$}};
\node at (1.9,-3.5) {\large  {\bf $m$}};
\draw[ultra thick,red,<->] (1.05,0.95) -- (1.95,0.05);
\node[red,rotate=315] at (1.75,0.65) {{\large {\bf {$\widehat{a}_1$}}}};
\draw[ultra thick,red,<->] (2.05,-0.05) -- (2.95,-0.95);
\node[red,rotate=315] at (2.75,-0.35) {{\large {\bf {$\widehat{a}_2$}}}};
\draw[ultra thick,red,<->] (3.05,-1.05) -- (3.95,-1.95);
\node[red,rotate=315] at (3.75,-1.35) {{\large {\bf {$\widehat{a}_3$}}}};
\draw[dashed] (-0.7,-1.7) -- (-1.2,-2.2);
\draw[dashed] (3,-3) -- (1.25,-4.75);
\draw[ultra thick,red,<->] (1.3,-4.6) -- (-1.1,-2.2);
\node[red] at (-0.2,-3.7) {{\large {\bf {$S$}}}};
\end{tikzpicture}}}
\caption{\sl Web diagram of $X_{3,1}$.}
\label{Fig:31web}
\vspace{-0.55cm}
\end{center}
\end{wrapfigure} 

\noindent
presenting $X_{3,1}$, which is relevant for the $N=3$ free energy is show in \figref{Fig:31web}. In addition to the K\"ahler parameters shown in the figure, we also have
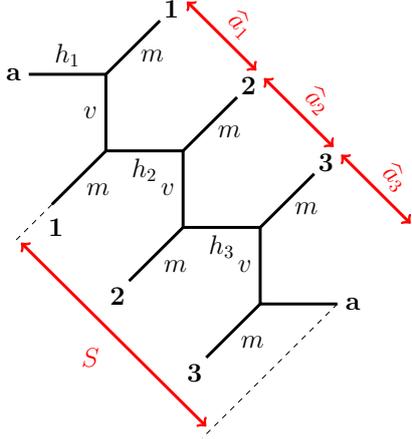
\begin{align}
&\rho=\widehat{a}_1+\widehat{a}_2+\widehat{a}_3\,,&&
R-3S=m-2v\,,
\end{align}
From the partition function $\mathcal{Z}_{3,1}$ we can compute the free energy
\begin{align}
\mathcal{F}_{3,1}(\widehat{a}_{1,2,3},S,R,\epsilon_{1,2})=\log\mathcal{Z}_{3,1}(\widehat{a}_{1,2,3},S,R,\epsilon_{1,2})\,,\nonumber
\end{align}
As in the case $N=2$, we focus exclusively on the NS-limit. In this case, following eq.~(\ref{ResumP}), we can decompose the free energy in terms of $H^{(r),\underline{n}}_{(2s,0)}$, where $\underline{n}$ can be either of the following triples
\begin{align}
&\{0,0,0\}\,,&&\{n,0,0\}\,, &&\{n,n,0\}\,, &&\{n_1+n_2,n_1,0\}\,,&&\text{with} &&n,n_1,n_2\in\mathbb{N}\,,
\end{align}
more concretely, we can write the following (a priori formal) decomposition
\begin{align}
&P^{(r)}_{3,(2s,0)}(\widehat{a}_{1,2,3},S)=H^{(r),\{0,0,0\}}_{(2s,0)}(\rho,S)+\sum_{n=1}^\infty H^{(r),\{n,0,0\}}_{(2s,0)}(\rho,S)\,\left(Q_{\widehat{a}_1}^n+Q_{\widehat{a}_2}^n+\frac{Q_\rho^n}{Q_{\widehat{a}_1}^nQ_{\widehat{a}_2}^n}\right)\nonumber\\
&\hspace{0.5cm}+\sum_{n=1}^\infty H^{(r),\{n,n,0\}}_{(2s,0)}(\rho,S)\left(Q^n_{\widehat{a}_1}Q^n_{\widehat{a}_2}+\frac{Q_\rho^n}{Q^n_{\widehat{a}_1}}+\frac{Q_\rho^n}{Q^n_{\widehat{a}_2}}\right)\nonumber\\
&\hspace{0.5cm}+\sum_{n_1,n_2=1}^\infty H^{(r),\{n_1+n_2,n_1,0\}}_{(2s,0)}(\rho,S)\left(Q_{\widehat{a}_1}^{n_1+n_2}Q_{\widehat{a}_2}^{n_1}+\frac{Q_{\widehat{a}_1}^{n_2}Q_\rho^{n_1}}{Q_{\widehat{a}_2}^{n_1}}+\frac{Q_\rho^{n_1+n_2}}{Q_{\widehat{a}_1}^{n_1+n_2}Q_{\widehat{a}_2}^{n_2}}+(\widehat{a}_1\leftrightarrow \widehat{a}_2)\right)
\,.\label{DefPN3Ord}
\end{align}
Comparing with an explicit expansion of the free energy (\ref{FreeEnergyExpansion}) we observe that the coefficients $H^{(r),\underline{n}}_{(2s,0)}$ can be written in the form
{\allowdisplaybreaks
\begin{align}
H^{(r),\{n,0,0\}}_{(2s,0)}(\rho,S)&=\frac{1}{1-Q_\rho^n}\sum_{k=1}^{rs+1}\,n^{2k-1}\,\mathfrak{f}^{(r)}_{k,(2s)}(\rho,S)+\frac{Q_\rho^n}{(1-Q_\rho^n)^2}\sum_{k=1}^{rs+1}\,n^{2k}\,\mathfrak{g}^{(r)}_{k,(2s)}(\rho,S)\,,\nonumber\\
H^{(r),\{n,n,0\}}_{(2s,0)}(\rho,S)&=\frac{1}{1-Q_\rho^n}\sum_{k=1}^{rs+1}\,n^{2k-1}\,\mathfrak{f}^{(r)}_{k,(2s)}(\rho,S)+\frac{1}{(1-Q_\rho^n)^2}\sum_{k=1}^{rs+1}\,n^{2k}\,\mathfrak{g}^{(r)}_{k,(2s)}(\rho,S)\,,\nonumber\\
H^{(r),\{n_1+n_2,n_1,0\}}_{(2s,0)}(\rho,S)&=\frac{n_2(n_2+2n_1)}{(1-Q_\rho^{n_1})(1-Q_\rho^{n_2})}\sum_{k=1}^{rs+1}\sum_{\ell}\,p^{(r)}_{\ell,k,(2s)}(n_1,-n_1-n_2)\,\mathfrak{j}^{(r)}_{\ell,k,(2s)}(\rho,S)\nonumber\\
&+\frac{(n_1^2-n_2^2)}{(1-Q_\rho^{n_1})(1-Q_\rho^{n_1+n_2})}\sum_{k=1}^{rs+1}\sum_{\ell}\,p^{(r)}_{\ell,k,(2s)}(n_1,n_2)\,\mathfrak{j}^{(r)}_{\ell,k,(2s)}(\rho,S)\,,\label{GenericFormN3}
\end{align}}
where $\mathfrak{f}^{(r)}_{k,(2s)}$ are (quasi) Jacobi forms of index $3r$ and weight $2s-2-2k$, $\mathfrak{g}^{(r)}_{k,(2s)}$ are (quasi) Jacobi forms of index $3r$ and weight $2s-4-2k$ and $\mathfrak{j}^{(r)}_{a,k,(2s)}$ are (quasi) Jacobi forms of index $3r$ and weight $2s-4-2k$. They can be written in the following fashion
\begin{align}
\mathfrak{f}^{(r)}_{k,(2s)}(\rho,S)&=-\sum_{a=0}^{3r}f^{(r)}_{a,k,(2s)}(\rho)\,\left(\phi_{-2,1}(\rho,S)\right)^a\,\left(\phi_{0,1}(\rho,S)\right)^{3r-a}\,,\nonumber\\
\mathfrak{g}^{(r)}_{k,(2s)}(\rho,S)&=-\sum_{a=0}^{3r}g^{(r)}_{a,k,(2s)}(\rho)\,\left(\phi_{-2,1}(\rho,S)\right)^a\,\left(\phi_{0,1}(\rho,S)\right)^{3r-a}\,,\nonumber\\
\mathfrak{j}^{(r)}_{\ell,k,(2s)}(\rho,S)&=-\sum_{a=0}^{3r}j^{(r)}_{a,\ell,k,(2s)}(\rho)\,\left(\phi_{-2,1}(\rho,S)\right)^a\,\left(\phi_{0,1}(\rho,S)\right)^{3r-a}\,,\nonumber
\end{align}
where $f^{(r)}_{a,k,(2s)}$, $g^{(r)}_{a,k,(2s)}$ and $j^{(r)}_{a,\ell,k,(2s)}$ are quasi-modular forms of weight $2s-2-2k+2a$, $2s-4-2k+2a$ and $2s-4-2k+2a$ respectively. Similarly, we can expand
\begin{align}
H^{(r),\{0,0,0\}}_{(2s,0)}(\rho,S)&=-\sum_{a=0}^{3r}d^{(r)}_{a,(2s)}(\rho)\,\left(\phi_{0,1}(\rho,S)\right)^a\,\left(\phi_{-2,1}(\rho,S)\right)^{3r-a}\,,
\end{align}
where $d^{(r)}_{a,(2s)}$ are quasi-Jacobi forms of weight $2s+2k$. Furthermore, $p^{(r)}_{\ell,k,(2s)}(n_1,n_2)$ in (\ref{GenericFormN3}) are homogeneous polynomials in $n_{1,2}$ of order $2(k-1)$, that are symmetric in the exchange of $n_1\leftrightarrow n_2$. Explicit expressions for $d^{(r)}_{a,k,(2s)}$, $f^{(r)}_{a,k,(2s)}$, $g^{(r)}_{a,k,(2s)}$ and $j^{(r)}_{a,k,(2s)}$ as well as $p^{(r)}_{k,(2s)}(n_1,n_2)$ for low values of $s$ for $r=1$ are tabulated in Table~\ref{Tab:CoeffsExpdr1}, Table~\ref{Tab:CoeffsExpfr1}, Table~\ref{Tab:CoeffsExpgr1} and Table~\ref{Tab:CoeffsExpjr1}  and for $r=2$ in Table~\ref{Tab:CoeffsExpdr2}, Table~\ref{Tab:CoeffsExpfr2}, Table~\ref{Tab:CoeffsExpgr2} and Table~\ref{Tab:CoeffsExpjr2}, respectively.

\begin{table}[htbp]
\begin{center}
\begin{tabular}{|c||c|c|c|c|}\hline
&&&&\\[-12pt]
$s$ & $d^{(r=1)}_{0,(2s)}$ & $d^{(r=1)}_{1,(2s)}$ & $d^{(r=1)}_{2,(2s)}$ & $d^{(r=1)}_{3,(2s)}$\\[4pt]\hline
&&&&\\[-12pt]
$0$ & $0$ & $\frac{1}{192}$ & $\frac{E_2 }{48}$ & $\frac{E_2 ^2}{48}$\\[4pt]\hline
&&&&\\[-12pt]
$1$ & $\frac{1}{18432}$ & $\frac{E_2 }{9216}$ & $\frac{2 E_4 -3 E_2 ^2}{4608}$ & $\frac{2 E_2  E_4 -3 E_2 ^3}{2304}$\\[4pt]\hline
&&&&\\[-12pt]
$2$ & $\frac{E_2 }{884736}$ & $\frac{45 E_2 ^2-43 E_4 }{4423680}$ & $\frac{8 E_6 -21 E_2  E_4 }{1105920}$ & $\frac{-45 E_2 ^4+21 E_2 ^2 E_4 +16 E_2  E_6 -10 E_4 ^2}{1105920}$\\[4pt]\hline
&&&&\\[-12pt]
$3$ & $\frac{17 E_4 -5 E_2 ^2}{424673280}$ & $\frac{315 E_2 ^3-63 E_2  E_4 -248 E_6 }{1486356480}$ & $\frac{315 E_2 ^4-819 E_2 ^2 E_4 -208 E_2  E_6 +468 E_4 ^2}{743178240}$ & $\frac{152 E_2 ^2 E_6-315 E_2 ^5-189 E_2 ^3 E_4  +300 E_2  E_4 ^2-112 E_4  E_6 }{371589120}$\\[4pt]\hline
\end{tabular}
\end{center}
\caption{Expansion coefficients $d^{(r=1)}_{a,(2s)}$.}
\label{Tab:CoeffsExpdr1}
\end{table}

\begin{table}[htbp]
\begin{center}
\begin{tabular}{|c||c||c|c|c|c|}\hline
&&&&&\\[-12pt]
$s$ & $k$ & $f^{(r=1)}_{0,k,(2s)}$ & $f^{(r=1)}_{1,k,(2s)}$ & $f^{(r=1)}_{2,k,(2s)}$ & $f^{(r=1)}_{3,k,(2s)}$\\[4pt]\hline\hline
&&&&&\\[-12pt]
$0$ & $1$ & $0$ & $0$ & $\frac{1}{12}$ & $\frac{E_2 }{6}$\\[4pt]\hline\hline
&&&&&\\[-12pt]
$1$ & $1$ & $0$ & $\frac{1}{576}$ & $0$ & $\frac{E_4 -3 E_2 ^2}{288} $\\[4pt]\hline
&&&&&\\[-12pt]
 & $2$ & $0$ & $0$ & $\frac{1}{72}$ & $\frac{E_2 }{36}$\\[4pt]\hline\hline
&&&&&\\[-12pt]
$2$ & $1$ & $\frac{-1}{110592}$ & $\frac{E_2 }{18432}$ & $\frac{15 E_2 ^2-17 E_4 }{92160}$ & $\frac{-45 E_2 ^3-9 E_2  E_4 +8 E_6 }{138240}$\\[4pt]\hline
&&&&&\\[-12pt]
 & $2$ & $0$ & $-\frac{1}{3456}$ & $0$ & $\frac{3 E_2 ^2-E_4 }{1728}$\\[4pt]\hline
&&&&&\\[-12pt]
 & $3$ & $0$ & $0$ & $-\frac{1}{1440}$ & $-\frac{E_2 }{720}$\\[4pt]\hline\hline
&&&&&\\[-12pt]
3 & 1  & $-\frac{E_2}{2654208}$ & $\frac{E_4}{442368}$ & $\frac{315 E_2^3 - 189 E_2 E_4 - 136 E_6}{46448640}$ & $-\frac{16 E_6 E_2- 255 E_4^2 + 504 E_2^2 E_4+315 E_2^4 }{46448640}$ \\
&&&&&\\[-12pt]\hline
&&&&&\\[-12pt]
 & 2 & $\frac{1}{663552}$  & $-\frac{E_2}{110592}$  & $\frac{17 E_4-15 E_2^2}{552960}$ & $-\frac{8 E_6-9E_4 E_2-45 E_2^3 }{829440}$ \\
 &&&&&\\[-12pt]\hline
&&&&&\\[-12pt]
 & 3  & $0$ & $\frac{1}{69120}$ & $0$ & $\frac{E_4-3 E_2^2 }{34560}$ \\
  &&&&&\\[-12pt]\hline
&&&&&\\[-12pt]
 & 4   & $0$ & $0$ & $\frac{1}{60480}$ & $\frac{E_2}{30240}$\\[2pt]\hline
\end{tabular}
\end{center}
\caption{Expansion coefficients $f^{(r=1)}_{a,k,(2s)}$.}
\label{Tab:CoeffsExpfr1}
\end{table}

\begin{table}[htbp]
\begin{center}
\begin{tabular}{|c||c||c|c|c|c|}\hline
&&&&&\\[-12pt]
$s$ & $k$ & $g^{(r=1)}_{0,k,(2s)}$ & $g^{(r=1)}_{1,k,(2s)}$ & $g^{(r=1)}_{2,k,(2s)}$ & $g^{(r=1)}_{3,k,(2s)}$\\[4pt]\hline\hline
&&&&&\\[-12pt]
$0$ & $1$ & $0$ & $0$ & $0$ & $1$\\[4pt]\hline\hline
&&&&&\\[-12pt]
$1$ & $1$ & $0$ & $0$ & $\frac{1}{32}$ & $-\frac{E_2 }{16}$\\[4pt]\hline
&&&&&\\[-12pt]
 & $2$ & $0$ & $0$ & $0$ & $\frac{1}{12}$\\[4pt]\hline\hline
&&&&&\\[-12pt]
$2$ & $1$ & $0$ & $-\frac{1}{3072}$ & $\frac{E_2 }{512}$ & $\frac{-15 E_2 ^2-13 E_4 }{7680}$\\[4pt]\hline
&&&&&\\[-12pt]
 & $2$ & $0$ & $0$ & $-\frac{1}{384}$ & $\frac{E_2 }{192}$\\[4pt]\hline
 &&&&&\\[-12pt]
 & $3$ & $0$ & $0$ & $0$ & $-\frac{1}{360}$\\[4pt]\hline\hline
 &&&&&\\[-12pt]
 3 & 1 & $\frac{1}{884736}$ & $-\frac{E_2}{49152}$ & $\frac{11E_4+15 E_2^2}{245760}$ & $-\frac{184 E_6+ 819 E_2 E_4 +315 E_2^3 }{7741440}$\\
&&&&&\\[-12pt]\hline
&&&&&\\[-12pt]
& 2  & $0$ & $\frac{1}{36864}$ & $-\frac{E2}{6144}$ & $\frac{13 E_4+15 E_2^2}{92160}$ \\
&&&&&\\[-12pt]\hline
&&&&&\\[-12pt]
& 3 & $0$ & $0$  & $\frac{1}{11520}$ & $-\frac{E_2}{5760}$ \\
&&&&&\\[-12pt]\hline
&&&&&\\[-12pt]
& 4 & $0$ & $0$ & $0$ & $\frac{1}{20160}$\\[4pt]\hline
\end{tabular}
\end{center}
\caption{Expansion coefficients $g^{(r=1)}_{a,k,(2s)}$.}
\label{Tab:CoeffsExpgr1}
\end{table}

\begin{table}[htbp]
\begin{center}
\begin{tabular}{|c||c||c||c|c|c|c||c|}\hline
&&&&&&&\\[-12pt]
$s$ & $k$ & $\ell$ & $j^{(r=1)}_{0,\ell,k,(2s)}$ & $j^{(r=1)}_{1,\ell,k,(2s)}$ & $j^{(r=1)}_{2,\ell,k,(2s)}$ & $j^{(r=1)}_{3,\ell,k,(2s)}$ & $p_{\ell,k,(2s)}^{(r)}$\\[4pt]\hline\hline
&&&&&&&\\[-12pt]
$0$ & $1$ & $1$ & $0$ & $0$ & $0$ & $1$ & $1$\\[4pt]\hline\hline
&&&&&&&\\[-12pt]
$1$ & $1$ & $1$ & $0$ & $0$ & $\frac{1}{32}$ & $-\frac{E_2 }{16}$ & $1$\\[4pt]\hline
&&&&&&&\\[-12pt]
 & $2$ & $1$ & $0$ & $0$ & $0$ & $\frac{1}{12}$ & $n_1^2+n_2^2$\\[4pt]\hline\hline
&&&&&&&\\[-12pt]
$2$ & $1$ & $1$ & $0$ & $-\frac{1}{3072}$ & $\frac{E_2 }{512}$ & $\frac{-15 E_2 ^2-13 E_4 }{7680}$ & $1$\\[4pt]\hline
&&&&&&&\\[-12pt]
 & $2$ & $1$ & $0$ & $0$ & $-\frac{1}{384}$ & $\frac{E_2 }{192}$ & $n_1^2+n_2^2$\\[4pt]\hline
 &&&&&&&\\[-12pt]
 & $3$ & $1$ & $0$ & $0$ & $0$ & $-\frac{1}{360}$ & $n_1^4+n_1^2n_2^2+n_2^4$\\[4pt]\hline\hline
 &&&&&&&\\[-12pt]
 3 & 1 & $1$ & $\frac{1}{884736}$ & $-\frac{E_2}{49152}$ & $\frac{11E_4+15 E_2^2}{245760}$ & $-\frac{184 E_6+ 819 E_2 E_4 +315 E_2^3 }{7741440}$ & $1$\\
&&&&&&&\\[-12pt]\hline
&&&&&&&\\[-12pt]
& 2 & $1$  & $0$ & $\frac{1}{36864}$ & $-\frac{E2}{6144}$ & $\frac{13 E_4+15 E_2^2}{92160}$ & $n_1^2+n_2^2$ \\
&&&&&&&\\[-12pt]\hline
&&&&&&&\\[-12pt]
& 3 & $1$ & $0$ & $0$  & $\frac{1}{11520}$ & $-\frac{E_2}{5760}$ & $n_1^4+n_1^2n_2^2+n_2^4$\\
&&&&&&&\\[-12pt]\hline
&&&&&&&\\[-12pt]
& 4 & $1$ & $0$ & $0$ & $0$ & $\frac{1}{20160}$ & $n_1^6+n_1^4n_2^2+n_1^2n_2^4+n_2^6$\\[4pt]\hline
\end{tabular}
\end{center}
\caption{Expansion coefficients $j^{(r=1)}_{a,k,(2s)}$.}
\label{Tab:CoeffsExpjr1}
\end{table}

\begin{table}[htbp]
\begin{center}
\begin{tabular}{|c||c|c|c|c|c|c|}\hline
&&&&&\\[-12pt]
$s$ & $d^{(r=2)}_{0,(2s)}$ & $d^{(r=2)}_{1,(2s)}$ & $d^{(r=2)}_{2,(2s)}$ & $d^{(r=2)}_{3,(2s)}$ & $d^{(r=2)}_{4,(2s)}$ \\[4pt]\hline\hline
&&&&&\\[-12pt]
$0$ & $0$ & $\frac{1}{1769472}$ & $\frac{E_2 }{221184}$ & $\frac{2 E_2 ^2+E_4 }{221184}$ & $\frac{3 E_2  E_4 -2 E_6 }{55296}$  \\[4pt]\hline
&&&&&\\[-12pt]
$1$ & $\frac{1}{512\cdot 24^4}$ & $\frac{E_2 }{42467328}$ & $\frac{5 E_4 -4 E_2 ^2}{14155776}$ & $\frac{7 E_2  E_4 -2 E_2 ^3-4 E_6 }{1769472}$ & $\frac{108 E_2 ^2 E_4 -208 E_2  E_6 +123 E_4 ^2}{10616832}$ \\[4pt]\hline
\end{tabular}
${}$\\[10pt]
\begin{tabular}{|c||c|c|}\hline
&&\\[-12pt]
$s$ & $d^{(r=2)}_{5,(2s)}$ & $d^{(r=2)}_{6,(2s)}$\\[4pt]\hline\hline
&&\\[-12pt]
$0$ & $\frac{24 E_2 ^2 E_4 -32 E_2  E_6 +9 E_4 ^2}{110592}$ & $\frac{-E_6 E_2 ^2+3 E_2  E_4 ^2-2 E_4  E_6 }{6912}$ \\[4pt]\hline
&&\\[-12pt]
$1$ & $\frac{67 E_2  E_4 ^2-24 E_2 ^3 E_4 -8 (E_6E_2 ^2 +4 E_6E_4) }{884736}$ & $\frac{128 E_6 ^2+48 E_2 ^3 E_6 +72 E_2 ^2 E_4 ^2-384 E_2  E_4  E_6 +141 E_4 ^3}{2654208}$ \\[4pt]\hline
\end{tabular}
\end{center}
\caption{Expansion coefficients $d^{(r=2)}_{a,(2s)}$.}
\label{Tab:CoeffsExpdr2}
\end{table}

\begin{table}[htbp]
\begin{center}
\begin{tabular}{|c||c||c|c|c|c|c|c|c|c|}\hline
&&&&&&&&\\[-12pt]
$s$ & $k$ & $f^{(r=2)}_{0,k,(2s)}$ & $f^{(r=2)}_{1,k,(2s)}$ & $f^{(r=2)}_{2,k,(2s)}$ & $f^{(r=2)}_{3,k,(2s)}$ & $f^{(r=2)}_{4,k,(2s)}$ & $f^{(r=2)}_{5,k,(2s)}$ & $f^{(r=2)}_{6,k,(2s)}$\\[4pt]\hline\hline
&&&&&&&&\\[-12pt]
$0$ & $1$ & $0$ & $0$ & $\frac{1}{55296}$ & $\frac{E_2 }{13824}$ & $\frac{E_4 }{4608}$ & $\frac{3 E_2  E_4 -2 E_6 }{1728}$ & $\frac{3 E_4 ^2-2 E_2  E_6 }{1728}$ \\[4pt]\hline
&&&&&&&&\\[-12pt]
& $2$ & $0$ & $0$ & $0$ & $\frac{1}{6912}$ & $\frac{E_2 }{1728}$ & $\frac{E_4 }{1728}$ & $\frac{E_2  E_4 -E_6 }{1440}$ \\[4pt]\hline
&&&&&&&&\\[-12pt]
& $3$ & $0$ & $0$ & $0$ & $0$ & $\frac{1}{13824}$ & $\frac{E_2 }{3456}$ & $\frac{E_4 }{3456}$ \\[4pt]\hline\hline
&&&&&&&&\\[-12pt]
$1$ & $1$ & $0$ & $\frac{1}{2654208}$ & $0$ & $\frac{2 E_4 -E_2 ^2}{110592}$ & $\frac{9 E_2  E_4 -8 E_6 }{82944}$ & $\frac{-12 E_2 ^2 E_4 -12 E_2  E_6 +19 E_4 ^2}{55296}$ & $\frac{E_2 ^2 E_6 +3 E_2  E_4 ^2-5 E_4  E_6 }{6912}$ \\[4pt]\hline
&&&&&&&&\\[-12pt]
 & $2$ & $0$ & $0$ & $\frac{1}{73728}$ & $\frac{7 E_2 }{165888}$ & $\frac{59 E_4 -30 E_2 ^2}{414720}$ & $\frac{120 E_2  E_4 -77 E_6 }{207360}$ & $\frac{-126 E_2 ^2 E_4 -274 E_2  E_6 +771 E_4 ^2}{1451520}$ \\[4pt]\hline
 &&&&&&&&\\[-12pt]
 & $3$ & $0$ & $0$ & $0$ & $\frac{13}{331776}$ & $\frac{25 E_2 }{165888}$ & $\frac{14 E_4 -3 E_2 ^2}{82944}$ & $\frac{33 E_2  E_4 -38 E_6 }{207360}$ \\[4pt]\hline
&&&&&&&&\\[-12pt]
 & $4$ & $0$ & $0$ & $0$ & $0$ & $\frac{1}{69120}$ & $\frac{E_2 }{17280}$ & $\frac{E_4 }{17280}$ \\[4pt]\hline
\end{tabular}
\end{center}
\caption{Expansion coefficients $f^{(r=2)}_{a,k,(2s)}$.}
\label{Tab:CoeffsExpfr2}
\end{table}

\begin{table}[htbp]
\begin{center}
\begin{tabular}{|c||c||c|c|c|c|c|c|c|c|}\hline
&&&&&&&&\\[-12pt]
$s$ & $k$ & $g^{(r=2)}_{0,k,(2s)}$ & $g^{(r=2)}_{1,k,(2s)}$ & $g^{(r=2)}_{2,k,(2s)}$ & $g^{(r=2)}_{3,k,(2s)}$ & $g^{(r=2)}_{4,k,(2s)}$ & $g^{(r=2)}_{5,k,(2s)}$ & $g^{(r=2)}_{6,k,(2s)}$\\[4pt]\hline\hline
&&&&&&&&\\[-12pt]
$0$ & $1$ & $0$ & $0$ & $0$ & $\frac{1}{2304}$ & $0$ & $\frac{E_4 }{96}$ & $-\frac{E_6 }{144}$ \\[4pt]\hline
&&&&&&&&\\[-12pt]
& $2$ & $0$ & $0$ & $0$ & $0$ & $\frac{7}{1152}$ & $0$ & $\frac{11 E_4 }{480}$ \\[4pt]\hline
&&&&&&&&\\[-12pt]
& $3$ & $0$ & $0$ & $0$ & $0$ & $0$ & $\frac{5}{576}$ & $0$ \\[4pt]\hline
&&&&&&&&\\[-12pt]
& $4$ & $0$ & $0$ & $0$ & $0$ & $0$ & $0$ & $\frac{1}{720}$ \\[4pt]\hline\hline
&&&&&&&&\\[-12pt]
$1$ & $1$ & $0$ & $0$ & $\frac{1}{73728}$ & $-\frac{E_2 }{18432}$ & $\frac{5 E_4 }{6144}$ & $\frac{-3 E_2  E_4 -5 E_6 }{2304}$ & $\frac{2 E_2  E_6 +9 E_4 ^2}{2304}$ \\[4pt]\hline
&&&&&&&&\\[-12pt]
 & $2$ & $0$ & $0$ & $0$ & $\frac{5}{13824}$ & $-\frac{7 E_2 }{9216}$ & $\frac{83 E_4 }{11520}$ & $-\frac{11 E_2  E_4 }{3840}-\frac{29 E_6 }{6048}$ \\[4pt]\hline
 &&&&&&&&\\[-12pt]
 & $3$ & $0$ & $0$ & $0$ & $0$ & $\frac{97}{55296}$ & $-\frac{5 E_2 }{4608}$ & $\frac{3 E_4 }{512}$ \\[4pt]\hline
  &&&&&&&&\\[-12pt]
 & $4$ & $0$ & $0$ & $0$ & $0$ & $0$ & $\frac{23}{17280}$ & $-\frac{E_2 }{5760}$ \\[4pt]\hline
   &&&&&&&&\\[-12pt]
 & $5$ & $0$ & $0$ & $0$ & $0$ & $0$ & $0$ & $\frac{1}{6048}$ \\[4pt]\hline
\end{tabular}
\end{center}
\caption{Expansion coefficients $g^{(r=2)}_{a,k,(2s)}$.}
\label{Tab:CoeffsExpgr2}
\end{table}

\begin{table}[htbp]
\begin{center}
\rotatebox{90}{\parbox{23.8cm}{\begin{tabular}{|c||c||c||c|c|c|c|c|c|c||c|}\hline
&&&&&&&&&&\\[-12pt]
$s$ & $k$ & $\ell$ & $j^{(r=2)}_{0,\ell,k,(2s)}$ & $j^{(r=2)}_{1,\ell,k,(2s)}$ & $j^{(r=2)}_{2,\ell,k,(2s)}$ & $j^{(r=2)}_{3,\ell,k,(2s)}$ & $j^{(r=2)}_{4,\ell,k,(2s)}$ & $j^{(r=2)}_{5,\ell,k,(2s)}$ & $j^{(r=2)}_{6,\ell,k,(2s)}$ & $p_{\ell,k,{(2s)}}^{(r=2)}$\\[4pt]\hline\hline
&&&&&&&&&&\\[-12pt]
$0$ & $1$ & $1$ & $0$ & $0$ & $0$ & $\frac{1}{4\cdot 24^2}$ & $0$ & $0$ & $0$ & $1$ \\[4pt]\hline
&&&&&&&&&&\\[-12pt]
$0$ & $1$ & $2$ & $0$ & $0$ & $0$ & $0$ & $0$ & $\frac{E_4}{96}$ & $0$ & $1$ \\[4pt]\hline
&&&&&&&&&&\\[-12pt]
$0$ & $1$ & $3$ & $0$ & $0$ & $0$ & $0$ & $0$ & $0$ & $-\frac{E_6}{144}$ & $1$ \\[4pt]\hline
&&&&&&&&&&\\[-12pt]
 & $2$ & $1$ & $0$ & $0$ & $0$ & $0$ & $\frac{1}{1152}$ & $0$ & $0$ & $7n_1^2+10n_1n_2+7n_2^2$ \\[4pt]\hline
 &&&&&&&&&&\\[-12pt]
 & $2$ & $1$ & $0$ & $0$ & $0$ & $0$ & $0$ & $0$ & $\frac{E_4}{480}$ & $11n_1^2+20n_1n_2+11n_2^2$ \\[4pt]\hline
 &&&&&&&&&&\\[-12pt]
 & $2$ & $1$ & $0$ & $0$ & $0$ & $0$ & $0$ & $\frac{1}{576}$ & $0$ & $\left(n_1 ^2+n_1  n_2 +n_2 ^2\right) \left(5 n_1 ^2+9 n_1 
   n_2 +5 n_2 ^2\right)$ \\[4pt]\hline
 &&&&&&&&&&\\[-12pt]
 & $3$ & $1$ & $0$ & $0$ & $0$ & $0$ & $0$ & $0$ & $\frac{1}{1440}$ & $(n_1 +n_2 )^2 \left(2 n_1 ^4+4 n_1 ^3 n_2 +9 n_1 ^2 n_2 ^2+4 n_1  n_2 ^3+2
   n_2 ^4\right)$ \\[4pt]\hline\hline
&&&&&&&&&&\\[-12pt]
$1$ & $1$ & $1$ & $0$ & $0$ & $\frac{1}{73728}$ & $-\frac{E_2 }{18432}$ & $\frac{5 E_4 }{6144}$ & $\frac{-3 E_2  E_4 -5 E_6 }{2304}$ & $\frac{2 E_2  E_6 +9 E_4 ^2}{2304}$ & $1$ \\[4pt]\hline
&&&&&&&&&&\\[-12pt]
 & $2$ & $1$ & $0$ & $0$ & $0$ & $\frac{5}{13824}$ & $0$ & $0$ & $0$ & $n_1^2+n_1n_2+n_2^2$ \\[4pt]\hline
&&&&&&&&&&\\[-12pt]
 & $2$ & $2$ & $0$ & $0$ & $0$ & $0$ & $-\frac{E_2 }{9216}$ & $0$ & $0$ & $7n_1^2+10n_1n_2+7n_2^2$ \\[4pt]\hline
&&&&&&&&&&\\[-12pt]
& $2$ & $3$ & $0$ & $0$ & $0$ & $0$ & $0$ & $\frac{E_4 }{11520}$ & $0$ & $83 n_1 ^2+121 n_1  n_2 +83 n_2 ^2$ \\[4pt]\hline
&&&&&&&&&&\\[-12pt]
 & $2$ & $4$ & $0$ & $0$ & $0$ & $0$ & $0$ & $0$ & $-\frac{E_2  E_4 }{3840}$ & $11 n_1 ^2+20 n_1  n_2 +11 n_2 ^2$ \\[4pt]\hline
&&&&&&&&&&\\[-12pt]
 & $2$ & $5$ & $0$ & $0$ & $0$ & $0$ & $0$ & $0$ & $-\frac{E_6 }{6048}$ & $29 n_1 ^2+48 n_1  n_2 +29 n_2 ^2$ \\[4pt]\hline
&&&&&&&&&&\\[-12pt]
 & $3$ & $1$ & $0$ & $0$ & $0$ & $0$ & $\frac{1}{55296}$ & $0$ & $0$ & $\left(n_1 ^2+n_1  n_2 +n_2 ^2\right) \left(97 n_1 ^2+141 n_1  n_2 +97 n_2 ^2\right)$ \\[4pt]\hline
&&&&&&&&&&\\[-12pt]
 & $3$ & $2$ & $0$ & $0$ & $0$ & $0$ & $0$ & $-\frac{E_2 }{4608}$ & $\frac{3 E_4 }{2560}$ & $\left(n_1 ^2+n_1  n_2 +n_2 ^2\right) \left(5 n_1 ^2+9 n_1  n_2 +5 n_2 ^2\right)$ \\[4pt]\hline
&&&&&&&&&&\\[-24pt]
 & $4$ & $1$ & $0$ & $0$ & $0$ & $0$ & $0$ & $\frac{1}{69120}$ & $0$ & \parbox{5cm}{\begin{align}92 n_1 ^6&+344 n_1 ^5 n_2+751 n_1 ^4 n_2 ^2+974 n_1 ^3 n_2 ^3 \nonumber\\
&+751 n_1 ^2 n_2 ^4+344
   n_1  n_2 ^5+92 n_2 ^6\nonumber\end{align}} \\[-10pt]\hline
   &&&&&&&&&&\\[-12pt]
 & $4$ & $2$ & $0$ & $0$ & $0$ & $0$ & $0$ & $0$ & $-\frac{E_2 }{11520}$ & $(n_1 +n_2 )^2 \big(2 n_1 ^4+4 n_1 ^3 n_2 +9 n_1 ^2 n_2 ^2+4 n_1  n_2 ^3+2
   n_2 ^4\big)$ \\[4pt]\hline\hline
&&&&&&&&&&\\[-24pt]
$0$ & $5$ & $1$ & $0$ & $0$ & $0$ & $0$ & $0$ & $0$ & $\frac{1}{120960}$ & \parbox{5cm}{\begin{align}(n_1 +n_2 )^2 \big(20 n_1 ^6+60 n_1 ^5 n_2 +165 n_1 ^4 n_2 ^2\nonumber\\
\hspace{0.2cm}+166 n_1 ^3 n_2 ^3+165n_1 ^2 n_2 ^4+60 n_1  n_2 ^5+20 n_2 ^6\big)\nonumber\\[-24pt]\nonumber\end{align}} \\[0pt]\hline
\end{tabular}}}
\end{center}
\caption{Expansion coefficients $j^{(r=2)}_{a,k,(2s)}$.}
\label{Tab:CoeffsExpjr2}
\end{table}

\subsection{Factorisation at Order $\mathcal{O}(Q_R)$}\label{FactorisationR1N3}
Following the results of \cite{Hohenegger:2019tii} for $N=2$, which have been reviewed in section~\ref{Sect:N2FactOrd1}, we expect that the free energy for the $N=3$ LSTs in the NS-limit to order $\mathcal{O}(Q_R)$ can be decomposed in terms of $H^{(1),\{0\}}_{(2s,0)}$, similar to eq.~(\ref{DecompN2MatrixFirst}). In the following we shall provide non-trivial evidence that such a decomposition indeed is possible.

From the Tables~\ref{Tab:CoeffsExpgr1} and \ref{Tab:CoeffsExpjr1} we first notice that $g^{(r=1)}_{a,k,(2s)}(\rho)=j^{(r=1)}_{a,k,(2s)}(\rho)$ and similarly, the polynomials $p_{\ell,k,{(2s)}}^{(r=1)}$ take the following simple form
\begin{align}
&p_{\ell=1,k,{(2s)}}^{(r=1)}(n_1,n_2)=\sum_{\alpha=0}^{k-1}n_1^{2k-2-2\alpha}n_2^{2\alpha}\,,&&\text{and} &&p_{\ell,k,{(2s)}}^{(r=1)}(n_1,n_2)=0\hspace{1cm} \forall \ell>1\,.
\end{align}
Furthermore, we can summarise Tables~\ref{Tab:CoeffsExpdr1} -- \ref{Tab:CoeffsExpjr1} through the following decompositions
{\allowdisplaybreaks
\begin{align}
\sum_{k=1}^{s+1}n^{2k-1}\mathfrak{f}_{k,(2s)}^{(r=1)}(\rho,S)&=-\sum_{a=0}^sW_{(2s-2a)}(\rho,S)\sum_{i,j=0}^aA_{ij}^{(a)}(n)\,H^{(1),\{0\}}_{(2i,0)}(\rho,S)\,H^{(1),\{0\}}_{(2j,0)}(\rho,S)\,,\nonumber\\
\sum_{k=1}^{s+1}\mathfrak{g}_{k,(2s)}^{(r=1)}(\rho,S)&=-\sum_{a=0}^sW_{(2s-2a)}(\rho,S)\sum_{i,j=0}^aB_{ij}^{(a)}(n)\,H^{(1),\{0\}}_{(2i,0)}(\rho,S)\,H^{(1),\{0\}}_{(2j,0)}(\rho,S)\,,\nonumber\\
\sum_{k=1}^{s+1}\sum_{\ell}p_{\ell,k,(2s)}^{(r=1)}(n_1,n_2)\,\mathfrak{j}_{\ell,k,(2s)}^{(r)}(\rho,S)&=-\sum_{a=0}^sW_{(2s-2a)}(\rho,S)\sum_{i,j=0}^aC_{ij}^{(a)}(n_1,n_2)\,H^{(1),\{0\}}_{(2i,0)}(\rho,S)\,H^{(1),\{0\}}_{(2j,0)}(\rho,S)\,,\label{DecompositionN3OrderR1}
\end{align}}
which we conjecture to hold for generic values of $s$ and where $A_{ij}^{(a)}$, $B_{ij}^{(a)}$ and $C_{ij}^{(a)}$ are $(a+1)\times (a+1)$ matrices, whose components are given by
{\allowdisplaybreaks
\begin{align}
A_{ij}^{(a)}&=-\frac{2(-1)^{a+i+j} n^{2a+1-2(i+j)}}{\Gamma(2a-2(i+j-1))}\,,\nonumber\\
B_{ij}^{(a)}&=\frac{2(-1)^{a+i+j} n^{2a+2-2(i+j)}}{\Gamma(2a+1-2(i+j-1))}\,\theta(a-i-j)\,,\nonumber\\
C_{ij}^{(a)}&=\frac{2(-1)^{a+i+j} }{\Gamma(2a+1-2(i+j-1))}\,\theta(a-i-j)\,\sum_{\alpha=0}^{a-(i+j)}n_1^{2\alpha}n_2^{2(a-1-\alpha)}\,,&&\forall\,\begin{array}{l}a\in\{0,\ldots,s\}\,,\\ i,j\in\{0,\ldots,a\}\,.\end{array}
\end{align}}
Explicitly, for low values of $a$ we have
{\allowdisplaybreaks
\begin{align}
&A^{(0)}=\left(-2n\right)\,,&&A^{(1)}=\left(\begin{array}{cc}\frac{n^3}{3} & -2n \\ -2n & 0\end{array}\right)\,,&&A^{(2)}=\left(\begin{array}{ccc}-\frac{n^5}{60} & \frac{n^3}{3} & -2n \\ \frac{n^3}{3} & -2n & 0 \\ -2n & 0 & 0 \end{array}\right)\,,\nonumber\\
&B^{(0)}=\left(-n^2\right)\,,&&B^{(1)}=\left(\begin{array}{cc}\frac{n^4}{12} & -n^2 \\ -n^2 & 0\end{array}\right)\,,&&B^{(2)}=\left(\begin{array}{ccc}-\frac{n^6}{360} & \frac{n^4}{12} & -n^2 \\ \frac{n^4}{12} & -n^2 & 0 \\ -n^2 & 0 & 0 \end{array}\right)\,,\nonumber\\
&C^{(0)}=\left(1\right)\,,&&C^{(1)}=\left(\begin{array}{cc}-\frac{n_1^2+n_2^2}{12} & 1 \\ 1 & 0\end{array}\right)\,,&&C^{(2)}=\left(\begin{array}{ccc}\frac{n_1^4+n_1^2n_2^2+n_2^4}{360} & -\frac{n_1^2+n_2^2}{12} & 1 \\ -\frac{n_1^2+n_2^2}{12} & 1 & 0 \\ 1 & 0 & 0 \end{array}\right)\,.
\end{align}}
Notice that these matrices are very closely related and satisfy for example $\partial_n B^{(a)}(n)= A^{(a)}(n)$. Moreover, (\ref{DecompositionN3OrderR1}) yields a complete decomposition of the free energy for the $N=3$ LSTs in the NS-limit, where the building blocks are only given by the free energy of the $N=1$ LST $H^{(1),\{0\}}_{(2s,0)}$ and the expansion coefficients of the NS-limit of the function (\ref{ExpansionFunctionW}) $W_{(2s)}(\rho,S)$.

\subsection{Hecke Structures}
Following the discussion in section~\ref{Sect:HeckeOrder2} for the case $N=2$, we will search for subsectors of the $N=3$ free energy which in the NS-limit are related via Hecke transformations. We shall be able to identify three different contributions that are defined via certain contour integral prescriptions.


Generalising eq.~(\ref{DecomposeN2HeckeSeed}) and eq.~(\ref{DecomposeN2HeckeSeed2}) to the case $N=3$, we can define the following three subsectors of the $N=3$ free energy
\begin{align}
\factDE{1}{r}{2s,0}(\rho,S)&:=\frac{1}{(2\pi i)^3}\sum_{\ell=0}^\infty Q_\rho^\ell\oint_0\frac{dQ_{\widehat{a}_1}}{Q_{\widehat{a}_1}^{1+\ell}}\oint_0\frac{dQ_{\widehat{a}_2}}{Q_{\widehat{a}_2}^{1+\ell}}\oint_0\frac{dQ_{\widehat{a}_3}}{Q_{\widehat{a}_3}^{1+\ell}}\,P_{2,(2s,0)}^{(r)}(\widehat{a}_1,\widehat{a}_2,\widehat{a}_3,S)\,,\label{FactN3Definition1}\\
\factDE{2}{r}{2s,0}(\rho,S)&:=\frac{1}{(2\pi i)^3\, r}\sum_{\ell=0}^\infty Q_\rho^{\ell}\oint_0 d\widehat{a}_1\,\widehat{a}_1 Q_{\widehat{a}_1}^{-\ell} \oint_0 \frac{dQ_{\widehat{a}_2}}{Q_{\widehat{a}_2}^{1+\ell}}\,\oint_0 \frac{dQ_{\widehat{a}_3}}{Q_{\widehat{a}_3}^{1+\ell}}\,P^{(r)}_{2,(2s,0)}(\widehat{a}_1,\widehat{a}_2,\widehat{a}_3,S)\,,\label{FactN3Definition2}\\
\factDE{3}{r}{2s,0}(\rho,S,\epsilon_1)&:=\frac{1}{(2\pi i)^2\,r^2}\oint_0 d\widehat{a}_1\,\widehat{a}_1\oint_{-\widehat{a}_1}d\widehat{a}_2(\widehat{a}_1+\widehat{a}_2)\, P^{(r)}_{2,(2s,0)}(\widehat{a}_1,\widehat{a}_2,\rho,S)\,.\label{FactN3Definition3}
\end{align}
Here the contour integral $\oint_z$ is along a small circle around the point $z\in\mathbb{Z}$, in such a way to extract the residue in a Laurent series. Furthermore, in the definition of $\factDE{3}{r}{2s,0}(\rho,S)$ we have implicitly used $\widehat{a}_3=\rho-\widehat{a}_1-\widehat{a}_2$. Finally, as in the case of $N=2$, we also introduce the following series in powers of $\epsilon_1$ 
\begin{align}
\factD{a}{r}(\rho,S,\epsilon_1)=\sum_{s=0}^\infty \epsilon_1^{2s-2a}\,\factDE{a}{r}{2s}(\rho,S)\,,&&\forall a=1,2.3\,,
\end{align}
In the case of $N=3$, the free energy counts BPS states of 3 M5-branes separated on a circle with multiple M2-branes stretched between them. The functions $\factD{a}{r}$ only receive contributions from certain such configurations, as is schematically shown in \figref{fig:M5braneConfigN3}:

\begin{figure}[h]
\begin{center}
\scalebox{0.69}{\parbox{7cm}{\begin{tikzpicture}[scale = 1]
\draw[very thick,fill=orange!25!white,xshift=-0.1cm, yshift=0.3cm] (1,-1.85) -- (1.75,-2.8) -- (1.75,-1.15) -- (1,-0.2) -- (1,-1.85);
\draw[very thick,fill=orange!25!white,yshift=0.3cm] (-3.9,-0.9) -- (-2.1,-0.9) -- (-2.1,0.9) -- (-3.9,0.9) -- (-3.9,-0.9);
\draw[very thick,fill=orange!25!white,xshift=-0.1cm, yshift=0.3cm] (1,1.85) -- (1.75,2.8) -- (1.75,1.15) -- (1,0.2) -- (1,1.85);
\node at (2,3) {\bf 1};
\node at (-3.65,1.55) {\bf 2};
\node at (2,-2.3) {\bf 3};
\draw [ultra thick,<->,domain=-57:65,yshift=0.7cm] plot ({3*cos(\x)}, {1.75*sin(\x)});
\node at (-1.3,2.6) {$\widehat{a}_1$}; 
\draw [thick,blue,domain=-56:65] plot ({3*cos(\x)}, {1.75*sin(\x)});
\draw [thick,blue,domain=-56:65,yshift=0.1cm] plot ({3*cos(\x)}, {1.75*sin(\x)});
\draw [thick,blue,domain=-56:65,yshift=0.2cm] plot ({3*cos(\x)}, {1.75*sin(\x)});
\node[blue] at (-0.3,1.5) {$\ell$}; 
\draw [ultra thick,<->,domain=161:70,yshift=0.7cm] plot ({3*cos(\x)}, {1.75*sin(\x)});
\node at (2.6,2) {$\widehat{a}_3$}; 
\draw [thick,blue,domain=137:73] plot ({3*cos(\x)}, {1.75*sin(\x)});
\draw [thick,blue,domain=140:73,yshift=0.1cm] plot ({3*cos(\x)}, {1.75*sin(\x)});
\draw [thick,blue,domain=145:73,yshift=0.2cm] plot ({3*cos(\x)}, {1.75*sin(\x)});
\node[blue] at (2.3,-1.5) {$\ell$}; 
\draw [ultra thick,<->,domain=-180:-63,yshift=0.7cm] plot ({3*cos(\x)}, {1.75*sin(\x)});
\node at (-1.3,-0.55) {$\widehat{a}_2$}; 
\draw [thick,blue,domain=-180:-63] plot ({3*cos(\x)}, {1.75*sin(\x)});
\draw [thick,blue,domain=-180:-63,yshift=0.1cm] plot ({3*cos(\x)}, {1.75*sin(\x)});
\draw [thick,blue,domain=-180:-63,yshift=0.2cm] plot ({3*cos(\x)}, {1.75*sin(\x)});
\node[blue] at (-1.3,-1.9) {$\ell$}; 
\node at (0,-3.6) {\large\text{\bf{(a)}}};
\end{tikzpicture}}}
\hspace{1cm}
\scalebox{0.69}{\parbox{7cm}{\begin{tikzpicture}[scale = 1]
\draw[very thick,fill=orange!25!white,xshift=-0.1cm, yshift=0.3cm] (1,-1.85) -- (1.75,-2.8) -- (1.75,-1.15) -- (1,-0.2) -- (1,-1.85);
\draw[very thick,fill=orange!25!white,yshift=0.3cm] (-3.9,-0.9) -- (-2.1,-0.9) -- (-2.1,0.9) -- (-3.9,0.9) -- (-3.9,-0.9);
\draw[very thick,fill=orange!25!white,xshift=-0.1cm, yshift=0.3cm] (1,1.85) -- (1.75,2.8) -- (1.75,1.15) -- (1,0.2) -- (1,1.85);
\node at (2,3) {\bf 1};
\node at (-3.65,1.55) {\bf 2};
\node at (2,-2.3) {\bf 3};
\draw [ultra thick,<->,domain=-57:65,yshift=0.7cm] plot ({3*cos(\x)}, {1.75*sin(\x)});
\node at (-1.3,2.6) {$\widehat{a}_1$}; 
\draw [thick,blue,domain=-56:65] plot ({3*cos(\x)}, {1.75*sin(\x)});
\draw [thick,blue,domain=-56:65,yshift=0.1cm] plot ({3*cos(\x)}, {1.75*sin(\x)});
\draw [thick,blue,domain=-56:65,yshift=0.2cm] plot ({3*cos(\x)}, {1.75*sin(\x)});
\node[red] at (-0.5,1.3) {$n\neq \ell$}; 
\draw [ultra thick,<->,domain=161:70,yshift=0.7cm] plot ({3*cos(\x)}, {1.75*sin(\x)});
\node at (2.6,2) {$\widehat{a}_3$}; 
\draw [thick,red,domain=137:73] plot ({3*cos(\x)}, {1.75*sin(\x)});
\draw [thick,red,domain=140:73,yshift=0.1cm] plot ({3*cos(\x)}, {1.75*sin(\x)});
\draw [thick,red,domain=145:73,yshift=0.2cm] plot ({3*cos(\x)}, {1.75*sin(\x)});
\draw [thick,red,domain=148:73,yshift=0.3cm] plot ({3*cos(\x)}, {1.75*sin(\x)});
\draw [thick,red,domain=152:73,yshift=0.4cm] plot ({3*cos(\x)}, {1.75*sin(\x)});
\draw [thick,red,domain=156:73,yshift=0.5cm] plot ({3*cos(\x)}, {1.75*sin(\x)});
\node[blue] at (2.3,-1.5) {$\ell$}; 
\draw [ultra thick,<->,domain=-180:-63,yshift=0.7cm] plot ({3*cos(\x)}, {1.75*sin(\x)});
\node at (-1.3,-0.55) {$\widehat{a}_2$}; 
\draw [thick,blue,domain=-180:-63] plot ({3*cos(\x)}, {1.75*sin(\x)});
\draw [thick,blue,domain=-180:-63,yshift=0.1cm] plot ({3*cos(\x)}, {1.75*sin(\x)});
\draw [thick,blue,domain=-180:-63,yshift=0.2cm] plot ({3*cos(\x)}, {1.75*sin(\x)});
\node[blue] at (-1.3,-1.9) {$\ell$}; 
\node at (0,-3.6) {\large\text{\bf{(b)}}};
\end{tikzpicture}}}
\hspace{1cm}
\scalebox{0.69}{\parbox{7cm}{\begin{tikzpicture}[scale = 1]
\draw[very thick,fill=orange!25!white,xshift=-0.1cm, yshift=0.3cm] (1,-1.85) -- (1.75,-2.8) -- (1.75,-1.15) -- (1,-0.2) -- (1,-1.85);
\draw[very thick,fill=orange!25!white,yshift=0.3cm] (-3.9,-0.9) -- (-2.1,-0.9) -- (-2.1,0.9) -- (-3.9,0.9) -- (-3.9,-0.9);
\draw[very thick,fill=orange!25!white,xshift=-0.1cm, yshift=0.3cm] (1,1.85) -- (1.75,2.8) -- (1.75,1.15) -- (1,0.2) -- (1,1.85);
\node at (2,3) {\bf 1};
\node at (-3.65,1.55) {\bf 2};
\node at (2,-2.3) {\bf 3};
\draw [ultra thick,<->,domain=-57:65,yshift=0.7cm] plot ({3*cos(\x)}, {1.75*sin(\x)});
\node at (-1.3,2.6) {$\widehat{a}_1$}; 
\draw [thick,red,domain=-56:65] plot ({3*cos(\x)}, {1.75*sin(\x)});
\draw [thick,red,domain=-56:65,yshift=0.1cm] plot ({3*cos(\x)}, {1.75*sin(\x)});
\draw [thick,red,domain=-56:65,yshift=0.2cm] plot ({3*cos(\x)}, {1.75*sin(\x)});
\node[blue] at (-0.4,1.4) {$n_1$};  
\draw [ultra thick,<->,domain=161:70,yshift=0.7cm] plot ({3*cos(\x)}, {1.75*sin(\x)});
\node at (2.6,2) {$\widehat{a}_3$}; 
\draw [thick,blue,domain=137:73] plot ({3*cos(\x)}, {1.75*sin(\x)});
\draw [thick,blue,domain=140:73,yshift=0.1cm] plot ({3*cos(\x)}, {1.75*sin(\x)});
\draw [thick,blue,domain=145:73,yshift=0.2cm] plot ({3*cos(\x)}, {1.75*sin(\x)});
\draw [thick,blue,domain=148:73,yshift=0.3cm] plot ({3*cos(\x)}, {1.75*sin(\x)});
\draw [thick,blue,domain=152:73,yshift=0.4cm] plot ({3*cos(\x)}, {1.75*sin(\x)});
\draw [thick,blue,domain=156:73,yshift=0.5cm] plot ({3*cos(\x)}, {1.75*sin(\x)});
\node[red] at (2.3,-1.5) {$n_3$}; 
\draw [ultra thick,<->,domain=-180:-63,yshift=0.7cm] plot ({3*cos(\x)}, {1.75*sin(\x)});
\node at (-1.3,-0.55) {$\widehat{a}_2$}; 
\draw [thick,green!50!black,domain=-180:-63] plot ({3*cos(\x)}, {1.75*sin(\x)});
\draw [thick,green!50!black,domain=-180:-63,yshift=0.1cm] plot ({3*cos(\x)}, {1.75*sin(\x)});
\draw [thick,green!50!black,domain=-180:-63,yshift=0.2cm] plot ({3*cos(\x)}, {1.75*sin(\x)});
\draw [thick,green!50!black,domain=-180:-63,yshift=0.3cm] plot ({3*cos(\x)}, {1.75*sin(\x)});
\draw [thick,green!50!black,domain=-180:-63,yshift=0.4cm] plot ({3*cos(\x)}, {1.75*sin(\x)});
\node[green!50!black] at (-1.3,-1.9) {$n_2$}; 
\node at (0,-3.6) {\large\text{\bf{(c)}}};
\end{tikzpicture}}}
\end{center} 
\caption{{\it Brane web configurations made up from $N=3$ M5-branes (drawn in orange) spaced out on a circle, with various M2-branes (drawn in red, blue and green) stretched between them. (a) An equal number $\ell$ of M2-branes is stretched between any two neighbouring M5-branes. Configurations of this type are relevant for $\factD{1}{r}$. (b) $\ell$ M2-branes are stretched between M5-branes 2 and 3 as well as 3 and 1, while $n\neq \ell$ M2-branes between M5-branes 1 and 2. Configurations of this type are relevant for $\factD{2}{r}$. (c) Different numbers $n_{1,2,3}$ (with $n_i\neq n_j$ for $i\neq j$) of M2-branes are stretched between any of the neighbouring M5-branes. Configurations of this type are relevant for $\factD{3}{r}$.}}
\label{fig:M5braneConfigN3}
\end{figure}

\begin{itemize}
\item Combination $\factDE{1}{r}{2s,0}$:\\
$\factDE{1}{r}{2s,0}$ can be described by extracting a particular class of terms in the Fourier expansion of $P_{3,(2,0s)}^{(r)}$ in powers of $Q_{\widehat{a}_{1,2,3}}$. Indeed, upon writing
\begin{align}
P^{(r)}_{3,(2s,0)}(\widehat{a}_1,\widehat{a}_2,\widehat{a}_3,S)=\sum_{n_1,n_2,n_3=0}^\infty Q_{\widehat{a}_1}^{n_1} Q_{\widehat{a}_2}^{n_2}Q_{\widehat{a}_3}^{n_3}\,P^{(r),\{n_1,n_2,n_3\}}_{(2s,0)}(S)\,,\label{FourierExpansionFreeN3}
\end{align}
the contour prescriptions in (\ref{FactN3Definition3}) are designed to extract only those terms with $n_1=n_2=n_3$. Therefore, $\factDE{1}{r}{2s,0}$ receives contributions only from those brane configurations, in which an equal number of M2-branes is stretched between any two adjacent M5-branes, as is visualised in \figref{fig:M5braneConfigN3}~(a). Following the definition of $H_{(2s,0)}^{(r),\underline{n}}$ in (\ref{DefAbstractH}), we find that $\factDE{1}{r}{2s,0}$ can equivalently be written as
\begin{align}
&\factDE{1}{r}{2s,0}(\rho,S)=H_{(2s,0)}^{(r),\{0,0,0\}}(\rho,S)\,,&&\factD{1}{r}(\rho,S,\epsilon_1)=\sum_{s=0}^\infty \epsilon_1^{2s-2}\,H_{(2s,0)}^{(r),\{0,0,0\}}(\rho,S)\,.\label{FactD1intoH}
\end{align}
It is in fact the reduced free energy for $N=3$ that was studied in \cite{Ahmed:2017hfr}. Explicit expansions of $\factDE{1}{r}{2s,0}$ for $r=1$ and $r=2$ can be recovered from Table~\ref{Tab:CoeffsExpdr1} and Table~\ref{Tab:CoeffsExpdr2}.
\item Combination $\factDE{2}{r}{2s,0}$:\\
The function $\factDE{2}{r}{2s,0}$ in (\ref{FactN3Definition2}) extracts specific coefficients in a mixed Fourier- and Laurent series expansion of the free energy. Starting from the Fourier expansion (\ref{FourierExpansionFreeN3}), $\factDE{2}{r}{2s,0}$ receives contributions only from coefficients with $n_1\neq n_2=n_3$. Physically, these correspond to configurations where an equal number $\ell$ of M2-branes is stretched between the second and third as well as third and first M5-branes, while a different number $n\neq \ell$ of M2-branes is stretched between the first and second M5-branes. Such configurations are schematically shown in~\figref{fig:M5braneConfigN3}~(b). Finally, the last contour integral in (\ref{FactN3Definition2}) over $\widehat{a}_1$ extracts the second order pole in the Laurent expansion.

With respect to the decomposition (\ref{DefPN3Ord}), the coefficients $\factDE{2}{r}{2s,0}$ can be written in the following form\footnote{We remark in passing that contributions with $n=\ell$ in \figref{fig:M5braneConfigN3}~(b) would give rise to terms with $H_{(2s,0)}^{(r),\{0,0,0\}}$ in (\ref{IntegD2}). The latter, however, only depends on $\rho$ and not $\widehat{a}_1$ and thus do not contribute to the contour integral over $\widehat{a}_1$ in $\factDE{2}{r}{2s}$.}
\begin{align}
\factDE{2}{r}{2s,0}(\rho,S)=\frac{1}{(2\pi i)}\oint_0 d\widehat{a}_1\,\widehat{a}_1\sum_{n=1}^\infty \left[H^{(r),\{n,0,0\}}_{(2s,0)}(\rho,S)\, Q_{\widehat{a}_1}^n+H_{(2s,0)}^{(r),\{n,n,0\}}(\rho,S)\,\frac{Q_\rho^n}{Q_{\widehat{a}_1}^n}\right]\,.\label{IntegD2}
\end{align}
In order to perform the final contour integration over $\widehat{a}_1$, we can use the conjectured form (\ref{GenericFormN3}) of $H^{(r),\{n,0,0\}}_{(2s,0)}$ and $H^{(r),\{n,n,0\}}_{(2s,0)}$ to write for the integrand
{\allowdisplaybreaks
\begin{align}
\mathcal{I}_{\factDE{2}{r}{2s,0}}&=\sum_{n=1}^\infty\left[H^{(r),\{n,0,0\}}_{(2s,0)}(\rho,S)\, Q_{\widehat{a}_1}^n+H_{(2s,0)}^{(r),\{n,n,0\}}(\rho,S)\,\frac{Q_\rho^n}{Q_{\widehat{a}_1}^n}\right]\nonumber\\
&=\sum_{n=1}^\infty\sum_{k=1}^{rs+1}\left[\frac{n^{2k-1}\,\mathfrak{f}^{(r)}_{k,(2s)}}{1-Q_\rho^n}\left(Q_{\widehat{a}_1}^n+\frac{Q_\rho^n}{Q_{\widehat{a}_1}^n}\right)+\frac{n^{2k}Q_\rho^n\,\mathfrak{g}^{(r)}_{k,(2s)}}{(1-Q_\rho^n)^2}\left(Q_{\widehat{a}_1}^n+\frac{1}{Q_{\widehat{a}_1}^n}\right)\right]\nonumber\\
&=\sum_{k=1}^{rs+1}\mathfrak{f}_{k,(2s)}^{(r)}\,\mathcal{I}_{k-1}(\rho,\widehat{a}_1)+\sum_{k=1}^{rs+1}\mathfrak{g}_{k,(2s)}^{(r)}\sum_{n=1}^\infty\frac{n^{2k}\, Q_\rho^n}{(1-Q_\rho^n)^2}\left(Q_{\widehat{a}_1}^n+Q_{\widehat{a}_1}^{-n}\right)\,,\label{ContourD2}
\end{align}}
where we have exchanged the order of summations 
and $\mathcal{I}_\alpha$ is defined in (\ref{DefIalpha}). With $\frac{x}{(1-x)^2}=\sum_{\ell=1}^\infty\ell\, x^\ell$, we can write for the sum over $n$ in the last term in~(\ref{ContourD2})
\begin{align}
\sum_{n=1}^\infty \sum_{\ell=1}^\infty n^{2k}\, \ell\,Q_\rho^{n\ell}\left(Q_{\widehat{a}_1}^n+Q_{\widehat{a}_1}^{-n}\right)=D_{\widehat{a}_1}^{2k}\sum_{n=1}^\infty Q_\rho^n\sum_{\ell|n}\frac{n}{\ell}\left(Q_{\widehat{a}_1}^\ell+Q_{\widehat{a}_1}^{-\ell}\right):=\mathcal{J}_{k}(\rho,\widehat{a}_1)=D_{\widehat{a}_1}^{2k}\mathcal{J}_{0}(\rho,\widehat{a}_1)\,.\nonumber
\end{align}
For $0<|Q_\rho|<1$ the function $\mathcal{J}_0$ is in fact regular at $\widehat{a}_1$, such that $\oint_0d\widehat{a}_1\,\widehat{a}_1\mathcal{J}_k(\rho,\widehat{a}_1)=0$ for $k\geq 0$. Using furthermore (\ref{ContourIalpha}), we get
\begin{align}
&\factDE{2}{r}{2s,0}(\rho,S)=\frac{1}{r}\,\mathfrak{f}^{(r)}_{k=1,(2s)}(\rho,S)\,,&&\factD{2}{r}(\rho,S,\epsilon_1)=\frac{1}{r}\sum_{s=0}^\infty\epsilon_1^{2s-1}\mathfrak{f}^{(r)}_{k=1,(2s)}(\rho,S)\,.
\end{align}
\item Combination $\factDE{3}{r}{2s,0}$:\\
The function $\factDE{3}{r}{2s,0}$ in (\ref{FactN3Definition3}) receives contributions from M-brane configurations with $n_i$ M2-branes stretched between the $i$th and $(i+1)$st M5-brane (with $n_i\neq n_j$ for $i\neq j$). The contour integrals, however, extract the second order poles for the successive limits $\widehat{a}_2\to -\widehat{a}_1$ and $\widehat{a}_1\to 0$. In the decomposition (\ref{DefPN3Ord}) only the terms with $\underline{n}=\{n_1+n_2,n_1,0\}$ for $n_{1,2}\geq 1$ contribute: 
\begin{align}
\factDE{3}{r}{2s,0}(\rho,S)=\frac{1}{(2\pi i)^2}\oint_0d\widehat{a}_1&\widehat{a}_1\oint d\widehat{a}_2(\widehat{a}_1+\widehat{a}_2) \sum_{n_1,n_2=1}^\infty H^{(r),\{n_1+n_2,n_1,0\}}_{(2s,0)}(\rho,S)\,X_{n_1,n_2}(\widehat{a}_1,\widehat{a}_2,\rho)\,,
\end{align}
with
\begin{align}
X_{n_1,n_2}(\widehat{a}_1,\widehat{a}_2,\rho)=Q_{\widehat{a}_1}^{n_1+n_2}Q_{\widehat{a}_2}^{n_1}+\frac{Q_{\widehat{a}_1}^{n_2}Q_\rho^{n_1}}{Q_{\widehat{a}_2}^{n_1}}+\frac{Q_\rho^{n_1+n_2}}{Q_{\widehat{a}_1}^{n_1+n_2}Q_{\widehat{a}_2}^{n_2}}+(\widehat{a}_1\leftrightarrow \widehat{a}_2)
\end{align}
Using the conjectured form (\ref{GenericFormN3}) of $H^{(r),\{n_1+n_2,n_1,0\}}_{(2s,0)}$ we can write
{\allowdisplaybreaks
\begin{align}
\factDE{3}{r}{2s,0}(\rho,S)=&\frac{1}{(2\pi i)^2}\oint_0d\widehat{a}_1\widehat{a}_1\oint_{-\widehat{a}_1} d\widehat{a}_2(\widehat{a}_1+\widehat{a}_2) \sum_{n_1,n_2=1}^\infty\sum_{k=1}^{rs+1}\sum_\ell\sum_{k_1,k_2=0} ^\infty X_{n_1,n_2}(\widehat{a}_1,\widehat{a}_2,\rho)\nonumber\\
&\times \bigg[Q_\rho^{k_1 n_1+k_2 n_2}\,n_2(n_2+2n_1)p_{\ell,k,(2s)}^{(r)}(n_1,-n_1-n_2)\,\mathfrak{j}^{(r)}_{\ell,k,(2s)}(\rho,S)\nonumber\\
&\hspace{1cm}+Q_\rho^{k_1 n_1+k_2 (n_1+n_2)}\,(n_1^2-n_2^2)p_{\ell,k,(2s)}^{(r)}(n_1,n_2)\,\mathfrak{j}^{(r)}_{\ell,k,(2s)}(\rho,S)\bigg]\,.\label{FormD3F}
\end{align}}
To further simplify this expression, let $\beta_{1,2}\in\mathbb{N}$ and consider
\begin{align}
\Upsilon=\frac{1}{(2\pi i)^2}\oint_0d\widehat{a}_1\widehat{a}_1\oint d\widehat{a}_2(\widehat{a}_1+\widehat{a}_2) \sum_{n_1,n_2=1}^\infty n_1^{\beta_1}n_2^{\beta_2}\,\sum_{k_1,k_2=0}^\infty  Q_\rho^{k_1 n_1+k_2 n_2} X_{n_1,n_2}(\widehat{a}_1,\widehat{a}_2,\rho)\,.
\end{align}
Assuming that $0<|Q_\rho|<1$, the factors $Q_\rho$ act as regulators for the sum over $n_{1,2}$ in the limit $Q_{\widehat{a}_{1,2}}\to 1$. The divergence that is relevant for the contour integrals therefore only stems from those terms where these factors are absent, namely for $k_1=k_2=0$\footnote{Notice, in order to have a pole for $\widehat{a}_2\to -\widehat{a}_1$ and $\widehat{a}_1\to 0$, both $k_1$ and $k_2$ have to vanish simultaneously.}
{\allowdisplaybreaks
\begin{align}
\Upsilon&=\frac{1}{(2\pi i)^2}\oint_0d\widehat{a}_1\widehat{a}_1\oint d\widehat{a}_2(\widehat{a}_1+\widehat{a}_2) \sum_{n_1,n_2=1}^\infty n_1^{\beta_1}n_2^{\beta_2}\,\left[(Q_{\widehat{a}_1}Q_{\widehat{a}_2})^{n_1}\,(Q_{\widehat{a}_1}^{n_2}+Q_{\widehat{a}_2}^{n_2})\right]\nonumber\\
&=\frac{1}{(2\pi i)^2}\oint_0d\widehat{a}_1\widehat{a}_1\oint d\widehat{a}_2(\widehat{a}_1+\widehat{a}_2) \bigg[D_{\widehat{a}_2}^{\beta_1}(D_{\widehat{a}_1}-D_{\widehat{a}_2})^{\beta_2}\sum_{n_1,n_2=1}^\infty (Q_{\widehat{a}_1}Q_{\widehat{a}_2})^{n_1}\,Q_{\widehat{a}_1}^{n_2}\nonumber\\
&\hspace{4.5cm}+D_{\widehat{a}_1}^{\beta_1}(D_{\widehat{a}_2}-D_{\widehat{a}_1})^{\beta_2}\sum_{n_1,n_2=1}^\infty (Q_{\widehat{a}_1}Q_{\widehat{a}_2})^{n_1}\,Q_{\widehat{a}_2}^{n_2}\bigg]
\end{align}}
Assuming that $0<|Q_{\widehat{a}_{1,2}}|<1$ we can perform the sum over $n_{1,2}$ to find
\begin{align}
\Upsilon&=\frac{1}{(2\pi i)^2}\oint_0d\widehat{a}_1\widehat{a}_1\oint d\widehat{a}_2(\widehat{a}_1+\widehat{a}_2) \bigg[D_{\widehat{a}_2}^{\beta_1}(D_{\widehat{a}_1}-D_{\widehat{a}_2})^{\beta_2}\frac{Q_{\widehat{a}_1}^2Q_{\widehat{a}_2}}{(1-Q_{\widehat{a}_1}Q_{\widehat{a}_2})(1-Q_{\widehat{a}_1})}\nonumber\\
&\hspace{4.5cm}+D_{\widehat{a}_1}^{\beta_1}(D_{\widehat{a}_2}-D_{\widehat{a}_1})^{\beta_2}\frac{Q_{\widehat{a}_1}Q_{\widehat{a}_2}^2}{(1-Q_{\widehat{a}_1}Q_{\widehat{a}_2})(1-Q_{\widehat{a}_2})}\bigg]\label{UpsilonContour}
\end{align}
From the explicit series expansions
{\allowdisplaybreaks
\begin{align}
&\frac{Q_{\widehat{a}_1}^2Q_{\widehat{a}_2}}{(1-Q_{\widehat{a}_1}Q_{\widehat{a}_2})(1-Q_{\widehat{a}_1})}=\frac{1}{(2\pi i)^2\,\widehat{a}_1(\widehat{a}_1+\widehat{a}_2)}\left[1+\pi i(2\widehat{a}_1+\widehat{a}_2)+\mathcal{O}(\widehat{a}_{1,2}^2)\right]\,,\nonumber\\
&\frac{Q_{\widehat{a}_1}Q_{\widehat{a}_2}^2}{(1-Q_{\widehat{a}_1}Q_{\widehat{a}_2})(1-Q_{\widehat{a}_2})}=\frac{1}{(2\pi i)^2\,\widehat{a}_2(\widehat{a}_1+\widehat{a}_2)}\left[1+\pi i(\widehat{a}_1+2\widehat{a}_2)+\mathcal{O}(\widehat{a}_{1,2}^2)\right]\,,\nonumber\\
&\hspace{0.5cm}=-\frac{\left(1+\frac{\widehat{a}_1+\widehat{a}_2}{\widehat{a}_1}+\frac{(\widehat{a}_1+\widehat{a}_2)^2}{\widehat{a}_1^2}+\mathcal{O}((\widehat{a}_1+\widehat{a}_2)^3)\right)}{(2\pi i)^2\widehat{a}_1(\widehat{a}_1+\widehat{a}_2)}\left[1+\pi i(\widehat{a}_1+2\widehat{a}_2)+\mathcal{O}(\widehat{a}_{1,2}^2)\right]\,,
\end{align}}
it follows that $\Upsilon$ is only non-vanishing for $\beta_1+\beta_2=2$. To understand why no higher derivatives may contribute, we define $\widehat{b}=\widehat{a}_1+\widehat{a}_2$ and consider respectively for the first and second term in (\ref{UpsilonContour})
\begin{align}
&\frac{Q_{\widehat{a}_{1,2}}}{1-Q_{\widehat{a}_{1,2}}}
\frac{Q_{\widehat{b}}}{1-Q_{\widehat{b}}}=\left[\frac{1}{2\pi i \widehat{a}_{1,2}}+\frac{1}{2}+\frac{i\pi \widehat{a}_{1,2}}{6}+\mathcal{O}(\widehat{a}_1^2)\right]\left[\frac{1}{2\pi i \widehat{b}}+\frac{1}{2}+\frac{i\pi \widehat{b}}{6}+\mathcal{O}(\widehat{b}^2)\right]\nonumber\\
&\hspace{0.5cm}=\left[\frac{1}{2\pi i \widehat{b}}+\frac{1}{2}+\frac{i\pi \widehat{b}}{6}+\mathcal{O}(\widehat{b}^2)\right]\times\left\{\begin{array}{l} \frac{1}{2\pi i \widehat{a}_{1}}+\frac{1}{2}+\frac{i\pi \widehat{a}_{1}}{6}+\mathcal{O}(\widehat{a}_1^2) \\[10pt] \frac{\left(1+\frac{\widehat{b}}{\widehat{a}_1}+\frac{\widehat{b}^2}{\widehat{a}_1^2}+\mathcal{O}(\widehat{b}^3)\right)}{2\pi i \widehat{a}_{1}}+\frac{1}{2}+\frac{i\pi (\widehat{b}-\widehat{a}_{1})}{6}+\mathcal{O}(\widehat{a}_1^2) \end{array}\right.
\end{align}
which only has poles of second order in $\widehat{a}_{1,2}$ and $\widehat{b}$ if hit with two derivatives. This implies that in (\ref{FormD3F}) only terms with $k=1$ (in which case $p^{(r)}_{\ell,k=1,(2s)}=\text{const.}$ are polynomials of order 0) contribute. Performing the explicit integrals, we obtain
\begin{align}
\factDE{3}{r}{2s,0}(\rho,S)&=\frac{1}{r^2}\sum_{\ell} p^{(r)}_{\ell,k=1,(2s)}\,\mathfrak{j}_{\ell,k=1,(2s)}^{(r)}(\rho,S)\,,\nonumber\\
\factD{3}{r}(\rho,S,\epsilon_1)&=\frac{1}{r^2}\sum_{s=0}^\infty \epsilon_1^{2s-4}\sum_{\ell} p^{(r)}_{\ell,k=1,(2s)}\,\mathfrak{j}_{\ell,k=1,(2s)}^{(r)}(\rho,S)\,.
\end{align}
\end{itemize}

\noindent
By comparing the explicit expressions for the contributions $\factDE{1,2,3}{r}{2s}$ for $r=1,2$ (and $r=3$) and $s$ up to 4, we find the that they are related through Hecke operators in the following fashion
\begin{align}
&\factDE{a}{r}{2s,0}(\rho,S)=\mathcal{H}_r\left[\factDE{1}{1}{2s,0}(\rho,S)\right]\,,&&\forall a=1,2,3\,.\label{HeckeN3}
\end{align}
The normalisation factors $1/r$ and $1/r^2$ appearing in the definition (\ref{FactN3Definition2}) and (\ref{FactN3Definition3}) were chosen to normalise the right hand side of (\ref{HeckeN3}).



\subsection{Decomposition of $\factD{i}{r}$}\label{Sect:FactorisationN3}

\subsubsection{Factorisation at Order $Q_R^1$}
Similar to the entire free energy $P^{(r)}_{3,(2s,0)}(\widehat{a}_{1,2,3},S)$ (see eq.~(\ref{DecompositionN3OrderR1})) also the functions $\factDE{1,2,3}{r=1}{2s}$ can be decomposed into small building blocks. Based on the examples provided in section~\ref{Sect:FreeEnergN3}, we find the following decomposition
{\allowdisplaybreaks
\begin{align}
\factDE{1}{r=1}{2s,0}&=3\sum_{a=0}^sH^{(1),\{0\}}_{(2s-2a)}\sum_{i,j=0}^a\delta_{a,i+j}\,W_{(2i,0)}\,W_{(2j,0)}\,,\nonumber\\
\factDE{2}{r=1}{2s,0}&=-2\sum_{a=0}^sH^{(1),\{0\}}_{(2s-2a)}(\rho,S)\sum_{i,j=0}^a\delta_{a,i+j}\,W_{(2i,0)}(\rho,S)\,H^{(1),\{0\}}_{(2j,0)}(\rho,S)\,,\nonumber\\
\factDE{3}{r=1}{2s,0}&=\sum_{a=0}^sH^{(1),\{0\}}_{(2s-2a)}\sum_{i,j=0}^a\delta_{a,i+j}\,H^{(1),\{0\}}_{(2i,0)}\,H^{(1),\{0\}}_{(2j,0)}\,.\label{R1FactorisationN3}
\end{align}}
These expressions (as similar equations in the remainder of this subsection) are understood to hold order by order in an expansion of $\epsilon_1$. Combining these expansion coefficients (in a series of $\epsilon_1$), we can equivalently write
\begin{align}
\factD{1}{r=1}(\rho,S,\epsilon_1)&=3\,\buildH{1}(\rho,S,\epsilon_1)\,\buildW{1}(\rho,S,\epsilon_1)\,\buildW{1}(\rho,S,\epsilon_1)\,,\nonumber\\
\factD{2}{r=1}(\rho,S,\epsilon_1)&=-2\epsilon_1\,\buildH{1}(\rho,S,\epsilon_1)\,\buildH{1}(\rho,S,\epsilon_1)\,\buildW{1}(\rho,S,\epsilon_1)\,,\nonumber\\
\factD{3}{r=1}(\rho,S,\epsilon_1)&=\epsilon_1^2\,\buildH{1}(\rho,S,\epsilon_1)\,\buildH{1}(\rho,S,\epsilon_1)\,\buildH{1}(\rho,S,\epsilon_1)\,,
\end{align}
where the coefficients $\buildH{1}$ and $\buildW{1}$ are defined in (\ref{DefBuildH}) and (\ref{DefBuildW}) respectively. 

\subsubsection{Factorisation at Order $Q_R^2$}\label{Sect:FactN3OrdR2}
Following the example of $N=2$ discussed in Section~\ref{Sect:N2HigherFactorisation}, we expect a decomposition of $\factD{1,2,3}{r}$ into more fundamental building blocks to hold also for $r>1$. Indeed, for $r=2$ we find
{\allowdisplaybreaks
\begin{align}
\factD{1}{2}&=\frac{4}{3}\,H_{N=1}^{(2)}\,W_{\text{NS}}^{(2)}\,W_{\text{NS}}^{(2)}+(H_{N=1}^{(1)})^6\mathfrak{T}^{(2)}_{1,(6,0)}+(H_{N=1}^{(1)})^4H_{N=1}^{(2)}\mathfrak{T}^{(2)}_{1,(4,1)}+(H_{N=1}^{(1)})^2\,(H_{N=1}^{(2)})^2\mathfrak{T}^{(2)}_{1,(2,2)}\,,\nonumber\\
\factD{2}{2}&=-\frac{8}{9}\,H_{N=1}^{(2)}\,H_{N=1}^{(2)}\,W_{\text{NS}}^{(2)}+(H_{N=1}^{(1)})^6\mathfrak{T}^{(2)}_{2,(6,0)}+(H_{N=1}^{(1)})^4H_{N=1}^{(2)}\mathfrak{T}^{(2)}_{2,(4,1)}+(H_{N=1}^{(1)})^2\,(H_{N=1}^{(2)})^2\mathfrak{T}^{(2)}_{2,(2,2)}\,,\nonumber\\
\factD{3}{2}&=\frac{4}{9}\,H_{N=1}^{(2)}\,H_{N=1}^{(2)}\,H_{N=1}^{(2)}+(H_{N=1}^{(1)})^6\mathfrak{T}^{(2)}_{3,(6,0)}+(H_{N=1}^{(1)})^4H_{N=1}^{(2)}\mathfrak{T}^{(2)}_{3,(4,1)}+(H_{N=1}^{(1)})^2\,(H_{N=1}^{(2)})^2\mathfrak{T}^{(2)}_{3,(2,2)}\,,\nonumber
\end{align}}
and for $r=3$
{\allowdisplaybreaks
\begin{align}
\factD{1}{3}&=\frac{27}{16}\,H_{N=1}^{(3)}\,W_{\text{NS}}^{(3)}\,W_{\text{NS}}^{(3)}+(H_{N=1}^{(1)})^9\,\mathfrak{T}^{(3)}_{1,(9,0,0)}+(H_{N=1}^{(1)})^7H_{N=1}^{(2)}\,\mathfrak{T}^{(3)}_{1,(7,1,0)}\nonumber\\
&\hspace{0.5cm}+(H_{N=1}^{(1)})^5(H_{N=1}^{(2)})^2\,\mathfrak{T}^{(3)}_{1,(5,2,0)}+(H_{N=1}^{(1)})^3(H_{N=1}^{(2)})^3\,\mathfrak{T}^{(3)}_{1,(3,3,0)}\nonumber\\
&\hspace{0.5cm}+H_{N=1}^{(1)}H_{N=1}^{(2)}(H_{N=1}^{(3)})^2\,\mathfrak{T}^{(3)}_{1,(1,1,2)}+(H_{N=1}^{(1)})^2(H_{N=1}^{(2)})^2H_{N=1}^{(3)}\,\mathfrak{T}^{(3)}_{1,(2,2,1)}\,,\nonumber\\
\factD{2}{3}&=-\frac{9}{8}\,H_{N=1}^{(3)}\,H_{N=1}^{(3)}\,W_{\text{NS}}^{(3)}+(H_{N=1}^{(1)})^9\,\mathfrak{T}^{(3)}_{2,(9,0,0)}+(H_{N=1}^{(1)})^7H_{N=1}^{(2)}\,\mathfrak{T}^{(3)}_{2,(7,1,0)}\nonumber\\
&\hspace{0.5cm}+(H_{N=1}^{(1)})^5(H_{N=1}^{(2)})^2\,\mathfrak{T}^{(3)}_{2,(5,2,0)}+(H_{N=1}^{(1)})^3(H_{N=1}^{(2)})^3\,\mathfrak{T}^{(3)}_{2,(3,3,0)}\nonumber\\
&\hspace{0.5cm}+H_{N=1}^{(1)}H_{N=1}^{(2)}(H_{N=1}^{(3)})^2\,\mathfrak{T}^{(3)}_{2,(1,1,2)}+(H_{N=1}^{(1)})^2(H_{N=1}^{(2)})^2H_{N=1}^{(3)}\,\mathfrak{T}^{(3)}_{2,(2,2,1)}\,,\nonumber\\
\factD{3}{3}&=\frac{9}{16}\,H_{N=1}^{(3)}\,H_{N=1}^{(3)}\,H_{N=1}^{(3)}+(H_{N=1}^{(1)})^9\,\mathfrak{T}^{(3)}_{3,(9,0,0)}+(H_{N=1}^{(1)})^7H_{N=1}^{(2)}\,\mathfrak{T}^{(3)}_{3,(7,1,0)}\nonumber\\
&\hspace{0.5cm}+(H_{N=1}^{(1)})^5(H_{N=1}^{(2)})^2\,\mathfrak{T}^{(3)}_{3,(5,2,0)}+(H_{N=1}^{(1)})^3(H_{N=1}^{(2)})^3\,\mathfrak{T}^{(3)}_{3,(3,3,0)}\nonumber\\
&\hspace{0.5cm}+H_{N=1}^{(1)}H_{N=1}^{(2)}(H_{N=1}^{(3)})^2\,\mathfrak{T}^{(3)}_{3,(1,1,2)}+(H_{N=1}^{(1)})^2(H_{N=1}^{(2)})^2H_{N=1}^{(3)}\,\mathfrak{T}^{(3)}_{3,(2,2,1)}\,.\nonumber
\end{align} }
Here $\mathfrak{T}^{(2)}_{a,(i_1,i_2)}$ are quasi modular forms that are independent of $S$, which satisfy
\begin{align}
&\mathfrak{T}^{(2)}_{3,(i_1,i_2)}=\frac{3}{\epsilon_1^2}\,\frac{\partial \mathfrak{T}^{(2)}_{2,(i_1,i_2)}}{\partial E_2(\rho)}=\frac{6}{\epsilon_1^4}\,\frac{\partial^2 \mathfrak{T}^{(2)}_{1,(i_1,i_2)}}{\partial (E_2(\rho))^2}\,,&&\mathfrak{T}^{(2)}_{2,(i_1,i_2)}=2\,\epsilon_1^2\,\frac{\partial \mathfrak{T}^{(2)}_{1,(i_1,i_2)}}{\partial E_2(\rho)}\,,\nonumber\\
&\mathfrak{T}^{(3)}_{3,(i_1,i_2,i_3)}=\frac{2}{\epsilon_1^2}\,\frac{\partial \mathfrak{T}^{(3)}_{2,(i_1,i_2,i_3)}}{\partial E_2(\rho)}=\frac{8}{3\epsilon_1^4}\,\frac{\partial^2 \mathfrak{T}^{(3)}_{1,(i_1,i_2,i_3)}}{\partial (E_2(\rho))^2}\,,&&\mathfrak{T}^{(3)}_{2,(i_1,i_2,i_3)}=\frac{4}{3\epsilon_1^2}\,\frac{\partial \mathfrak{T}^{(3)}_{1,(i_1,i_2,i_3)}}{\partial E_2(\rho)}\,,
\end{align}
and where $\mathfrak{T}^{(2)}_{3,\underline{i}}$ can be expanded in $\epsilon_1$ as follows \begin{align}
\frac{1}{\epsilon_1^6}\,\mathfrak{T}^{(2)}_{3,(6,0)}&=\frac{E_6 }{192}+\frac{E_4 ^2 \epsilon_1  ^2}{768}+\frac{11 E_4  E_6  \epsilon_1  ^4}{46080}+\frac{\epsilon_1  ^6
   \left(7 E_4 ^3+4 E_6 ^2\right)}{290304}+O\left(\epsilon_1  ^8\right)\,,\nonumber\\
\frac{1}{\epsilon_1^4}\,\mathfrak{T}^{(2)}_{3,(4,1)}&=\frac{E_4 }{8  }+\frac{E_6  \epsilon_1^2  }{48}+\frac{17 E_4 ^2 \epsilon_1  ^4}{5760}+\frac{31 E_4  E_6 
   \epsilon_1  ^6}{80640}+O\left(\epsilon_1  ^8\right)\,,\hspace{2cm}\mathfrak{T}^{(2)}_{3,(2,2)}=0\,,
\end{align}
and similarly $\mathfrak{T}^{(2)}_{3,\underline{i}}$ can be expanded as\footnote{To keep the length of this paper manageable, we refrein from explicitly writing $\mathfrak{T}^{(2,3)}_{1,2,\underline{i}}$.}
{\allowdisplaybreaks\begin{align}
\frac{1}{3\epsilon_1^{12}}\,\mathfrak{T}^{(3)}_{3,(9,0,0)}&=\frac{9 E_4 ^3+4 E_6 ^2}{62208}+\frac{19 E_4 ^2 E_6  \epsilon_1  ^2}{124416}+\frac{\epsilon_1  ^4 \left(493
   E_4 ^4+583 E_4  E_6 ^2\right)}{14929920}+O\left(\epsilon_1  ^6\right)\,,\nonumber\\
\frac{1}{3\epsilon_1^{10}}\,\mathfrak{T}^{(3)}_{3,(7,1,0)}&=\frac{5 E_4  E_6 }{1296  }+\frac{\epsilon_1^2   \left(13 E_4 ^3+7 E_6 ^2\right)}{7776}+\frac{215
   E_4 ^2 E_6  \epsilon_1  ^4}{186624}+O\left(\epsilon_1  ^6\right)\,,\nonumber\\
\frac{1}{3\epsilon_1^{8}}\,\mathfrak{T}^{(3)}_{3,(5,2,0)}&=\frac{5 E_4 ^2}{324 }+\frac{19 \epsilon_1^2 E_4  E_6 }{1944}+\epsilon_1  ^4 \left(\frac{2 E_4 ^3}{729}+\frac{73
   E_6 ^2}{46656}\right)+O\left(\epsilon_1  ^6\right)\,,\hspace{1cm}\mathfrak{T}^{(3)}_{3,(1,1,2)}=0\,.\nonumber\\
\frac{1}{3\epsilon_1^{6}}\,\mathfrak{T}^{(3)}_{3,(3,3,0)}&=\frac{20 E_6 }{243 }+\frac{\epsilon_1^2 E_4 ^2}{27  }+\frac{19 E_4  E_6  \epsilon_1^4
   }{1458}+O\left(\epsilon_1  ^6\right)\,,\hspace{1cm}\frac{1}{3\epsilon_1^{4}}\,\mathfrak{T}^{(3)}_{3,(2,2,1)}=\frac{E_4 }{3}+\frac{ \epsilon_1  ^2E_6 }{9 }+\frac{ \epsilon_1  ^4 E_4 ^2}{30}+O\left(\epsilon_1 ^6\right)\,.
\end{align}}
These examples suggest the following general form
\begin{align}
\left.\begin{array}{r}\factD{1}{r}\\[6pt] \factD{2}{r} \\[6pt] \factD{3}{r} \end{array}\right\}=\sum'_{i_1,\ldots,i_{r}}\mathfrak{T}^{(r)}_{a,(i_1,\ldots,i_{r})}\,(H_{N=1}^{(1)})^{i_1}\ldots (H_{N=1}^{(1)})^{i_{r}}+ \left(\frac{r}{\sigma_1(r)}\right)^2\left\{\begin{array}{lcl}3H_{N=1}^{(r)}\,W_{\text{NS}}^{(r)}\,W_{\text{NS}}^{(r)} & \text{for} & a=1\,, \\[6pt] -2H_{N=1}^{(r)}\,H_{N=1}^{(r)}\,W_{\text{NS}}^{(r)} & \text{for} & a=2\,, \\[6pt] H_{N=1}^{(r)}\,W_{\text{NS}}^{(r)}\,W_{\text{NS}}^{(r)} & \text{for} & a=3\,,\end{array}\right.\label{GenFormDecomposeN3R3}
\end{align}
which generalises (\ref{GenFormDecomposeN2R3}). Here the summation in (\ref{GenFormDecomposeN3R3}) is restricted to
\begin{align}
&\sum_{j=1}^{r}j i_j=3r\,,&&\text{and}&&i_1>0\,,
\end{align}
and the coefficients $\mathfrak{T}^{(r)}_{a,(i_1,\ldots,i_r)}$ satisfy
\begin{align}
&\frac{\partial \mathfrak{T}^{(r)}_{3,(i_1,\ldots,i_r)}}{\partial E_2(\rho)}=0\,,&&\mathfrak{T}^{(r)}_{3,(i_1,\ldots,i_3)}=\frac{6}{r\epsilon_1^2}\,\frac{\partial \mathfrak{T}^{(r)}_{2,(i_1,\ldots,i_r)}}{\partial E_2(\rho)}\,,&&\mathfrak{T}^{(r)}_{2,(i_1,\ldots,i_r)}=\frac{4}{r\epsilon_1^2}\,\frac{\partial \mathfrak{T}^{(r)}_{1,(i_1,\ldots,i_r)}}{\partial E_2(\rho)}\,,&&\forall r>1\,. 
\end{align}
The first equation in fact implies that $\mathfrak{T}^{(3)}_{3,(i_1,\ldots,i_r)}$ are (holomorphic) Jacobi forms.
\section{Hecke Structure for $N=4$}\label{Sect:CaseN4}
In this section we present some partial results for the LST with $N=4$. Since in this case the free energy is much more complicated than for $N=2$ or $N=3$, we shall not be able to achieve a full characterisation. However, the partial results we manage to extract fall in line with the patterns we have seen in the previous sections.
\subsection{Decomposition of the Free Energy}\label{Sect:FreeEnergN3}
As in the previous cases, the starting point is to compute the decomposition of the free energy.\\[-28pt]
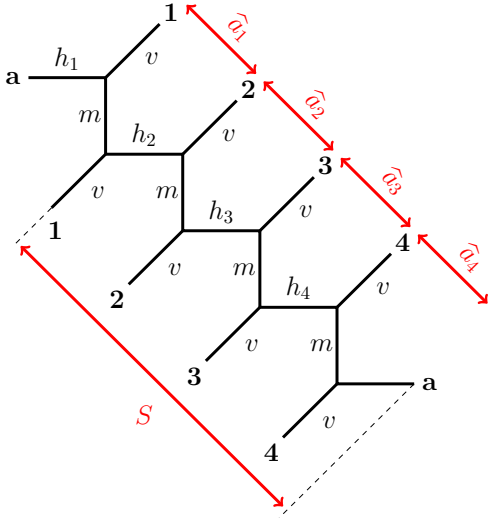
\begin{wrapfigure}{l}{0.45\textwidth}
\begin{center}
\vspace{0cm}
\scalebox{0.7}{\parbox{9.6cm}{\begin{tikzpicture}[scale = 1.45]
\draw[ultra thick] (-1,0) -- (0,0) -- (0,-1) -- (1,-1) -- (1,-2) -- (2,-2) -- (2,-3) -- (3,-3) -- (3,-4) -- (4,-4);
\draw[ultra thick] (0,0) -- (0.7,0.7);
\draw[ultra thick] (1,-1) -- (1.7,-0.3);
\draw[ultra thick] (2,-2) -- (2.7,-1.3);
\draw[ultra thick] (3,-3) -- (3.7,-2.3);
\draw[ultra thick] (0,-1) -- (-0.7,-1.7);
\draw[ultra thick] (1,-2) -- (0.3,-2.7);
\draw[ultra thick] (2,-3) -- (1.3,-3.7);
\draw[ultra thick] (3,-4) -- (2.3,-4.7);
\node at (-1.2,0) {\large {\bf $\mathbf a$}};
\node at (4.2,-4) {\large {\bf $\mathbf a$}};
\node at (0.85,0.85) {\large {$\mathbf 1$}};
\node at (1.85,-0.15) {\large {$\mathbf 2$}};
\node at (2.85,-1.15) {\large {$\mathbf 3$}};
\node at (3.85,-2.15) {\large {$\mathbf 4$}};
\node at (-0.65,-2) {\large {$\mathbf{1}$}};
\node at (0.15,-2.9) {\large {$\mathbf{2}$}};
\node at (1.15,-3.9) {\large {$\mathbf{3}$}};
\node at (2.15,-4.9) {\large {$\mathbf{4}$}};
\node at (-0.5,0.25) {\large  {\bf $h_1$}};
\node at (0.5,-0.75) {\large  {\bf $h_2$}};
\node at (1.5,-1.75) {\large  {\bf $h_3$}};
\node at (2.5,-2.75) {\large  {\bf $h_4$}};
\node at (-0.2,-0.5) {\large  {\bf $m$}};
\node at (0.8,-1.5) {\large  {\bf $m$}};
\node at (1.8,-2.5) {\large  {\bf $m$}};
\node at (2.8,-3.5) {\large  {\bf $m$}};
\node at (0.6,0.25) {\large  {\bf $v$}};
\node at (1.6,-0.75) {\large  {\bf $v$}};
\node at (2.6,-1.75) {\large  {\bf $v$}};
\node at (3.6,-2.75) {\large  {\bf $v$}};
\node at (-0.1,-1.5) {\large  {\bf $v$}};
\node at (0.9,-2.5) {\large  {\bf $v$}};
\node at (1.9,-3.5) {\large  {\bf $v$}};
\node at (2.9,-4.5) {\large  {\bf $v$}};
\draw[ultra thick,red,<->] (1.05,0.95) -- (1.95,0.05);
\node[red,rotate=315] at (1.75,0.65) {{\large {\bf {$\widehat{a}_1$}}}};
\draw[ultra thick,red,<->] (2.05,-0.05) -- (2.95,-0.95);
\node[red,rotate=315] at (2.75,-0.35) {{\large {\bf {$\widehat{a}_2$}}}};
\draw[ultra thick,red,<->] (3.05,-1.05) -- (3.95,-1.95);
\node[red,rotate=315] at (3.75,-1.35) {{\large {\bf {$\widehat{a}_3$}}}};
\draw[ultra thick,red,<->] (4.05,-2.05) -- (4.95,-2.95);
\node[red,rotate=315] at (4.75,-2.35) {{\large {\bf {$\widehat{a}_4$}}}};
\draw[dashed] (-0.7,-1.7) -- (-1.2,-2.2);
\draw[dashed] (4,-4) -- (2.25,-5.75);
\draw[ultra thick,red,<->] (2.3,-5.6) -- (-1.1,-2.2);
\node[red] at (0.5,-4.4) {{\large {\bf {$S$}}}};
\end{tikzpicture}}}
\caption{\sl Web diagram of $X_{4,1}$.}
\label{Fig:41web}
\vspace{-0.5cm}
\end{center}
\end{wrapfigure} 

\noindent
The web diagram representing $X_{4,1}$, which is relevant for the $N=4$ free energy is show in \figref{Fig:41web}. In addition to the K\"ahler parameters shown in the figure, we also have
\begin{align}
&\rho=\widehat{a}_1+\widehat{a}_2+\widehat{a}_3+\widehat{a}_4\,,&&R-4S=v-3m\,,
\end{align}
From the partition function $\mathcal{Z}_{4,1}$ we can compute the free energy
\begin{align}
\mathcal{F}_{4,1}(\widehat{a}_{1,2,3,4},S,R,\epsilon_{1,2})=\log\mathcal{Z}_{4,1}(\widehat{a}_{1,2,3,4},S,R,\epsilon_{1,2})\,,\nonumber
\end{align}
As in the cases $N=2$ and $N=3$, we focus exclusively on the NS-limit. In this case, following eq.~(\ref{ResumP}), we can decompose the free energy in terms of $H^{(r),\underline{n}}_{(2s,0)}$, where $\underline{n}$ can be either of the following combinations
{\allowdisplaybreaks
\begin{align}
&\{0,0,0,0\}\,,&&\{n,0,0,0\}\,, &&\{n,n,0,0\}\,,\nonumber\\ 
&(n,0,n,0)\,,&&(n,n,n,0)\,,&&\{n_1+n_2,n_1,0,0\}\,,\nonumber\\
&(n_1+n_2,0,n_1,0)\,,&&(n_1+n_2,n_1,n_1,0)\,,&&(n_1+n_2,n_1,0,n_1)\,,\nonumber\\
&(n_1+n_2,n_1+n_2,n_1,0)\,,&&(n_1+n_2,n_1,n_1+n_2,0)\,,&&(n_1+n_2+n_3,n_1+n_2,n_1,0)\,,\nonumber\\
&(n_1+n_2+n_3,n_1,n_1+n_2,0)\,,&&(n_1+n_2+n_3,n_1+n_2,0,n_1)\,,\label{N4Combinations}
\end{align}}
with $n,n_1,n_2,n_3\in\mathbb{N}$ as well as all combination that can be obtained from (\ref{N4Combinations}) through the action of the dihedral group $\text{Dih}_4$,  together with their cyclically permutations. The full free energy can then be written in the following somewhat symbolic fashion
\begin{align}
&P^{(r)}_{4,(2s,0)}(\widehat{a}_{1,2,3,4},S)=\sum_{\underline{m}}H_{(2s,0)}^{(r),\underline{m}}\sum_{\text{Dih}_4}Q_{\widehat{a}_1}^{m_1}Q_{\widehat{a}_2}^{m_2}Q_{\widehat{a}_3}^{m_3}Q_{\widehat{a}_4}^{m_4}\,.\label{DefPN4Ord}
\end{align}
Here the first sum $\underline{m}=(m_1,m_2,m_3,m_4)$ is over all combinations appearing in (\ref{N4Combinations}), while the second sum is over distinct orbits of $\text{Dih}_4$ acting on $(\widehat{a}_1,\widehat{a}_2,\widehat{a}_3,\widehat{a}_4)$. 

Conjectures for the $H_{(2s,0)}^{(r=1),\underline{m}}$ for all $\underline{m}$ and $s=0$ have been presented in \cite{Bastian:2019wpx}, which are of the form
\begin{align}
H_{(2s,0)}^{(r=1),\underline{m}}=-\sum_{a=0}^4 w^{(r=1),\underline{m}}_{a,(2s,0)}(\rho) \left(\phi_{-2,1}(\rho,S)\right)^a\left(\phi_{0,1}(\rho,S)\right)^{4-a}
\end{align}
For the readers convenience, we recall some of the $g^{(r=1),\underline{m}}_{(2s,0)}(\rho)$ in Table~\ref{Tab:CoeffsExpN4r1} below, which turn out to be relevant for the further discussion

\begin{table}[htbp]
\begin{center}
\begin{tabular}{|c||c|c|c|c|c|}\hline
&&&&&\\[-12pt]
$\underline{m}$ & $a=0$ & $a=1$ & $a=2$ & $a=3$ & $a=4$\\[4pt]\hline
&&&&&\\[-12pt]
$(0,0,0,0)$ & $0$ & $\frac{1}{6\cdot 24^2}$ & $\frac{E_2}{576}$ & $\frac{E_2^2}{288}$ & $\frac{E_2^3}{432}$\\[4pt]\hline
&&&&&\\[-12pt]
$(n,0,0,0)$ & $0$ & $0$ & $\frac{n}{288(1-Q_\rho^n)}$ & $\frac{nE_2}{72(1-Q_\rho^n)}+\frac{n^2 Q_\rho^n}{12(1-Q_\rho^n)^2}$ & $\frac{nE_2^2}{72(1-Q_\rho^n)}+\frac{n^2 Q_\rho^n E_2}{6(1-Q_\rho^n)^2}$\\[6pt]\hline
&&&&&\\[-12pt]
$(n,n,n,0)$ & $0$ & $0$ & $\frac{n}{288(1-Q_\rho^n)}$ & $\frac{nE_2}{72(1-Q_\rho^n)}+\frac{n^2 }{12(1-Q_\rho^n)^2}$ & $\frac{nE_2^2}{72(1-Q_\rho^n)}+\frac{n^2 E_2 }{6(1-Q_\rho^n)^2}$\\[6pt]\hline
\end{tabular}
\end{center}
\caption{Expansion coefficients $w^{(r=1),\underline{m}}_{a,(0,0)}$.}
\label{Tab:CoeffsExpN4r1}
\end{table}

\noindent
While the structure of the $H_{(2s,0)}^{(r=2),\underline{m}}$ is in general more complicated, we have managed to identify particular patterns in some of them, which allow us to conjecture the following expressions
{\allowdisplaybreaks
\begin{align}
H_{(0,0)}^{(r),(0,0,0,0)}&=\sum_{a=0}^8 v^{(r=2),(0,0,0,0)}_{a,(0,0)}(\rho) \left(\phi_{-2,1}(\rho,S)\right)^a\left(\phi_{0,1}(\rho,S)\right)^{8-a}\,,\nonumber\\
H_{(0,0)}^{(r),(n,0,0,0)}&=\frac{1}{1-Q_\rho^n}\sum_{k=1}^{3}\,n^{2k-1}\,\mathfrak{v}^{(2),(n,0,0,0)}_{1,k,(0)}(\rho,S)+\frac{Q_\rho^n}{(1-Q_\rho^n)^2}\sum_{k=1}^{4}\,n^{2k-2}\,\mathfrak{v}^{(2),(n,0,0,0)}_{2,k,(0)}(\rho,S)\nonumber\\
&\hspace{1cm}+\frac{Q_\rho^n(1+Q_\rho^n)}{(1-Q_\rho^n)^3}\sum_{k=1}^{3}\,n^{2k+1}\,\mathfrak{v}^{(2),(n,0,0,0)}_{3,k,(0)}(\rho,S)\,,\nonumber\\
H_{(0,0)}^{(r),(n,n,n,0)}&=\frac{1}{1-Q_\rho^n}\sum_{k=1}^{3}\,n^{2k-1}\,\mathfrak{v}^{(2),(n,0,0,0)}_{1,k,(0)}(\rho,S)+\frac{1}{(1-Q_\rho^n)^2}\sum_{k=1}^{4}\,n^{2k-2}\,\mathfrak{v}^{(2),(n,0,0,0)}_{2,k,(0)}(\rho,S)\nonumber\\
&\hspace{1cm}+\frac{1+Q_\rho^n}{(1-Q_\rho^n)^3}\sum_{k=1}^{3}\,n^{2k+1}\,\mathfrak{v}^{(2),(n,0,0,0)}_{3,k,(0)}(\rho,S)\,,\label{CoefSelectN4}
\end{align}}
where
\begin{align}
\mathfrak{v}^{(2),(n,0,0,0)}_{a,k,(0)}=\sum_{i=2}^8v^{(r=2),(n,0,0,0)}_{i,a,k,(0,0)}(\rho) \left(\phi_{-2,1}(\rho,S)\right)^i\left(\phi_{0,1}(\rho,S)\right)^{8-i}\,,
\end{align}
where the coefficients $v^{(r=2),(0,0,0,0)}_{a,(0,0)}$ are tabulated in Table~\ref{Tab:CoeffsExpN4r2v1} and $v^{(r=2),(0,0,0,0)}_{a,k,(0,0)}$ in Table~\ref{Tab:CoeffsExpN4r2v2}.

\begin{table}[htbp]
\begin{center}
\begin{tabular}{|c||c|}\hline
&\\[-12pt]
$a$ & $v^{(r=2),(0,0,0,0)}_{a,(0,0)}$ \\[4pt]\hline
&\\[-12pt]
$0$ & $0$ \\[4pt]\hline
&\\[-12pt]
$1$ & $-\frac{1}{4\cdot 24^6}$ \\[4pt]\hline
&\\[-12pt]
$2$ & $-\frac{E_2 }{8\cdot 24^5}$ \\[4pt]\hline
\end{tabular}
\hspace{0.2cm}
\begin{tabular}{|c||c|}\hline
&\\[-12pt]
$a$ & $\mathfrak{v}^{(r=2),(0,0,0,0)}_{a,(0,0)}$ \\[4pt]\hline
&\\[-12pt]
$3$ & $\frac{-4 E_2 ^2-E_4 }{8\cdot 24^5}$ \\[4pt]\hline
&\\[-12pt]
$4$ & $\frac{3 E_6 -2 E_2 ^3-6 E_2  E_4 }{3\cdot 24^5}$ \\[4pt]\hline
&\\[-12pt]
$5$ & $\frac{32 E_2  E_6 -28 E_2 ^2 E_4 -9 E_4 ^2}{2\cdot 24^5}$ \\[4pt]\hline
\end{tabular}
\hspace{0.2cm}
\begin{tabular}{|c||c|}\hline
$a$ & $\mathfrak{v}^{(r=2),(0,0,0,0)}_{a,(0,0)}$ \\[4pt]\hline
&\\[-12pt]
$6$ & $\frac{8 E_4  E_6 -16 E_2 ^3 E_4 +28 E_2 ^2 E_6 -21 E_2  E_4 ^2}{12\cdot 24^4}$ \\[4pt]\hline
&\\[-12pt]
$7$ & $\frac{64 E_2 ^3 E_6 -216 E_2 ^2 E_4 ^2+192 E_2  E_4  E_6 -9 E_4 ^3-32 E_6 ^2}{36\cdot 24^4}$ \\[4pt]\hline
&\\[-12pt]
$8$ & $\frac{-4 E_2 ^3 E_4 ^2+14 E_2 ^2 E_4  E_6 -9 E_2  E_4 ^3-8 E_2  E_6 ^2+7 E_4 ^2 E_6 }{36\cdot 24^3}$ \\[4pt]\hline
\end{tabular}
\end{center}
\caption{Expansion coefficients $\mathfrak{v}^{(r=2),(0,0,0,0)}_{a,(0,0)}$.}
\label{Tab:CoeffsExpN4r2v1}
\end{table}

\begin{table}[htbp]
\begin{center}
\begin{tabular}{|c||c||c|c|c|c|c|c|}\hline
&&&&&&\\[-12pt]
$a$ & $k$ & $v^{(r=2),(n,0,0,0)}_{2,a,k,(0,0)}$ & $v^{(r=2),(n,0,0,0)}_{3,a,k,(0,0)}$ & $v^{(r=2),(n,0,0,0)}_{4,a,k,(0,0)}$ & $v^{(r=2),(n,0,0,0)}_{5,a,k,(0,0)}$ & $v^{(r=2),(n,0,0,0)}_{6,a,k,(0,0)}$  \\[4pt]\hline\hline
&&&&&&\\[-12pt]
$1$ & $1$ & $-\frac{1}{4\cdot24^5}$ & $-\frac{E_2}{12\cdot 24^4}$ & $\frac{-E_2 ^2-E_4 }{6\cdot 24^4}$ & $\frac{4 E_6 -7 E_2  E_4 }{3\cdot 24^4}$ & $\frac{56 E_2  E_6 -3 E_4  \left(16 E_2 ^2+7 E_4 \right)}{6\cdot 24^4}$\\[4pt]\hline 
&&&&&&\\[-12pt]
 & $2$ & $0$ & $-\frac{1}{12\cdot 24^4}$ & $-\frac{E_2}{36\cdot 24^3}$ & $\frac{-2 E_2 ^2-E_4 }{36\cdot 24^3}$ & $\frac{3 E_6 -13 E_2  E_4 }{90\cdot 24^3}$\\[4pt]\hline 
 &&&&&&\\[-12pt]
 & $3$ & $0$ & $0$ & $-\frac{1}{24^5}$ & $-\frac{E_2}{3\cdot 24^4}$ & $\frac{-2 E_2 ^2-E_4 }{3\cdot 24^4}$\\[4pt]\hline\hline
&&&&&&\\[-10pt]
$2$ & $1$ & $0$ & $-\frac{1}{2\cdot 24^4}$ & $-\frac{E_2}{12\cdot 24^3}$ & $-\frac{7E_4}{12\cdot 24^3}$ & $\frac{27 E_6 -52 E_2  E_4 }{15\cdot 24^3}$\\[4pt]\hline 
&&&&&&\\[-10pt]
& $2$ & $0$ & $0$ & $-\frac{7}{24^4}$ & $-\frac{7E_2}{6\cdot 24^3}$ & $-\frac{17E_4}{180\cdot 24^2}$\\[4pt]\hline 
&&&&&&\\[-10pt]
& $3$ & $0$ & $0$ & $0$ & $-\frac{5}{12\cdot 24^3}$ & $-\frac{5E_2}{3\cdot 24^3}$\\[4pt]\hline 
&&&&&&\\[-10pt]
& $3$ & $0$ & $0$ & $0$ & $0$ & $-\frac{1}{15\cdot 24^3}$\\[4pt]\hline\hline
&&&&&&\\[-10pt]
3 & $1$ & $0$ & $0$ & $0$ & $0$ & $-\frac{\text{E4}}{4320}$\\[4pt]\hline 
&&&&&&\\[-10pt]
 & $2$ & $0$ & $0$ & $0$ & $0$ & $-\frac{1}{13824}$\\[4pt]\hline
  &&&&&&\\[-10pt]
 & $3$ & $0$ & $0$ & $0$  & $0$ & $-\frac{7}{34560}$ \\[4pt]\hline 
   &&&&&&\\[-10pt]
 & $4$ & $0$ & $0$ & $0$  & $0$ & $0$ \\[4pt]\hline 
    &&&&&&\\[-10pt]
 & $5$ & $0$ & $0$ & $0$  & $0$ & $0$ \\[4pt]\hline 
\end{tabular}

${}$\\[10pt]

\begin{tabular}{|c||c||c|c|}\hline
&&&\\[-12pt]
$a$ & $k$ & $v^{(r=2),(n,0,0,0)}_{7,a,k,(0,0)}$ & $v^{(r=2),(n,0,0,0)}_{8,a,k,(0,0)}$ \\[4pt]\hline\hline
&&&\\[-12pt]
$1$ & $1$ & $\frac{4 E_6  \left(E_2 ^2+E_4 \right)-9 E_2  E_4 ^2}{9\cdot 24^3}$ & $\frac{-3 E_4 ^2 \left(4 E_2 ^2+3 E_4 \right)+28 E_2  E_4  E_6 -8 E_6 ^2}{18\cdot 24^3}$\\[4pt]\hline
&&&\\[-12pt]
 & $2$ & $\frac{-12 E_2 ^2 E_4 +12 E_2  E_6 -5 E_4 ^2}{90\cdot 24^3}$ & $\frac{E_4  (E_6 -E_2  E_4 )}{180\cot 24^2}$\\[4pt]\hline
 &&&\\[-12pt]
 & $3$ & $-\frac{E_2  E_4 }{18\cdot 24^3}$ & $-\frac{E_4 ^2}{36\cdot 24^3}$\\[4pt]\hline\hline
&&&\\[-12pt] 
2 & $1$ & $\frac{40 E_2  E_6 -47 E_4 ^2}{252\cdot 24^2}$ & $\frac{E_4  (27 E_6 -22 E_2  E_4 )}{90\cdot 24^2}$\\[4pt]\hline
&&&\\[-12pt]
 & $2$ & $\frac{10 E_6 -21 E_2  E_4 }{60\cdot 24^2}$ & $\frac{20 E_2  E_6 -97 E_4 ^2}{420\cdot 24^2}$\\[4pt]\hline
&&&\\[-12pt]
 & $3$ & $-\frac{5E_4}{3\cdot 24^3}$ & $\frac{E_6 -E_2  E_4 }{15\cdot 24^2}$\\[4pt]\hline
 &&&\\[-12pt]
 & $4$ & $-\frac{E_2}{90\cdot 24^2}$ & $-\frac{E_4}{90\cdot 24^2}$\\[4pt]\hline\hline
  &&&\\[-12pt]
3 & $1$ & $\frac{\text{E6}}{1512}$ & $-\frac{\text{E4}^2}{1080}$\\[4pt]\hline
  &&&\\[-12pt]
 & $2$ & $-\frac{5 \text{E4}}{3456}$ & $\frac{19 \text{E6}}{18144}$\\[4pt]\hline
   &&&\\[-12pt]
 & $3$ & $0$ & $-\frac{\text{E4}}{1440}$\\[4pt]\hline
    &&&\\[-12pt]
 & $4$ & $-\frac{1}{21\cdot 24^2}$ & $0$\\[4pt]\hline
     &&&\\[-12pt]
 & $5$ & $0$ & $-\frac{1}{315\cdot 24^2}$\\[4pt]\hline
\end{tabular}

\end{center}
\caption{Expansion coefficients $v^{(2),(n,0,0,0)}_{i,a,k,(0,0)}$.}
\label{Tab:CoeffsExpN4r2v2}
\end{table}

\noindent
We have found evidence that other configurations in (\ref{N4Combinations}) afford similar expansions. However, due to the increased complexity, it is difficult to make conjectures based on the limited expansion of the free energy\footnote{At order $Q_R^2$ and for $s=0$, we managed to compute coefficients up to $\mathcal{O}(Q_{\widehat{a}_i}^{24})$.}.

\subsection{Hecke Structures}
Since the factorisation of the free energy for $N=4$ at order $Q_R$ and $s=0$ in the fundamental building blocks $\buildH{1}$ and $\buildW{1}$ was already commented on in \cite{Hohenegger:2019tii}, we directly turn to the extraction of contributions, that are related through Hecke transformations. 

\subsubsection{Contour Prescription}
Similar to eq.~(\ref{DecomposeN2HeckeSeed}) and (\ref{DecomposeN2HeckeSeed2}) for $N=2$ and eq.~(\ref{FactN3Definition1}), (\ref{FactN3Definition2}) and (\ref{FactN3Definition3}) for $N=3$, we define the following three subsectors of the $N=4$ free energy
{\allowdisplaybreaks
\begin{align}
\factNE{i}{r}{2s,0}(\rho,S)=&\frac{1}{(2\pi i)^4\,r^{i-1}}\sum_{\ell=0}^\infty Q_\rho^\ell \oint_0 d\widehat{a}_1\,\widehat{a}_1\oint_{-\widehat{a}_1}d\widehat{a}_2\,(\widehat{a}_1+\widehat{a}_2)\ldots\oint_{-\widehat{a}_1-\ldots-\widehat{a}_{i-2}} d_{\widehat{a}_{i-1}}(\widehat{a}_1+\ldots+\widehat{a}_{i-1})\nonumber\\
&\times \oint_0\frac{dQ_{\widehat{a}_{i}}}{Q_{\widehat{a}_{i}}^{1+\ell}}\ldots \oint_0\frac{dQ_{\widehat{a}_{4}}}{Q_{\widehat{a}_{4}}^{1+\ell}}\,P_{4,(2s,0)}^{(r)}(\widehat{a}_1,\widehat{a}_2,\widehat{a}_3,\widehat{a}_4,S)\,,\hspace{2cm}\forall i=1,2,3,4.\label{FactN4Definition}
\end{align}}
In the following, we shall exclusively focus on $\factNE{1}{r}{2s,0}$ and $\factNE{2}{r}{2s,0}$, for which the functions presented in (\ref{CoefSelectN4}) are relevant:

\begin{figure}[h]
\begin{center}
\scalebox{0.8}{\parbox{6.1cm}{\begin{tikzpicture}[scale = 1]
\draw[very thick,fill=orange!25!white] (1.5,-1.25) -- (2.75,-2.5) -- (2.75,-1) -- (1.5,0.25) -- (1.5,-1.25);
\draw[very thick,fill=orange!25!white] (-1.5,-1.25) -- (-2.75,-2.5) -- (-2.75,-1) -- (-1.5,0.25) -- (-1.5,-1.25);
\draw[very thick,fill=orange!25!white] (1.2,1.95) -- (2.25,3) -- (2.25,1.5) -- (1.2,0.45) -- (1.2,1.95);
\draw[very thick,fill=orange!25!white] (-1.2,1.95) -- (-2.25,3) -- (-2.25,1.5) -- (-1.2,0.45) -- (-1.2,1.95);
\draw [thick,blue,domain=-35:55,yshift=0.3cm] plot ({3*cos(\x)}, {1.75*sin(\x)});
\draw [thick,blue,domain=-37:55,yshift=0.4cm] plot ({3*cos(\x)}, {1.75*sin(\x)});
\draw [thick,blue,domain=-39:55,yshift=0.5cm] plot ({3*cos(\x)}, {1.75*sin(\x)});
\node[blue] at (2.6,0.6) {$\ell$}; 
%
\draw [thick,blue,domain=-135:-45,yshift=0.3cm] plot ({3*cos(\x)}, {1.75*sin(\x)});
\draw [thick,blue,domain=-135:-45,yshift=0.4cm] plot ({3*cos(\x)}, {1.75*sin(\x)});
\draw [thick,blue,domain=-135:-45,yshift=0.5cm] plot ({3*cos(\x)}, {1.75*sin(\x)});
\node[blue] at (0,-1.8) {$\ell$}; 
%
\draw [thick,blue,domain=125:215,yshift=0.3cm] plot ({3*cos(\x)}, {1.75*sin(\x)});
\draw [thick,blue,domain=125:217,yshift=0.4cm] plot ({3*cos(\x)}, {1.75*sin(\x)});
\draw [thick,blue,domain=125:219,yshift=0.5cm] plot ({3*cos(\x)}, {1.75*sin(\x)});
\node[blue] at (-2.6,0.6) {$\ell$}; 
%
\draw [thick,blue,domain=66:113,yshift=0.3cm] plot ({3*cos(\x)}, {1.75*sin(\x)});
\draw [thick,blue,domain=66:115,yshift=0.4cm] plot ({3*cos(\x)}, {1.75*sin(\x)});
\draw [thick,blue,domain=64:116,yshift=0.5cm] plot ({3*cos(\x)}, {1.75*sin(\x)});
\node[blue] at (0,1.6) {$\ell$}; 
\node at (2.6,2.9) {\bf 1};
\node at (-2.6,2.9) {\bf 2};
\node at (-3.1,-2.4) {\bf 3};
\node at (3.1,-2.4) {\bf 4};
\draw [ultra thick,<->,domain=-44:55,yshift=0.9cm] plot ({3*cos(\x)}, {1.75*sin(\x)});
\node at (2.8,2.2) {$\widehat{a}_4$}; 
\draw [ultra thick,<->,domain=60:121,yshift=0.9cm] plot ({3*cos(\x)}, {1.75*sin(\x)});
\node at (0.1,3) {$\widehat{a}_1$}; 
\draw [ultra thick,<->,domain=125:223,yshift=0.9cm] plot ({3*cos(\x)}, {1.75*sin(\x)});
\node at (-2.8,2.2) {$\widehat{a}_2$}; 
\draw [ultra thick,<->,domain=230:310,yshift=0.9cm] plot ({3*cos(\x)}, {1.75*sin(\x)});
\node at (0,-0.5) {$\widehat{a}_3$}; 
\node at (0,-3.6) {\large\text{\bf{(a)}}};
\end{tikzpicture}}}
\hspace{4cm}
\scalebox{0.8}{\parbox{6.1cm}{\begin{tikzpicture}[scale = 1]
\draw[very thick,fill=orange!25!white] (1.5,-1.25) -- (2.75,-2.5) -- (2.75,-1) -- (1.5,0.25) -- (1.5,-1.25);
\draw[very thick,fill=orange!25!white] (-1.5,-1.25) -- (-2.75,-2.5) -- (-2.75,-1) -- (-1.5,0.25) -- (-1.5,-1.25);
\draw[very thick,fill=orange!25!white] (1.2,1.95) -- (2.25,3) -- (2.25,1.5) -- (1.2,0.45) -- (1.2,1.95);
\draw[very thick,fill=orange!25!white] (-1.2,1.95) -- (-2.25,3) -- (-2.25,1.5) -- (-1.2,0.45) -- (-1.2,1.95);
\draw [thick,blue,domain=-35:55,yshift=0.3cm] plot ({3*cos(\x)}, {1.75*sin(\x)});
\draw [thick,blue,domain=-37:55,yshift=0.4cm] plot ({3*cos(\x)}, {1.75*sin(\x)});
\draw [thick,blue,domain=-39:55,yshift=0.5cm] plot ({3*cos(\x)}, {1.75*sin(\x)});
\node[blue] at (2.6,0.6) {$\ell$}; 
%
\draw [thick,blue,domain=-135:-45,yshift=0.3cm] plot ({3*cos(\x)}, {1.75*sin(\x)});
\draw [thick,blue,domain=-135:-45,yshift=0.4cm] plot ({3*cos(\x)}, {1.75*sin(\x)});
\draw [thick,blue,domain=-135:-45,yshift=0.5cm] plot ({3*cos(\x)}, {1.75*sin(\x)});
\node[blue] at (0,-1.8) {$\ell$}; 
%
\draw [thick,blue,domain=125:215,yshift=0.3cm] plot ({3*cos(\x)}, {1.75*sin(\x)});
\draw [thick,blue,domain=125:217,yshift=0.4cm] plot ({3*cos(\x)}, {1.75*sin(\x)});
\draw [thick,blue,domain=125:219,yshift=0.5cm] plot ({3*cos(\x)}, {1.75*sin(\x)});
\node[blue] at (-2.6,0.6) {$\ell$}; 
\draw [thick,red,domain=62:117,yshift=0.6cm] plot ({3*cos(\x)}, {1.75*sin(\x)});
\draw [thick,red,domain=60:119,yshift=0.7cm] plot ({3*cos(\x)}, {1.75*sin(\x)});
\draw [thick,red,domain=66:113,yshift=0.3cm] plot ({3*cos(\x)}, {1.75*sin(\x)});
\draw [thick,red,domain=66:115,yshift=0.4cm] plot ({3*cos(\x)}, {1.75*sin(\x)});
\draw [thick,red,domain=64:116,yshift=0.5cm] plot ({3*cos(\x)}, {1.75*sin(\x)});
\node[red] at (0,1.6) {$n\neq\ell$}; 
\node at (2.6,2.9) {\bf 1};
\node at (-2.6,2.9) {\bf 2};
\node at (-3.1,-2.4) {\bf 3};
\node at (3.1,-2.4) {\bf 4};
\draw [ultra thick,<->,domain=-44:55,yshift=0.9cm] plot ({3*cos(\x)}, {1.75*sin(\x)});
\node at (2.8,2.2) {$\widehat{a}_4$}; 
\draw [ultra thick,<->,domain=60:121,yshift=0.9cm] plot ({3*cos(\x)}, {1.75*sin(\x)});
\node at (0.1,3) {$\widehat{a}_1$}; 
\draw [ultra thick,<->,domain=125:223,yshift=0.9cm] plot ({3*cos(\x)}, {1.75*sin(\x)});
\node at (-2.8,2.2) {$\widehat{a}_2$}; 
\draw [ultra thick,<->,domain=230:310,yshift=0.9cm] plot ({3*cos(\x)}, {1.75*sin(\x)});
\node at (0,-0.5) {$\widehat{a}_3$}; 
\node at (0,-3.6) {\large\text{\bf{(b)}}};
\end{tikzpicture}}}
\end{center} 
\caption{{\it Brane web configurations made up from $N=4$ M5-branes (drawn in orange) spaced out on a circle, with various M2-branes (drawn in red and blue) stretched between them. (a) An equal number $\ell$ of M2-branes is stretched between any two neighbouring M5-branes. Configurations of this type are relevant for the computation of $\factN{1}{r}$. (b) $\ell$ M2-branes are stretched between M5-branes 2 and 3, 3 and 4 as well as 4 and 1, while $n\neq \ell$ M2-branes between M5-branes 1 and 2. Configurations of this type are relevant for the computation of $\factN{2}{r}$. }}
\label{fig:M5braneConfigN4}
\end{figure}


\begin{itemize}
\item Combination $\factNE{1}{r}{2s,0}$\\
As before, $\factNE{1}{r}{2s,0}$ can be described by extracting a particular class of terms in the Fourier expansion of $P_4^{(r)}$ in powers of $Q_{\widehat{a}_{1,2,3,4}}$. Indeed, upon writing
\begin{align}
P^{(r)}_{4,(2s,0)}(\widehat{a}_1,\widehat{a}_2,\widehat{a}_3,\widehat{a}_4,S)=\sum_{n_1,n_2,n_3,n_4=0}^\infty Q_{\widehat{a}_1}^{n_1} Q_{\widehat{a}_2}^{n_2}Q_{\widehat{a}_3}^{n_3}Q_{\widehat{a}_4}^{n_4}\,P^{(r),\{n_1,n_2,n_3,n_4\}}_{(2s,0)}(S)\,,\label{FourierExpansionFreeN4}
\end{align}
the contour prescriptions for $\factN{1}{r}$ in (\ref{FactN4Definition}) are designed to extract only those terms with $n_1=n_2=n_3=n_4$. Therefore, $\factDE{1}{r}{2s,0}$ receives contributions only from those brane configurations, in which an equal number of M2-branes is stretched between any two adjacent M5-branes, as is visualised in \figref{fig:M5braneConfigN4}~(a). Following the definition of $H_{(2s,0)}^{(r),\underline{n}}$ in (\ref{DefAbstractH}), we find that $\factDE{1}{r}{2s,0}$ can equivalently be written as
\begin{align}
&\factNE{1}{r}{2s,0}(\rho,S)=H_{(2s,0)}^{(r),\{0,0,0,0\}}(\rho,S)\,,&&\factN{1}{r}(\rho,S,\epsilon_1)=\sum_{s=0}^\infty \epsilon_1^{2s-1}\,H_{(2s,0)}^{(r),\{0,0,0,0\}}(\rho,S)\,.\label{FactN1intoH}
\end{align}
This is in fact the reduced free energy for $N=4$ that was studied in \cite{Ahmed:2017hfr}. Explicit expansions of $\factN{1}{r}$ for $r=1$ and $r=2$ can be recovered from Table~\ref{Tab:CoeffsExpN4r1} and Table~\ref{Tab:CoeffsExpN4r2v1}.
\item Combination $\factN{2}{r}$:\\
The function $\factN{2}{r}$ in (\ref{FactN4Definition}) extracts specific coefficients in a mixed Fourier- and Laurent series expansion of the free energy. Starting from the Fourier expansion (\ref{FourierExpansionFreeN4}) $\factN{2}{r}$ receives contributions only from coefficients with $n_1\neq n_2=n_3=n_4$. From the brane web picture, these correspond to configurations where an equal number $\ell$ of M2-branes is stretched between the M5 branes 2 and 3, 3 and 4 as well as 4 and 1, while a different number $n\neq \ell$ of M2-branes is stretched between the first and second M5-branes. Such configurations are schematically shown in~\figref{fig:M5braneConfigN4}~(b). Finally, the last contour integral in (\ref{FactN3Definition2}) over $\widehat{a}_1$ extracts the second order pole in the Laurent expansion.

With respect to the decomposition (\ref{DefPN4Ord}), the coefficients $\factDE{2}{r}{2s,0}$ can be written in the following form
\begin{align}
\factNE{2}{r}{2s,0}(\rho,S)=\frac{1}{2\pi i}\oint_0 d\widehat{a}_1\,\widehat{a}_1\sum_{n=1}^\infty \left[H^{(r),\{n,0,0,0\}}_{(2s,0)}(\rho,S)\, Q_{\widehat{a}_1}^n+H_{(2s,0)}^{(r),\{n,n,n,0\}}(\rho,S)\,\frac{Q_\rho^n}{Q_{\widehat{a}_1}^n}\right]\,.\label{IntegN2}
\end{align}
In order to perform the final contour integration over $\widehat{a}_1$, we can use the conjectured form (\ref{CoefSelectN4}) of $H^{(r),\{n,0,0,0\}}_{(0,0)}$ and $H^{(r),\{n,n,n,0\}}_{(0,0)}$ for $s=0$ and $r=1$ and $r=2$ to write for the integrand
{\allowdisplaybreaks
\begin{align}
\mathcal{I}_{\factNE{2}{r}{0}}&=\sum_{n=1}^\infty\left[H^{(r),\{n,0,0,0\}}_{(0,0)}(\rho,S)\, Q_{\widehat{a}_1}^n+H_{(0,0)}^{(r),\{n,n,n,0\}}(\rho,S)\,\frac{Q_\rho^n}{Q_{\widehat{a}_1}^n}\right]\nonumber\\
&=\sum_{n=1}^\infty\sum_{k=1}^{3}\frac{n^{2k-1}\mathfrak{v}^{(r),(n,0,0,0)}_{1,k,(0)}}{1-Q_\rho^n}\,\left(Q_{\widehat{a}_1}^n+\frac{Q_\rho^n}{Q^n_{\widehat{a}_1}}\right)+\sum_{n=1}^\infty\sum_{k=1}^{4}\frac{n^{2k-2} Q_\rho^n \mathfrak{v}^{(r),(n,0,0,0)}_{2,k,(0)}}{(1-Q_\rho^n)^2}\,\left(Q_{\widehat{a}_1}^n+\frac{1}{Q^n_{\widehat{a}_1}}\right)\nonumber\\
&\hspace{1cm}+\sum_{n=1}^\infty\sum_{k=1}^{3}\frac{ n^{2k+1} Q_\rho^n(1+Q_\rho^n)}{(1-Q_\rho^n)^3}\,\mathfrak{v}^{(r),(n,0,0,0)}_{3,k,(0)}\left(Q_{\widehat{a}_1}^n+\frac{1}{Q^n_{\widehat{a}_1}}\right)\nonumber\\
&=\sum_{k=1}^{3}\mathfrak{v}^{(r),(n,0,0,0)}_{2,k,(0)}\,\mathcal{I}_{k-1}(\rho,\widehat{a}_1)+\sum_{k=1}^{4}\mathfrak{v}^{(r),(n,0,0,0)}_{2,k,(0)}\sum_{n=1}^\infty\frac{n^{2k-2} Q_\rho^n}{(1-Q_\rho^n)^2}(Q_{\widehat{a}_1}^n+Q_{\widehat{a}_1}^{-n})\nonumber\\
&\hspace{1cm}+\sum_{k=1}^{3}\mathfrak{v}^{(r),(n,0,0,0)}_{3,k,(0)}\sum_{n=1}^\infty\frac{n^{2k+1} Q_\rho^n (1+Q_\rho^n)}{(1-Q_\rho^n)^3}(Q_{\widehat{a}_1}^n+Q_{\widehat{a}_1}^{-n})\,,\label{ContourN2}
\end{align}}
where we have exchanged the summation over $k$ and $n$ and $\mathcal{I}_\alpha$ is defined in (\ref{DefIalpha}). Using the geometric series
\begin{align}
&\frac{x}{(1-x)^2}=\sum_{\ell=1}^\infty\ell\, x^\ell\,,&&\text{and}&&\frac{x(1+x)}{(1-x)^3}=\sum_{\ell=1}^\infty \ell^2 x^\ell\,,&&\text{for} &&|x|<1\,,
\end{align}
we can write for the sum over $n$ in the last two terms in (\ref{ContourN2})
\begin{align}
&\sum_{n=1}^\infty\frac{n^{2k-2} Q_\rho^n}{(1-Q_\rho^n)^2}(Q_{\widehat{a}_1}^n+Q_{\widehat{a}_1}^{-n})=\sum_{n=1}^\infty\sum_{\ell=1}^\infty n^{2k-2}\ell Q_{\rho}^{n\ell}(Q_{\widehat{a}_1}^n+Q_{\widehat{a}_1}^{-n})\nonumber\\
&\hspace{1cm}=D_{\widehat{a}_1}^{2k-2}\sum_{n=1}^\infty Q_\rho^n\sum_{\ell|n}\,\frac{n}{\ell}(Q_{\widehat{a}_1}^\ell+Q_{\widehat{a}_1}^{-\ell})=
D_{\widehat{a}_1}^{2k-2}\buildInt{1}{0}(\rho,\widehat{a}_1)\,,\\
&\sum_{n=1}^\infty\frac{n^{2k+1} Q_\rho^n(1+Q_\rho^n)}{(1-Q_\rho^n)^3}(Q_{\widehat{a}_1}^n+Q_{\widehat{a}_1}^{-n})=\sum_{n=1}^\infty\sum_{\ell=1}^\infty n^{2k+1}\ell^2 Q_{\rho}^{n\ell}(Q_{\widehat{a}_1}^n+Q_{\widehat{a}_1}^{-n})\nonumber\\
&\hspace{1cm}=D_{\widehat{a}_1}^{2k+1}\sum_{n=1}^\infty Q_\rho^n\sum_{\ell|n}\,\left(\frac{n}{\ell}\right)^2(Q_{\widehat{a}_1}^\ell+Q_{\widehat{a}_1}^{-\ell})=
D_{\widehat{a}_1}^{2k+1}\buildInt{2}{0}(\rho,\widehat{a}_1)\,.
\end{align}
For $0<|Q_\rho|<1$ the functions $\buildInt{1}{0}$ and $\buildInt{2}{0}$ are fact regular at $\widehat{a}_1=0$, such that $\oint_0d\widehat{a}_1\,\widehat{a}_1\mathcal{D}_{\widehat{a}_1}^k\buildInt{1,2}{0}(\rho,\widehat{a}_1)=0$ for $k\geq 0$. Using furthermore (\ref{ContourIalpha}), we get
\begin{align}
&\factNE{2}{r}{2s=0}(\rho,S)=\frac{1}{r}\,\mathfrak{v}^{(r),(n,0,0,0)}_{1,1,(0)}\,,&&\factN{2}{r}(\rho,S,\epsilon_1)=\frac{1}{r}\sum_{s=0}^\infty\epsilon_1^{2s-1}\mathfrak{v}^{(r),(n,0,0,0)}_{1,1,(0)}(\rho,S)\,.
\end{align}
\end{itemize}
While we leave the study of the other functions to future work, we remark in passing that the $\widehat{a}_i$ of which we extract the pole $Q_{\widehat{a}_i}^{-2}$ in (\ref{FactN4Definition}) are consecutive. An interesting question is if it makes sense to define more general functions, as for example
\begin{align}
\factNp{3}{r}{2s,0}(\rho,S)=&\frac{1}{r^{i-1}}\sum_{\ell=0}^\infty Q_\rho^\ell \oint_0 d\widehat{a}_1\,\widehat{a}_1\oint_{-\widehat{a}_1}d\widehat{a}_3\,(\widehat{a}_1+\widehat{a}_3) \oint_0\frac{dQ_{\widehat{a}_{1}}}{Q_{\widehat{a}_{1}}^{1+\ell}}\oint_0\frac{dQ_{\widehat{a}_{4}}}{Q_{\widehat{a}_{4}}^{1+\ell}}\,P_{4,(2s,0)}^{(r)}(\widehat{a}_{1,\ldots,4},S)\,,\label{FactN4pDefinition}
\end{align}

\begin{figure}[h]
\begin{center}
\scalebox{0.8}{\parbox{6.1cm}{\begin{tikzpicture}[scale = 1]
\draw[very thick,fill=orange!25!white] (1.5,-1.25) -- (2.75,-2.5) -- (2.75,-1) -- (1.5,0.25) -- (1.5,-1.25);
\draw[very thick,fill=orange!25!white] (-1.5,-1.25) -- (-2.75,-2.5) -- (-2.75,-1) -- (-1.5,0.25) -- (-1.5,-1.25);
\draw[very thick,fill=orange!25!white] (1.2,1.95) -- (2.25,3) -- (2.25,1.5) -- (1.2,0.45) -- (1.2,1.95);
\draw[very thick,fill=orange!25!white] (-1.2,1.95) -- (-2.25,3) -- (-2.25,1.5) -- (-1.2,0.45) -- (-1.2,1.95);
\draw [thick,blue,domain=-35:55,yshift=0.3cm] plot ({3*cos(\x)}, {1.75*sin(\x)});
\draw [thick,blue,domain=-37:55,yshift=0.4cm] plot ({3*cos(\x)}, {1.75*sin(\x)});
\draw [thick,blue,domain=-39:55,yshift=0.5cm] plot ({3*cos(\x)}, {1.75*sin(\x)});
\node[blue] at (2.6,0.6) {$\ell$}; 
%
\draw [thick,blue,domain=-135:-45,yshift=0.3cm] plot ({3*cos(\x)}, {1.75*sin(\x)});
\draw [thick,blue,domain=-135:-45,yshift=0.4cm] plot ({3*cos(\x)}, {1.75*sin(\x)});
\draw [thick,blue,domain=-135:-45,yshift=0.5cm] plot ({3*cos(\x)}, {1.75*sin(\x)});
\node[blue] at (0,-1.8) {$\ell$}; 
\draw [thick,green!50!black,domain=125:221,yshift=0.6cm] plot ({3*cos(\x)}, {1.75*sin(\x)});
\draw [thick,green!50!black,domain=125:215,yshift=0.3cm] plot ({3*cos(\x)}, {1.75*sin(\x)});
\draw [thick,green!50!black,domain=125:217,yshift=0.4cm] plot ({3*cos(\x)}, {1.75*sin(\x)});
\draw [thick,green!50!black,domain=125:219,yshift=0.5cm] plot ({3*cos(\x)}, {1.75*sin(\x)});
\node[green!50!black] at (-2.6,0.6) {$n_2$}; 
\draw [thick,red,domain=62:117,yshift=0.6cm] plot ({3*cos(\x)}, {1.75*sin(\x)});
\draw [thick,red,domain=60:119,yshift=0.7cm] plot ({3*cos(\x)}, {1.75*sin(\x)});
\draw [thick,red,domain=66:113,yshift=0.3cm] plot ({3*cos(\x)}, {1.75*sin(\x)});
\draw [thick,red,domain=66:115,yshift=0.4cm] plot ({3*cos(\x)}, {1.75*sin(\x)});
\draw [thick,red,domain=64:116,yshift=0.5cm] plot ({3*cos(\x)}, {1.75*sin(\x)});
\node[red] at (0,1.6) {$n_1$}; 
\node at (2.6,2.9) {\bf 1};
\node at (-2.6,2.9) {\bf 2};
\node at (-3.1,-2.4) {\bf 3};
\node at (3.1,-2.4) {\bf 4};
\draw [ultra thick,<->,domain=-44:55,yshift=0.9cm] plot ({3*cos(\x)}, {1.75*sin(\x)});
\node at (2.8,2.2) {$\widehat{a}_4$}; 
\draw [ultra thick,<->,domain=60:121,yshift=0.9cm] plot ({3*cos(\x)}, {1.75*sin(\x)});
\node at (0.1,3) {$\widehat{a}_1$}; 
\draw [ultra thick,<->,domain=125:223,yshift=0.9cm] plot ({3*cos(\x)}, {1.75*sin(\x)});
\node at (-2.8,2.2) {$\widehat{a}_2$}; 
\draw [ultra thick,<->,domain=230:310,yshift=0.9cm] plot ({3*cos(\x)}, {1.75*sin(\x)});
\node at (0,-0.5) {$\widehat{a}_3$}; 
\node at (0,-3.6) {\large\text{\bf{(a)}}};
\end{tikzpicture}}}
\hspace{4cm}
\scalebox{0.8}{\parbox{6.1cm}{\begin{tikzpicture}[scale = 1]
\draw[very thick,fill=orange!25!white] (1.5,-1.25) -- (2.75,-2.5) -- (2.75,-1) -- (1.5,0.25) -- (1.5,-1.25);
\draw[very thick,fill=orange!25!white] (-1.5,-1.25) -- (-2.75,-2.5) -- (-2.75,-1) -- (-1.5,0.25) -- (-1.5,-1.25);
\draw[very thick,fill=orange!25!white] (1.2,1.95) -- (2.25,3) -- (2.25,1.5) -- (1.2,0.45) -- (1.2,1.95);
\draw[very thick,fill=orange!25!white] (-1.2,1.95) -- (-2.25,3) -- (-2.25,1.5) -- (-1.2,0.45) -- (-1.2,1.95);
\draw [thick,blue,domain=-35:55,yshift=0.3cm] plot ({3*cos(\x)}, {1.75*sin(\x)});
\draw [thick,blue,domain=-37:55,yshift=0.4cm] plot ({3*cos(\x)}, {1.75*sin(\x)});
\draw [thick,blue,domain=-39:55,yshift=0.5cm] plot ({3*cos(\x)}, {1.75*sin(\x)});
\node[blue] at (2.6,0.6) {$\ell$}; 
\draw [thick,green!50!black,domain=-135:-45,yshift=0.2cm] plot ({3*cos(\x)}, {1.75*sin(\x)});
\draw [thick,green!50!black,domain=-135:-45,yshift=0.3cm] plot ({3*cos(\x)}, {1.75*sin(\x)});
\draw [thick,green!50!black,domain=-135:-45,yshift=0.4cm] plot ({3*cos(\x)}, {1.75*sin(\x)});
\draw [thick,green!50!black,domain=-135:-45,yshift=0.5cm] plot ({3*cos(\x)}, {1.75*sin(\x)});
\node[green!50!black] at (0,-1.8) {$n_3$}; 
%
\draw [thick,blue,domain=125:215,yshift=0.3cm] plot ({3*cos(\x)}, {1.75*sin(\x)});
\draw [thick,blue,domain=125:217,yshift=0.4cm] plot ({3*cos(\x)}, {1.75*sin(\x)});
\draw [thick,blue,domain=125:219,yshift=0.5cm] plot ({3*cos(\x)}, {1.75*sin(\x)});
\node[blue] at (-2.6,0.6) {$\ell$}; 
\draw [thick,red,domain=62:117,yshift=0.6cm] plot ({3*cos(\x)}, {1.75*sin(\x)});
\draw [thick,red,domain=60:119,yshift=0.7cm] plot ({3*cos(\x)}, {1.75*sin(\x)});
\draw [thick,red,domain=66:113,yshift=0.3cm] plot ({3*cos(\x)}, {1.75*sin(\x)});
\draw [thick,red,domain=66:115,yshift=0.4cm] plot ({3*cos(\x)}, {1.75*sin(\x)});
\draw [thick,red,domain=64:116,yshift=0.5cm] plot ({3*cos(\x)}, {1.75*sin(\x)});
\node[red] at (0,1.6) {$n_1$}; 
\node at (2.6,2.9) {\bf 1};
\node at (-2.6,2.9) {\bf 2};
\node at (-3.1,-2.4) {\bf 3};
\node at (3.1,-2.4) {\bf 4};
\draw [ultra thick,<->,domain=-44:55,yshift=0.9cm] plot ({3*cos(\x)}, {1.75*sin(\x)});
\node at (2.8,2.2) {$\widehat{a}_4$}; 
\draw [ultra thick,<->,domain=60:121,yshift=0.9cm] plot ({3*cos(\x)}, {1.75*sin(\x)});
\node at (0.1,3) {$\widehat{a}_1$}; 
\draw [ultra thick,<->,domain=125:223,yshift=0.9cm] plot ({3*cos(\x)}, {1.75*sin(\x)});
\node at (-2.8,2.2) {$\widehat{a}_2$}; 
\draw [ultra thick,<->,domain=230:310,yshift=0.9cm] plot ({3*cos(\x)}, {1.75*sin(\x)});
\node at (0,-0.5) {$\widehat{a}_3$}; 
\node at (0,-3.6) {\large\text{\bf{(b)}}};
\end{tikzpicture}}}
\end{center} 
\caption{{\it Brane web configurations made up from $N=4$ M5-branes (drawn in orange) spaced out on a circle, with various M2-branes (drawn in red and blue) stretched between them. (a) An equal number $\ell$ of M2-branes is stretched between the neighbouring M5-branes 3 and 4 as well as 4 and 1, while different numbers of M2-branes $n_1\neq n_2\neq \ell$ are stretched between the M5-branes 1 and 2 as well as 2 and 3. Configurations of this type are relevant for the computation of $\factN{3}{r}$. (b) An equal number $\ell$ of M2-branes is stretched between the neighbouring M5-branes 2 and 3 as well as 4 and 1, while different numbers of M2-branes $n_1\neq n_3\neq \ell$ are stretched between the M5-branes 1 and 2 as well as 3 and 4. Configurations of this type are relevant for the computation of $\factNp{3}{r}{2s,0}$.}}
\label{fig:M5braneConfigpN4}
\end{figure}
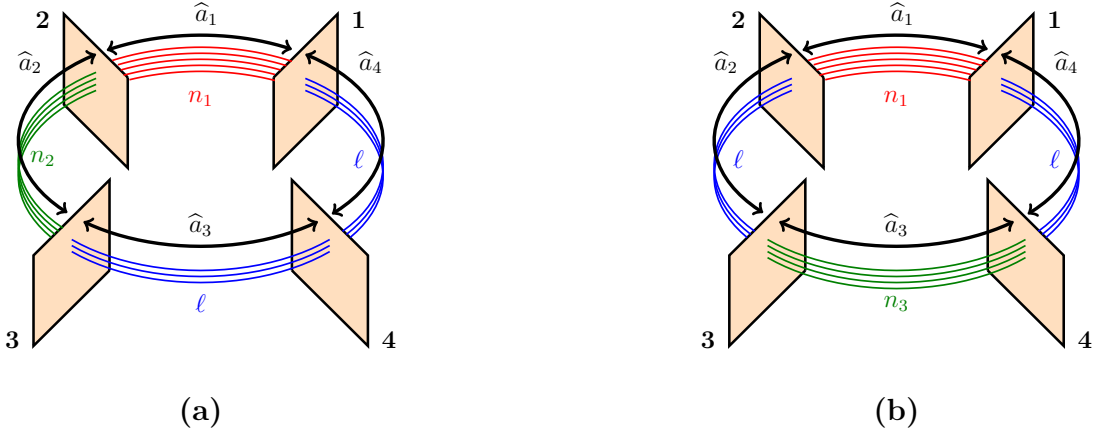
 
Both $\factN{3}{r}$ and $\factNp{3}{r}{2s,0}$ receive contributions from slightly different M2-brane configurations (they are schematically shown in \figref{fig:M5braneConfigpN4}). With respect to the list in (\ref{N4Combinations}), the precise configurations are respectively
\begin{align}
&\factNE{3}{r}{2s,0}:\,&&\{n_1+n_2,n_1,0,0\}\,,\hspace{0.5cm}\{n_1+n_2,n_1,n_1,0\}\hspace{0.5cm}\{n_1+n_2,n_1+n_2,n_1,0\}\,,\hspace{0.5cm}\{n,n,0,0\}\,,\nonumber\\
&\factNp{3}{r}{2s,0}:\,&&\{n_1+n_2,0,n_1,0\}\,,\hspace{0.5cm}\{n_1+n_2,n_1,0,n_1\}\,,\hspace{0.5cm}\{n_1+n_2,n_1,n_1+n_2,0\}\,,\hspace{0.5cm}\{n,0,n,0\}\,.\nonumber
\end{align}
From \cite{Bastian:2019wpx}, one can see that the corresponding $H^{(r=1),\underline{n}}_{(0,0)}$ in the case of $\factNp{3}{r}{2s,0}$ all involve polynomials of the $\underline{n}$ or order 3 or higher (while there are contribuions with polynomials of order $2$ in the case of $\factN{3}{r}$). Following the experience of the previous sections, this suggests that (at least to leading order in $Q_R$) $\factNp{3}{r=1}{2s,0}$ in fact may be vanishing.

\subsubsection{Factorisation and Hecke Relations}
In the cases $N=2$ and $N=3$ we have observed that the functions $\fact{a}{r=1}$ and $\factD{a}{r=1}$ can be factorised as in eq.~(\ref{DecomposeN2HeckeSeed2}) and eq.~(\ref{R1FactorisationN3}) respectively. Analysing $\factNE{i}{r=1}{2s,0}$ for $s>0$ is very complicated, however, based on the expansions presented above, we find
\begin{align}
\factNE{1}{r=1}{0,0}(\rho,S)&=4\,H^{(1),\{0\}}_{(0,0)}(\rho,S)\,\left(W_{(0,0)}(\rho,S)\right)^3\,,\nonumber\\
\factNE{2}{r=1}{0,0}(\rho,S)&=\left(H^{(1),\{0\}}_{(0,0)}(\rho,S)\right)^2\,\left(W_{(0,0)}(\rho,S)\right)^2\,,
\end{align}
which are indeed in agreement with the general conjectured form (\ref{FactorisationO1}). Moreover, by comparing the explicit expressions for the contributions $\factNE{1}{r}{0,0}$ and $\factNE{2}{r}{0,0}$ for $r=1$ and $r=2$ to the free energy, we find that they satisfy the following recursion relation
\begin{align}
&\factNE{1}{r=2}{0,0}(\rho,S)=\mathcal{H}_2\left[\factNE{1}{1}{0,0}(\rho,S)\right]\,,&&\factNE{2}{r=2}{0,0}(\rho,S)=\mathcal{H}_2\left[\factNE{2}{1}{0,0}(\rho,S)\right]\,.\label{HeckeN4}
\end{align}
which generalises the relations (\ref{HeckeN2}) and (\ref{HeckeN3}) to $N=4$. In view of the results of the previous sections, we conjecture that this result in fact generalises not only for $r>2$ and $s>0$, but also to all functions $\factN{i}{r}$ for $i=1,\ldots,4$.

\section{Conclusions and Interpretation}\label{Sect:Conclusion}
Although the observations of the previous sections were only for the specific cases $N=2$ and $N=3$ (as well as partially for $N=4$) and for limited values of the order of $Q_R$ (indicated by the index $r$) as well as $\epsilon_1$ (indicated by the index $2s$), the fact that they exhibit a rather clear cut pattern leads us to believe that they hold in general (\emph{i.e.} for generic $N$ and generic values of $r$ and $s$). To be concrete, we therefore conjecture that for given $N$, to any instanton order\footnote{Here we are taking the point of view of the $U(N)$ gauge theory that is engineered from the Calabi-Yau threefold $X_{N,1}$,} $r$ we can extract at every order $\epsilon_1^{2s-2}$ (for $s\in\mathbb{N}$) $N$ different functions $\factGE{i}{r}{2s,0}(\rho,S)$ (see eq.~(\ref{FactGenDefinition}) for the definitions) for $i=1,\ldots,N$ from the NS-limit of the free energy $P_N^{(r)}(\widehat{a}_{1,\ldots,N})$ that count very specific BPS states from the perspective of the M-brane webs. Indeed, focusing on configurations where the same number of M2-branes is stretched between $N-i$ neighbouring M5-branes, they extract a particular polar part of the free energy when the remaining M5-branes are collapsed on top of each other. Viewed order by order in $Q_R$, the formal series $\factG{i}{r}(\rho,S,\epsilon_1)$ for different values of $r$ are related through Hecke transformations (see eq.~(\ref{HeckeNCon})). This generalises the observation made in \cite{Ahmed:2017hfr}, which in our language is the specific case $\factG{i=1}{r}(\rho,S,\epsilon_1)$. Furthermore, following the logic put forward in \cite{Ahmed:2017hfr,Bastian:2019hpx}, the Hecke relation (\ref{HeckeNCon}) suggests that the BPS-states counted by $\factG{i=1}{r}(\rho,S,\epsilon_1)$ can be arranged into the form of a symmetric torus orbifold CFTs and we can define the corresponding CFT partition functions,
\begin{align}
Z^{(N)}_{i}(R,\rho,S,\epsilon_1)=\mbox{exp}\Big(\sum_{r\geq 1}Q_{R}^{N\,r}{\cal C}^{N,(r)}_{i}(\rho,S,\epsilon_1)\Big)\,.\label{ConclusionPartitionFunction}
\end{align}
The relation (\ref{HeckeNCon}) ${\cal C}^{N,(r)}_{i}(\rho,S,\epsilon_1)={\cal H}_{r}\Big({\cal C}^{N,(1)}_{i}(\rho,S,\epsilon_1)\Big)$ then implies \cite{Dijkgraaf:1996xw},
\begin{align}
Z^{(N)}_{i}(R,\rho,S,\epsilon_1)&=\mbox{exp}\Big(\sum_{r\geq 1}Q_{R}^{N\,r}{\cal H}_{r}\Big({\cal C}^{N,(1)}_{i}(\rho,S,\epsilon_1)\Big)\Big)\\
&=\sum_{r\geq }Q_{R}^{r}\,\chi_{\text{ell}}(\mbox{Sym}^{r}({\cal M}_{i}))\,.
\label{ReducedPartitionFunction}
\end{align}
Here $Q_{R}$ keeps track of the symmetric products and we conjecture the existence of spaces ${\cal M}_{i}$ with equivariant elliptic genus ${\cal C}^{N,(1)}_{i}(\rho,S,\epsilon_1)$. We only defined the terms with instanton order, in the language of the dual supersymmetric gauge theory, $\mathcal{O}(Q_R^r)$ with $r>0$, \emph{i.e.} we have not included the terms coming from the little string partition function with $Q_{R}=0$ which correspond to perturbative corrections in the dual gauge theory. Furthermore, to make a paramodular symmetry of the CFT partition function more manifest \cite{Bastian:2019hpx}, we have used $Q_{R}^N$ as the generating function parameter rather than $Q_{R}$ in (\ref{ReducedPartitionFunction}): indeed, the Hecke structure of eq.(\ref{HeckeN2}) implies that $Z^{(N)}_{i}(R,\rho,S,\epsilon_1)$ is the partition function of a symmetric orbifold conformal field theory on the torus that is invariant under the paramodular group $\Sigma_{N}^*\subset Sp(4,\mathbb{R})$ (see appendix~\ref{App:Paramodular} for the definition). To make invariance under $\Sigma_N^*$ more manifest, we remark that the Hecke structure of eq.(\ref{HeckeN2}) in $Z^{(N)}_{i}$ can be expressed in a product form~\cite{Dijkgraaf:1996xw},
\bea
Z^{(N)}_{i}=\prod_{r,k,\ell}\Big(1-Q_{R}^{N\,r}\,Q_{\rho}^{k}Q_{S}^{\ell}\,q^{p}\Big)^{-c_{i}(k\,r,\ell,p)}\,,
\eea
where $c_{i}(k,\ell,p)$ are the Fourier coefficients of the 'seed function' ${\cal C}^{N,(1)}_{i}(\rho,S,\epsilon_1)$,
\bea
{\cal C}^{N,(1)}_{i}(\rho,S,\epsilon_1)=\sum_{k,\ell,p}c_{i}(k,\ell,p)\,Q_{\rho}^{k}Q_{S}^{\ell}\,q^p\,.
\eea
Thus, the partition function $Z^{(N)}_{i}(R,\rho,S,\epsilon_1)$ is an exponential lift of the Jacobi form ${\cal C}^{N,(1)}_{i}(\rho,S,\epsilon_1)$ and related\footnote{If the terms with $Q_{R}=0$ are included in the definition (\ref{ReducedPartitionFunction}) of the reduced partition function then $Z^{(N=2)}_{i}$ is precisely the paramodular form for $\Sigma^*_N$.} to a paramodular form of the group $\Sigma_N^*$ satisfying the property \cite{Belin:2018oza},
\bea
Z^{(N)}_{i}(R,\rho,S,\epsilon_1)=Z^{(N)}_{i}(\tfrac{\rho}{N},N\,R,S,\epsilon_1)\,.
\eea
Finally, we remark that $\Sigma_N^*$ acts on $\mathbb{H}_{2}$ the space of $2\times 2$ matrices with positive imaginary part as in (\ref{ActionSp}). The quotient $\Sigma_{N}^{*}\setminus\mathbb{H}_{2}$ is the moduli space of abelian surfaces with polarization $(1,N)$ \cite{P1,P2}.  These abelian surfaces are precisely the ones appearing in the F-theory forming the fibers of the double elliptically fibered Calabi-Yau threefolds \cite{Kanazawa:2016tnt}.
It would be very interesting to have a clearer geometric interpretation of this result, for example understanding the target space of this CFT. We leave this question for future work.

The functions $\factG{i}{r=1}(\rho,S,\epsilon_1)$ at leading instanton order $\mathcal{O}(Q_R)$ exhibit a factorisation into simpler building blocks which go beyond the known self-similarity and recursive structure (see section~\ref{Sect:LSTFree} for a review of both) of the free energy and extend preliminary results in \cite{Hohenegger:2019tii}: indeed $\factG{i}{r=1}(\rho,S,\epsilon_1)$ can be written as the product (\ref{FactorisationO1}) where the building blocks $\buildH{1}(\rho,S,\epsilon_1)$ and $\buildW{1}(\rho,S,\epsilon_1)$ stem either from the expansion of the free energy for $N=1$ or govern the BPS-counting of a single M5-brane with single M2-branes attached to it on either side (for a review see appendix~\ref{App:BuildingBlock}). To higher order in $Q_R$, remnants of such a factorisation persist, but new elements appear as well (see eq.~(\ref{FactorisationOr})). It is difficult to conjecture a closed form expression of the latter, however, we have succeeded to show for $N=2$ and $N=3$ that they are governed by differential equations that are very similar to holomorphic anomaly equations. 

The $\factG{i}{r=1}(\rho,S,\epsilon_1)$ discussed in this work are specific contributions to the BPS-free energy of LSTs of type A. It would be very interesting to understand the geometric reason that makes these states special compared to others, such that they can be interpreted as part of the spectrum of a symmetric torus orbifold. This could give us the key to understanding if there are further sectors in the spectrum of the LSTs of A-type which exhibit similar properties. Furthermore, this may also give us a hint whether or not these various orbifold CFTs can be in any way connected via a duality transformation.

Another interesting observation is the fact that the $\factG{i}{r=1}(\rho,S,\epsilon_1)$ (except for $i=1$) are obtained through contour integrals from the free energy $P_N^{(r)}(\widehat{a}_{1,\ldots,N})$ that select the coefficient of (a) pole(s) in $\widehat{a}_{1,\ldots,N-1}$. In \cite{Dabholkar:2012nd} the BPS counting of supersymmetric black holes has been discussed. It has been pointed out that the phenomenon of {\emph{wall-crossing}} can be attributed to the polar part of a meromorphic Jacobi form that counts multi-centered black holes whose number can jump when crossing a wall. It would be interesting to analyse if a similar phenomenon takes place for the BPS counting functions discussed in this paper when we cross the loci $\widehat{a}_i=0$. In the dual $U(1)^N$ gauge theory $\widehat{a}_{i}$ are inverse coupling constants for each of the $U(1)$ factors and crossing the $\widehat{a}_{i}=0$ locus corresponds to passing through infinite coupling region \cite{AHK, Ferlito:2017xdq}. It would be interesting to understand what happens in this case to the BPS-states that are counted by $\factG{i}{r=1}(\rho,S,\epsilon_1)$. We leave this question for future work.

\section*{Acknowledgements}
We would like to thank Brice Bastian for collaboration on earlier projects that 
are related to this work. A.I. would like to acknowledge the “2019 Simons Summer Workshop on Mathematics and Physics” for hospitality during part of this work.
\appendix
\section{Modular Forms}\label{App:ModularStuff}
Throughout this work we are using various different modular objects. This appendix compiles the definitions of all objects that are used in the main body of this article, as well as additional useful information and identities. For a more comprehensive review, we relegate the reader to the literature, \emph{e.g.} \cite{EichlerZagier,Lang,Stein}.

A weak Jacobi form of the modular group $\Gamma\cong SL(2,\mathbb{Z})$ of index $m\in\mathbb{Z}$ and weight $w\in\mathbb{Z}$ is a holomorphic function of the type  \begin{align}
\phi:\,\,\mathbb{H}\times\mathbb{C}&\longrightarrow\mathbb{C}\,\nonumber\\
(\rho,z)&\longmapsto \phi(\rho;z)\,,
\end{align}
(where $\mathbb{H}$ is the upper complex plane) which behaves in the following manner under transformations of $\Gamma$
\begin{align}
\phi\left(\frac{a\rho+b}{c\rho+d};\frac{z}{c\rho+d}\right)&=(c\rho+d)^w\,e^{\frac{2\pi i m c z^2}{c\rho+d}}\,\phi(\rho;z)\,,&&\forall\,\left(\begin{array}{cc}a & b \\ c & d\end{array}\right)\in\Gamma\,,\nonumber\\
\phi(\rho;z+\ell_1 \rho+\ell_2)&=e^{-2\pi i m(\ell_1^2\rho+2\ell_1 z)}\,\phi(\rho;z)\,,&&\forall\,\ell_{1,2}\in\mathbb{N}\,,\label{JacobiFormGen}
\end{align}
Such functions allow a Fourier expansion of the form
\begin{align}
&\phi(z,\rho)=\sum_{n= 0}^\infty\sum_{\ell\in\mathbb{Z}}c(n,\ell)\,Q_\rho^n\,e^{2\pi i z \ell}\,,&&\text{with} && Q_\rho=e^{2\pi i\rho}\,.
\end{align}
The Jacobi forms encountered throughout this work can be decomposed in terms of two basis functions, \emph{i.e.} for index $m$ and weight $w$, we can write
\begin{align}
\phi(\rho;z)=\sum_{a=0}^m f_a(\rho)\,\left(\phi_{0,1}(\rho,z)\right)^a\,\left(\phi_{-2,1}(\rho,z)\right)^{m-a}\,.\label{DecomposeJacobiBasic}
\end{align}
\noindent
Here $\phi_{-2,1}$ and $\phi_{0,1}$ are Jacobi forms of index $1$ and weight $-2$ and $0$ respectively, which are defined as \footnote{$\phi_{0,1}(\rho,z)$ defined below differs by a factor of $2$ from its usual definition in the literature \cite{EichlerZagier}. As defined it is equal to the elliptic genus of $K3$. }
\begin{align}
&\phi_{0,1}(\rho,z)=8\sum_{a=2}^4\frac{\theta_a^2(z;\rho)}{\theta_a^2(0,\rho)}\,,&&\text{and}&&\phi_{-2,1}(\rho,z)=\frac{\theta_1^2(z;\rho)}{\eta^6(\rho)}\,,\label{DefPhiFuncts}
\end{align}
with $\theta_{a=1,2,3,4}(z;\rho)$ the Jacobi theta functions and $\eta(\rho)$ the Dedekind eta function. Furthermore, the $f_a(\rho)$ in (\ref{DecomposeJacobiBasic}) are modular forms of weight $w+2a$. In practice, the $f_a(\rho)$ can be written as homogeneous polynomials in the Eisenstein series $E_{2n}$, which are modular forms of weight $2n$ and which are defined as
\begin{align}
&E_{2k}(\rho)=1-\frac{4k}{B_{2k}}\sum_{n=1}^\infty \sigma_{2k-1}(n)\,Q_\rho^n\,,&&\forall\,k\in\mathbb{N}\,,\label{DefEisenStein}
\end{align}
where $B_{2k}$ are the Bernoulli numbers, while $\sigma_k(n)$ is the divisor function. We shall sometimes also use the differently normalised functions
\begin{align}
G_{2k}(\rho)=2\zeta(2k)+2\frac{(2\pi i)^{2k}}{(2k-1)!}\sum_{n=1}^\infty\sigma_{2k-1}(n)\,Q_\rho^n=2\zeta(2k)E_{2k}(\rho)\,.\label{NormEisenstein}
\end{align}
The holomorphic Eisenstein series (\emph{i.e.} the $E_{2n}$ for $n>1$) form a ring, which is generated by $\{E_4,E_6\}$. Furthermore, most of the examples we encounter in this paper are in fact quasi-Jacobi forms, in the sense that the $f_a(\rho)$ in their decomposition (\ref{DecomposeJacobiBasic}) also depend on the Eisenstein series $E_2$: the latter is strictly speaking not a modular form. However, one can define the following non-holomorphic object
\begin{align}
\widehat{E}_2(\rho,\bar{\rho})=E_2(\rho)-\frac{6i}{\pi(\rho-\bar{\rho})}\,,\label{NonholEisensteinD}
\end{align}
which transforms with weight $2$ under modular transformations.

Another object we will encounter in the main body of this paper is the Weierstrass elliptic function
\begin{align}
\wp(z;\rho)=\frac{1}{z^2}+\sum_{k=1}^\infty(2k+1)G_{2k+2}(\rho)\,z^{2k}\,.\label{DefWeierstrass}
\end{align}
which has a pole of order $2$ in $z$.

Finally, many of the results found in this paper use Hecke operators:  these are maps from the space $J_{w,m}(\Gamma)$ of Jacobi forms of index $m$ and weight $w$ into the space $J_{w,km}(\Gamma)$ of Jacobi forms of index $km$ and weight $w$ for $k\in\mathbb{N}$:
\begin{align}
\mathcal{H}_k:\,\,J_{w,m}(\Gamma)&\longrightarrow J_{w,km}(\Gamma)\nonumber\\
\phi(\rho;z)&\longmapsto \mathcal{H}_k(\phi(\rho;z)) =k^{w-1}\sum_{{d|k}\atop{b\text{ mod }d}}d^{-w}\,\phi\left(\frac{k\rho+bd}{d^2};\frac{kz}{d}\right)\,.\label{DefHecke}
\end{align}
Hecke transformations of this type can also be extended to Jacobi forms that depend on more than one variable: let $f_{w,\vec{m}}(\rho,\vec{z}):\,\mathbb{H}\times \mathbb{C}^n\rightarrow\mathbb{C}$ be a Jacobi form with index-vector $\vec{m}$. We then define
\begin{align}
\mathcal{H}_k:\,\,f_{w,\vec{m}}(\rho,\vec{z})&\longmapsto \mathcal{H}_k(f_{w,\vec{m}}(\rho,\vec{z})) =k^{w-1}\sum_{{d|k}\atop{b\text{ mod }d}}d^{-w}\,f_{w,\vec{m}}\left(\frac{k\rho+bd}{d^2};\frac{k\vec{z}}{d}\right)\,.\label{DefHeckeMultiple}
\end{align}
For further use, we consider the case that $f$ allows for a Laurent series expansion in one of the variables: let $(z_1,\ldots,z_n)=(\vec{z},z_n)$ (where $\vec{z}\in\mathbb{C}^{n-1}$) and let $(m_1,\ldots,m_n)=(\vec{m},m_n)$ be the index vector of a Jacobi form that affords the following (convergent) Laurent series
\begin{align}
f_{w,(m_1,\ldots,m_n)}(\rho;\vec{z},z_n)=\sum_{a}z_n^{a}\,f_{w+a,\vec{m}}(\rho,\vec{z})\,,
\end{align}
where $f_{w+a,\vec{m}}(\rho,\vec{z})$ are Jacobi forms of weight $w+a$ and index vector $\vec{m}$. We then find for the action of the Hecke operator 
\begin{align}
\mathcal{H}_k&\left(f_{w,(m_1,\ldots,m_n)}(\rho;\vec{z},z_n)\right)=k^{w-1}\sum_{{d|k}\atop{b\text{ mod }d}}d^{-w}\,f_{w,(m_1,\ldots,m_n)}\left(\frac{k\rho+bd}{d^2};\frac{k\vec{z}}{d},\frac{kz_n}{d}\right)\nonumber\\
&=\sum_a z_n^a\,k^{w+a-1}\sum_{{d|k}\atop{b\text{ mod }d}}d^{-w-a}\,f_{w+a,\vec{m}}\left(\frac{k\rho+bd}{d^2};\frac{k\vec{z}}{d}\right)=\sum_az_n^a\,\mathcal{H}_k\left(f_{w+a;\vec{m}}(\rho,\vec{z})\right)\,.
\end{align}
\section{$(N,1)$ Partition Functions}\label{App:N1PartitionFunction}
The topological string partition function of the Calabi-Yau threefold $X_{N,1}$ is given by \cite{Haghighat:2013gba,Hohenegger:2013ala,Hohenegger:2015btj},
\begin{align}
&{\cal Z}_{N,1}(\tau,\mathbf{\widehat{a}},m,\epsilon_{1,2})=\sum_{\lambda_{1}\cdots \lambda_{N}}Q_{\tau}^{|\lambda_{1}|+\cdots+|\lambda_{N}|}Z_{\lambda_{1}\cdots\lambda_{N}}(\mathbf{\widehat{a}},m,\epsilon_{1,2})\,,&&\text{with} &&\mathbf{\widehat{a}}=\{\widehat{a}_1,\ldots,\widehat{a}_N\}\,,\label{TopStringAppend}
\end{align}
and where the sum is over $N$-tuples of partitions of non-negative integers. The parts of the partition $\lambda_{\alpha}$ are denoted by $\lambda_{\alpha,i}$ with $\lambda_{\alpha,1}\geq \lambda_{\alpha,2}\geq \lambda_{\alpha,3}\geq \cdots$. Each partition $\lambda_{\alpha}$ corresponds to a Young diagram which is obtained by putting $\lambda_{\alpha,i}$ boxes in the $i$-th column such that a box in the Young diagram can be assigned a coordinate $(i,j)$ as long as $1\leq i\leq \ell(\lambda_{\alpha}), 1\leq j\leq \lambda_{\alpha,i}$. The transpose of a partition $\lambda_{\alpha}$ is denoted by $\lambda^{t}_{\alpha}$ and is defined as the partition corresponding to the Young diagram obtained by interchanging rows and columns of the Young diagram corresponding to $\lambda_{\alpha}$.  If we denote by $\ell(\lambda_{\alpha})$ the total number of non-zero parts of the partition $\lambda_{\alpha}$ then we define,
\begin{align}
&|\lambda_{\alpha}|=\sum_{i=1}^{\ell(\lambda_{\alpha})}\lambda_{\alpha,i}\,,&&||\lambda_{\alpha}||^2=\sum_{i=1}^{\ell(\lambda_{\alpha})}\lambda_{\alpha,i}^2\,.
\end{align}
As discussed in the main body of the paper, the topological string partition function (\ref{TopStringAppend}) also captures the partition function of a supersymmetric gauge theories. Furthermore, from a geometric point of view, the instanton part of ${\cal Z}_{N,1}$ is the generating function of equivariant elliptic genera of the instanton moduli space $M(N,k)$,
\bea
{\cal Z}_{N,1}=Z_{0}\sum_{k}Q_{\tau}^{k}\,\chi_{\text{ell}}\Big(M(N,k)\Big)\,,\label{DefZAppGenZ0}
\eea
where $\chi_{\text{ell}}(X)$ denotes the equivariant elliptic genus of any manifold $X$,
\bea
\chi_{\text{ell}}(X)=\mbox{Tr}_{{\cal H}(X)}(-1)^{F_{L}+F_{R}}\,y^{F_{L}}\,q^{H}\,e^{2\pi i {\bf \widehat{a}}\cdot {\bf h}}\,.
\eea
Here the trace is over the RR sector, $F_{L,R}$ are the left and the right moving fermion numbers and $h_{i}$ are the Cartan generators of the symmetry group $G$ which acts on $X$ (and $\mathbf{\widehat{a}}\cdot {\bf h}=\sum_{i=1}^N\widehat{a}_i\, h_i$).  The path integral representation of the above reduces to an index calculation,
\bea
\chi_{\text{ell}}(X)=\int_{X}\text{ch}(E_{Q_{\tau},y})\,\mbox{Td}(X)=\int_{X}\,\prod_{i=1}^{\text{dim}(X)}2\pi i \xi_{i}\,\frac{\vartheta(\tau,m+\xi_{i})}{\vartheta(\tau,\xi_i)}
\eea
where ($y=e^{2\pi i m}$),
\bea
E_{Q_{\tau},y}=y^{-\frac{d}{2}}\otimes_{\ell\geq 1}\Big[\wedge_{-yQ_{\tau}^{\ell-1}}T_{X}\otimes \wedge_{-y^{-1}Q_{\tau}^{\ell}}\overline{T_{X}}\otimes S_{Q_{\tau}^{\ell}}T_{X}\otimes S_{Q_{\tau}^{\ell}}\overline{T_X}\Big]
\eea
and $x_{i}$ are the formal roots of the Chern polynomial. The relation between $Z_{\lambda_{1}\cdots \lambda_{N}}$ and $\chi_{\text{ell}}(M(N,k))$ is given by,
\bea
\chi_{\text{ell}}(M(N,k))=\sum_{|\lambda_{1}|+\cdots+|\lambda_N|=k}Z_{\lambda_{1}\cdots\lambda_{N}}/Z_{0}\,.
\eea
The function $Z_{\lambda_{1}\cdots \lambda_{N}}({\bf \widehat{a}},m,\epsilon_{1,2})$ in (\ref{TopStringAppend}) is defined as,
\bea
Z_{\lambda_{1}\cdots \lambda_{N}}({\bf \widehat{a}},m,\epsilon_{1,2})=Z_{0}\prod_{\alpha=1}^{N}\frac{\vartheta_{\lambda_{\alpha}\lambda_{\alpha}}(Q_{m})}
{\vartheta_{\lambda_{\alpha}\lambda_{\alpha}}(\sqrt{\tfrac{t}{q})}}\prod_{1\leq \alpha<\beta\leq N}\frac{\vartheta_{\lambda_{\alpha}\lambda_{\beta}}(Q_{\alpha\beta}Q_{m})\vartheta_{\lambda_{\alpha}\lambda_{\beta}}(Q_{\alpha\beta}Q_{m}^{-1})}{\vartheta_{\lambda_{\alpha}\lambda_{\beta}}(Q_{\alpha\beta}\sqrt{\tfrac{t}{q}})\vartheta_{\lambda_{\alpha}\lambda_{\beta}}(Q_{\alpha\beta}\sqrt{\tfrac{q}{t}})}\,,
\label{Zlambdas}
\eea
where $Q_{\alpha\beta}=e^{2\pi i(\widehat{a}_{\alpha}-\widehat{a}_{\beta})}$ and
\bea\label{thetalambdamu}
\vartheta_{\lambda\mu}(\rho,z)=\prod_{(i,j)\in \lambda}\theta_{1}\Big(\rho; z^{-1}\,t^{-\mu^{t}_{j}+i-\tfrac{1}{2}}\,q^{-\lambda_{i}+j-\tfrac{1}{2}}\Big)\,\prod_{(i,j)\in \mu}\theta_{1}\Big(\rho; z^{-1}\,t^{\lambda^{t}_{j}-i+\tfrac{1}{2}}\,q^{\mu_{i}-j+\tfrac{1}{2}}\Big)
\eea
with $\theta_{1}(\rho,z)$ the Jacobi theta function and $\rho=\sum_{\alpha=1}^{N}\widehat{a}_{\alpha}$. The factor $Z_{0}$ in eq.~(\ref{DefZAppGenZ0}) and (\ref{Zlambdas}) is independent of $Q_{\tau}$ and is given by,
\bea\label{Z0Perturbative}
Z_{0}=\prod_{n=1}^{\infty}(1-Q_{\rho}^n)^{-1}\,\Big[\prod_{1\leq \alpha<\beta\leq N}F_{\alpha\beta}\Big]\,\Big[\prod_{\alpha,\beta=1}^{N}H_{\alpha\beta}\Big]
\eea
where (with $\widetilde{Q}_{\alpha\beta}=Q_{1}Q_{2}\cdots Q_{a}Q_{1}^{-1}\cdots Q_{b}^{-1}\,Q_{m}^{a-b}$)
\bea\label{Z0Factors}
F_{\alpha\beta}&=&\prod_{i,j=1}^{\infty}\frac{(1-Q_{\alpha\beta}Q_{m}^{-1}\,t^{i-\tfrac{1}{2}}\,q^{j-\tfrac{1}{2}})(1-Q_{\alpha\beta}Q_{m}t^{i-\tfrac{1}{2}}\,q^{j-\tfrac{1}{2}})}{(1-Q_{\alpha\beta}t^{i}q^{j-1})(1-Q_{\alpha\beta}t^{i-1}q^{j})}\\\nonumber
H_{\alpha\beta}&=&\prod_{n,i,j=1}^{\infty}\frac{(1-Q_{\rho}^{n}\widetilde{Q}_{\alpha\beta}Q_{m}^{-1}t^{i-\tfrac{1}{2}}\,q^{j-\tfrac{1}{2}})(1-Q_{\rho}^{n}\widetilde{Q}_{\alpha\beta}Q_{m}t^{i-\tfrac{1}{2}}q^{j-\tfrac{1}{2}})}{(1-Q_{\rho}^{n}\widetilde{Q}_{\alpha\beta}t^{i}q^{j-1})(1-Q_{\rho}^{n}\widetilde{Q}_{\alpha\beta}t^{i-1}q^{j})}\,.
\eea

\subsection{Modular Transformation}
To understand how the partition function ${\cal Z}_{N,1}$ transforms under the modular transformation,
\bea
\left(\begin{array}{cc}a & b \\ c & d\end{array}\right)\in PSL(2,\mathbb{Z}):\hspace{1cm}\Big(\rho,m,\epsilon_{1,2},\widehat{a}_{\alpha\beta}\Big)\mapsto \Big(\frac{a\rho+b}{c\rho+d},\frac{m}{c\rho+d},\frac{\epsilon_{1,2}}{c\rho+d},\frac{\widehat{a}_{\alpha\beta}}{c\rho+d}\Big)\,,\label{ModularTrafoForm}
\eea
which are generated by
\begin{align}
&T=\left(\begin{array}{cc}1 & 1 \\ 0 & 1\end{array}\right)\,,&&\text{and} &&S=\left(\begin{array}{cc}0 & -1 \\ 1 & 0\end{array}\right)\,,
\end{align}
we need to determine the transformation properties of $\vartheta_{\lambda\mu}(\rho,z)$. Although $\vartheta_{\lambda\mu}(\rho,z)$ is not invariant under the $T$ transformation, it is easy to see that the ratio $\frac{\vartheta_{\lambda\mu}(\rho,z_{1})}{\vartheta_{\lambda\mu}(\rho,z_{2})}$ is invariant for any $z_{1,2}$. In view of the structure of (\ref{Zlambdas}), this implies that ${\cal Z}_{N,1}$ is invariant under the $T$ transformation. The ratio $\frac{\vartheta_{\lambda\mu}(\rho,z_{1})}{\vartheta_{\lambda\mu}(\rho,z_{2})}$, however, is not invariant under the $S$ transformation,
\bea
\frac{\vartheta_{\lambda\mu}(-\tfrac{1}{\rho},\tfrac{z_1}{\rho})}{\vartheta_{\lambda\mu}(-\tfrac{1}{\rho},\tfrac{z_2}{\rho})}&=&e^{\tfrac{2\pi i K_{\lambda\mu}}{\tau}}\,\frac{\vartheta_{\lambda\mu}(\rho,z_1)}{\vartheta_{\lambda\mu}(\rho,z_2)}\nonumber\\
K_{\lambda\mu}(h_{1},h_{2})&=&\tfrac{1}{2}(h_{1}^2-h_{2}^2)(|\lambda|+|\mu|)+(h_{2}-h_{1})\Big(\sum_{(i,j)\in \lambda}(\epsilon_{2}\,(\mu^{t}_{j}-i+\tfrac{1}{2})-\epsilon_{1}\,(\lambda_{i}-j+\tfrac{1}{2}))\nonumber\\
&+&\sum_{(i,j)\in \mu}
(-\epsilon_{2}\,(\lambda^{t}_{j}-i+\tfrac{1}{2})+\epsilon_{1}\,(\mu_{i}-j+\tfrac{1}{2}))\Big)\,,
\label{ModularTransformation}
\eea
where $z_{1,2}=e^{2\pi i h_{1,2}}$. Using the following identities,
\begin{align}
&\sum_{(i,j)\lambda}\mu_{j}^{t}=\sum_{(i,j)\in \mu}\lambda_{j}^{t}\,,&&\text{and}&&\sum_{(i,j)\lambda}(\lambda_{i}-j+\tfrac{1}{2})=\tfrac{||\lambda||^2}{2}\,,
\end{align}
$K_{\lambda\mu}(h_{1},h_{2})$ appearing in eq.(\ref{ModularTransformation}) can be simplified,
\bea
K_{\lambda\mu}(h_1,h_2)&=&\tfrac{1}{2}(h_{1}^2-h_{2}^2)(|\lambda|+|\mu|)+(h_{1}-h_{2})\Big[\epsilon_{1}\,\tfrac{||\lambda||^2-||\mu||^2}{2}+\epsilon_{2}\,\tfrac{||\lambda^t||^2-||\mu^t||^2}{2}\Big]\,.
\eea
Notice that in the unrefined case $(\epsilon_{2}=-\epsilon_{1}=\epsilon)$ $K_{\lambda\mu}(h_1,h_2)$ simplifies
\bea
K_{\lambda\mu}(h_{1},h_{2})=\tfrac{1}{2}(h_{1}^2-h_{2}^2)(|\lambda|+|\mu|)+(h_{1}-h_{2})\,\epsilon\,\Big[\kappa(\lambda)-\kappa(\mu)\Big]\,
\eea
where $\kappa(\lambda)=\tfrac{||\lambda||^2-||\lambda^t||^2}{2}$.

\noindent
Thus the function $Z_{\lambda_{1}\cdots\lambda_{N}}$ given in Eq.(\ref{Zlambdas}) transforms as,
\bea
Z_{\lambda_{1}\cdots \lambda_{N}}\mapsto e^{\frac{2\pi i K}{\rho}}\,Z_{\lambda_{1}\cdots \lambda_{N}},
\eea
with
\bea\nonumber
K_{\lambda_{1}\cdots\lambda_{N}}(\widehat{a}_{\alpha\beta},m,\epsilon_{+})=\sum_{\alpha=1}^{N}K_{\lambda_{\alpha}\lambda_{\alpha}}(m,-\epsilon_{+})+\sum_{1\leq \alpha<\beta\leq N}&&\Big[K_{\lambda_{\alpha}\lambda_{\beta}}(\widehat{a}_{\alpha\beta}+m,\widehat{a}_{\alpha\beta}-\epsilon_{+})\\\nonumber
+&&K_{\lambda_{\alpha}\lambda_{\beta}}(\widehat{a}_{\alpha\beta}-m,\widehat{a}_{\alpha\beta}+\epsilon_{+})\Big]
\eea
Here we have defined $\widehat{a}_{\alpha\beta}=\widehat{a}_\alpha-\widehat{a}_\beta$. Thus the partition function is not invariant under modular transformations (\ref{ModularTrafoForm}) but can be made invariant at the expense of introducing a holomorphic anomaly \cite{Bershadsky:1993cx}.

\subsection{Singularities}
The function $\vartheta_{\lambda\mu}(\rho, z)$ has some interesting properties. In the unrefined case it becomes proportional to a Kronecker delta function for $z=1$ \cite{ambreenthesis} and $t=q$,
\begin{align}
\vartheta_{\lambda\mu}(\rho,1)&=\delta_{\lambda\mu}\,\prod_{(i,j)\in \lambda}\theta_{1}(\rho,q^{h(i,j)})\theta_{1}(\rho,q^{-h(i,j)})=(-1)^{|\lambda|}\,\delta_{\lambda\mu}\,\prod_{(i,j)\in \lambda}\theta_{1}(\rho,q^{h(i,j)})^2\,.
\end{align}
Since the partition function ${\cal Z}_{N,1}$ is a sum over all partitions, from Eq.(\ref{Zlambdas}) and Eq.(\ref{thetalambdamu}) it follows that the partition function will have a pole whenever $a_{\alpha\beta}\in {\cal S}_{\lambda_{\alpha}\lambda_{\beta}}^{1}\cup{\cal S}_{\lambda_{\alpha}\lambda_{\beta}}^{2}$,
\bea\label{sing}
{\cal S}_{\lambda_{\alpha}\lambda_{\beta}}^{1}&=&\{\epsilon_{1}\Big(-\lambda_{\alpha,i}+j-\tfrac{1}{2}\Big)+\epsilon_{2}\Big(\lambda^{t}_{\beta,j}-i+\tfrac{1}{2}\Big)\pm \epsilon_{+}\,~|~(i,j)\in \lambda_{\alpha}\}\\\nonumber
{\cal S}_{\lambda_{\alpha}\lambda_{\beta}}^{2}&=& \{\epsilon_{1}\Big(\lambda_{\beta,i}-j+\tfrac{1}{2}\Big)+\epsilon_{2}\Big(-\lambda^{t}_{\alpha,j}+i-\tfrac{1}{2}\Big)\pm \epsilon_{+}\,~|~(i,j)\in \lambda_{\beta}\}\,.
\eea
Thus the total order of the poles in $\widehat{a}_{\alpha\beta}$ (counting with possible multiplicity) is $2(|\lambda_{\alpha}|+|\lambda_{\beta}|)$. The poles in the variable $\widehat{a}_{\alpha\beta}$ only depend on the shape of the pair of partitions $(\lambda_{\alpha},\lambda_{\beta})$ and therefore the pole structure for $N>2$ in the variables $\widehat{a}_{\alpha\beta}$ follows from the pole structure for the $N=2$ case in the variable $\widehat{a}_{12}=\widehat{a}$. 

The poles in the variable $\widehat{a}$ for the $N=2$ case form a nested sequence i.e., the set of poles at order $Q_{R}^{k}$ are  contained in the set of poles at order $Q_{R}^{k+1}$. To see this, consider a pair of partitions $(\lambda_{1},\lambda_{2})$, with $|\lambda_{1}|+|\lambda_{2}|=k$, giving the set of poles ${\cal S}_{\lambda_{1}\lambda_{2}}$.  
For the case of $N=2$ consider the pair of partitions $(\lambda_{1},\lambda_{2})=((k_{1}),1^{k-k_{1}})$ which contribute to the coefficient of $Q_{\rho}^{k}$ for all $k_{1}=0,1,\cdots,k$. With this choice of the partitions the the set of possible poles in Eq.(\ref{sing}) becomes $(\sigma=0,1)$,
\bea
\widehat{a}_{12}=\widehat{a}\in 
&&\{-(k_{1}-1)\epsilon_{1}+k_{2}\,\epsilon_{2}-2\sigma\epsilon_{+}\}\cup \{-j\,\epsilon_{1}-2\sigma\epsilon_{+}~|~j=0,\cdots,k_{1}-2\}\nonumber\\
&& \{(i-2)\epsilon_{2}\,+ 2\epsilon_{+}\sigma\,~|~i=1,\cdots,k-k_{1}\}\,.
\eea
The free energy $\ln({\cal Z}_{N,1})$ is a power series in $\epsilon_{1,2}$ with coefficients which can are refined genus $g$ amplitudes. Once the expansion in $\epsilon_{1,2}$ has been carried out the coefficients, refined genus $g$ amplitudes, now have poles at $\widehat{a}_{\alpha\beta}=0$. In the paper we study the poles of the refined genus $g$ amplitudes at $\widehat{a}_{\alpha\beta}=0$ rather than the poles of the partition function which occur at various locations in the $(\epsilon_{1},\epsilon_{2})$ plane.

\noindent
{\bf Example:} Let us consider the case $N=2$ to first order in $Q_{\tau}$. The free energy (\ref{FullFreeEnergy}) is given by
\begin{align}
{\cal F}_{2,1}&=\mbox{ln}(Z_0)\nonumber\\
&\hspace{0.5cm}+Q_{\tau}\,\frac{\vartheta_{(1)(1)}(Q_{m})}{\vartheta_{(1)(1)}(\sqrt{\tfrac{t}{q}})}\,\Big[\frac{\vartheta_{(1)(0)}(Q_{12}Q_{m})\vartheta_{(1)(0)}(Q_{12}Q_{m}^{-1})}{\vartheta_{(1)(0)}(Q_{12}\sqrt{\tfrac{t}{q}})\vartheta_{(1)(0)}(Q_{12}\sqrt{\tfrac{q}{t}})}+\frac{\vartheta_{(0)(1)}(Q_{12}Q_{m})\vartheta_{(0)(1)}(Q_{12}Q_{m}^{-1})}{\vartheta_{(0)(1)}(Q_{12}\sqrt{\tfrac{t}{q}})\vartheta_{(0)(1)}(Q_{12}\sqrt{\tfrac{q}{t}})}\Big]+\cdots\nonumber\\
&=\mbox{ln}(Z_0)+Q_{\tau}\frac{\theta_{1}(\rho,m+\epsilon_{-})\theta_{1}(m-\epsilon_{-})}
{\theta_{1}(\rho,\epsilon_1)\theta_{1}(\rho,\epsilon_2)}\,\Big[\frac{\theta_{1}(\rho,\widehat{a}_1+m+\epsilon_{+})\theta_{1}(\rho,\widehat{a}_1-m+\epsilon_{+})}{\theta_{1}(\rho,\widehat{a}_1)\theta_{1}(\rho,\widehat{a}_1+2\epsilon_{+})}\nonumber\\
&\hspace{0.5cm}+\frac{\theta_{1}(\rho,\widehat{a}_1+m-\epsilon_{+})\theta_{1}(\widehat{a}_1-m-\epsilon_{+})}{\theta_{1}(\rho,\widehat{a}_1-2\epsilon_{+})\theta_{1}(\widehat{a}_1)}\Big]+\cdots
\end{align}
From Eq.(\ref{Z0Perturbative}) and Eq.(\ref{Z0Factors}) we see that as a function of $\widehat{a}_1$ 
\bea
\mbox{ln}(Z_{0})=A\,\mbox{ln}(\widehat{a}_1)+\cdots
\eea
where $A$ is independent of $\widehat{a}_1$. Thus the free energy diverges like,
\bea\nonumber
{\cal F}_{2,1}&=&A\,\mbox{ln}(\widehat{a}_1)-Q_{\tau}\,\frac{\theta_{1}(\rho,m+\epsilon_{-})\theta_{1}(m-\epsilon_{-})\theta_{1}(\rho,m+\epsilon_{+})\theta_{1}(\rho,m-\epsilon_{+})}
{\theta_{1}(\rho,\epsilon_1)\theta_{1}(\rho,\epsilon_2)}\times\\ \label{AppendixBEq}
&&\Big[\frac{2}{(\widehat{a}_1+2\epsilon_{+})(\widehat{a}_1-2\epsilon_{+})}+\cdots\Big]\,.
\eea
Thus we see that there is a pole at $\widehat{a}_1=\pm\,2\epsilon_{+}$. However, if we first expand in $\epsilon_{1,2}$ then we get a single pole $\widehat{a}_1=0$ of order two. This persists at higher order in $Q_{\tau}$ and we see poles at $\widehat{a}_1=0$ of various even orders.

\section{Expansion Coefficients of the Basic Building Blocks}\label{App:BuildingBlock}
In this appendix we collect explicit expressions for the expansions of the free energy for $N=1$, $N=2$ and $N=3$.
\subsection{Coefficients of the $N=1$ Free Energy}\label{App:N1Expansion}
Due to their frequent use throughout the main body of this article, we tabulate the coefficients $H^{(r),\{0\}}_{(2s,0)}(\rho,S)$ that appear in the expansion of the free energy to leading orders n $r$ and $s$. To this end, we decompose the former in the following fashion
\begin{align}
H^{(r),\{0\}}_{(2s,0)}(\rho,S)=\sum_{i=0}^r\mathfrak{b}^{(r)}_{i,(2s,0)}(\rho)\,(\phi_{-2,1}(\rho,S))^{i}(\phi_{0,1}(\rho,S))^{r-i}\,,
\end{align}
where $\mathfrak{b}^{(r)}_{i,(2s,0)}$ is a quasi-modular form of weight $2s+2i-2$, which can be written as a polynomial in the Eisenstein series $\{E_2,E_4,E_6\}$. For $r=1$, $r=2$ and $r=3$ the expansion coefficients are tabulated in \ref{Tab:Expansionb1}, \ref{Tab:Expansionb2} and \ref{Tab:Expansionb3} respectively. 

Following \cite{Ahmed:2017hfr}, the coefficients $H^{(r),\{0\}}_{(2s,0)}$ with $r>1$ can be recovered from those with $r=1$ through Hecke transformations, \emph{i.e.}
\begin{align}
H^{(r),\{0\}}_{(s,0)}(\rho,S)=\mathcal{H}_r\left(H^{(1),\{0\}}_{(s,0)}(\rho,S)\right)\,.\label{HeckeN1}
\end{align}
The relations (\ref{HeckeN2}) for $N=2$, (\ref{HeckeN3}) for $N=3$ and (\ref{HeckeN4}) for $N=4$ can be understood as generalisations of (\ref{HeckeN1}). Finally, for further use throughout the main body of this paper, we also introduce
\begin{align}
\buildH{r}(\rho,S,\epsilon_1)=\sum_{s=0}^\infty\epsilon_1^{2s-2}\,H^{(r),\{0\}}_{(s,0)}(\rho,S)\,.\label{DefBuildH}
\end{align}

\begin{table}[htbp]
\begin{center}
\begin{tabular}{|c||c|c|}\hline
&&\\[-12pt]
$s$ & $\mathfrak{b}^{(r=1)}_{0,(2s,0)}$ & $\mathfrak{b}^{(r=1)}_{1,(2s,0)}$\\[4pt]\hline\hline
&&\\[-12pt]
$0$ & $0$ & $-1$\\[4pt]\hline
&&\\[-12pt]
$1$ & $\tfrac{1}{96}$ & $-\tfrac{E_2}{48}$\\[4pt]\hline
&&\\[-12pt]
$2$ & $\tfrac{E_2 }{4608}$ & $-\tfrac{5 E_2 ^2+13 E_4 }{23040}$\\[4pt]\hline
&&\\[-12pt]
$3$ & $\tfrac{5 E_2 ^2+7 E_4 }{2211840}$ & $-\tfrac{35 E_2 ^3+273 E_2  E_4 +184 E_6 }{23224320}$\\[4pt]\hline
&&\\[-12pt]
$4$ & $\tfrac{35 E_2 ^3+147 E_2  E_4 +124 E_6 }{2229534720}$ & $-\frac{175 E_2 ^4+2730 E_2 ^2 E_4 +3680 E_2  E_6 +5583 E_4 ^2}{22295347200}$\\[4pt]\hline
\end{tabular}
\end{center}
\caption{Coefficients in the expansion of $\mathfrak{b}^{(r=1)}_{i,(2s,0)}(\rho,S)$.}
\label{Tab:Expansionb1}
\end{table}

\begin{table}
\begin{center}
\begin{tabular}{|c||c|c|c|}\hline
&&&\\[-12pt]
$s$ & $\mathfrak{b}^{(r=2)}_{0,(2s,0)}$ & $\mathfrak{b}^{(r=2)}_{1,(2s,0)}$ & $\mathfrak{b}^{(r=2)}_{2,(2s,0)}$\\[4pt]\hline\hline
&&&\\[-12pt]
$0$ & $0$ & $-\tfrac{1}{16}$ & $0$\\[4pt]\hline
&&&\\[-12pt]
$1$ & $\tfrac{1}{1536}$ & $-\tfrac{E_2 }{384}$ & $\tfrac{5 E_4 }{384}$\\[4pt]\hline
&&&\\[-12pt]
$2$ & $\frac{E_2 }{36864}$ & $-\frac{5 E_2 ^2+27 E_4 }{92160}$ & $\frac{5 (E_2  E_4 +2 E_6 )}{9216}$\\[4pt]\hline
&&&\\[-12pt]
$3$ & $\frac{5 E_2 ^2+13 E_4 }{8847360}$ & $-\frac{35 E_2 ^3+567 E_2  E_4 +1066 E_6 }{46448640}$ & $\frac{5 E_2 ^2 E_4 +20 E_2  E_6 +53 E_4 ^2}{442368}$\\[4pt]\hline
&&&\\[-12pt]
$4$ & $\frac{70 E_2 ^3+546 E_2  E_4 +1067 E_6 }{8918138880}$ & $-\frac{175 E_2 ^4+5670 E_2 ^2 E_4 +21320 E_2  E_6 +54303 E_4 ^2}{22295347200}$ & $\frac{70 E_2 ^3 E_4 +420 E_2 ^2 E_6 +2226 E_2  E_4 ^2+5393 E_4  E_6 }{445906944}$\\[4pt]\hline
\end{tabular}
\end{center}
\caption{Coefficients in the expansion of $\mathfrak{b}^{(r=2)}_{i,(2s,0)}(\rho,S)$.}
\label{Tab:Expansionb2}
\end{table}

\begin{table}
\begin{center}
\begin{tabular}{|c||c|c|c|c|}\hline
&&&&\\[-12pt]
$s$ & $\mathfrak{b}^{(r=3)}_{0,(2s,0)}$ & $\mathfrak{b}^{(r=3)}_{1,(2s,0)}$ & $\mathfrak{b}^{(r=3)}_{2,(2s,0)}$ & $\mathfrak{b}^{(r=3)}_{3,(2s,0)}$ \\[4pt]\hline\hline
&&&&\\[-12pt]
$0$ & $0$ & $-\tfrac{1}{432}$ & $0$ & $-\tfrac{E_4 }{36}$\\[4pt]\hline
&&&&\\[-12pt]
$1$ & $\frac{1}{41472}$ & $-\frac{E_2 }{6912}$ & $\frac{E_4 }{384}$ & $-\frac{9 E_2  E_4 +32 E_6 }{5184}$\\[4pt]\hline
&&&&\\[-12pt]
$2$ & $\frac{E_2 }{663552}$ & $-\frac{15 E_2 ^2+151 E_4 }{3317760}$ & $\frac{27 E_2  E_4 +88 E_6 }{165888}$ & $-\frac{45 E_2 ^2 E_4 +320 E_2  E_6 +1333 E_4 ^2}{829440}$\\[4pt]\hline
&&&&\\[-12pt]
$3$ & $\frac{5 E_2 ^2+23 E_4 }{106168320}$ & $\frac{-105 E_2 ^3-3171 E_2  E_4 -10088 E_6 }{1114767360}$ & $\frac{405 E_2 ^2 E_4 +2640 E_2  E_6 +10103 E_4 ^2}{79626240}$ & $\frac{-315 E_2 ^3 E_4 -3360 E_2 ^2 E_6 -27993 E_2  E_4 ^2-103400 E_4  E_6 }{278691840}$\\[4pt]\hline
\end{tabular}
\end{center}
\caption{Coefficients in the expansion of $\mathfrak{b}^{(r=3)}_{i,(2s,0)}(\rho,S)$.}
\label{Tab:Expansionb3}
\end{table}

\subsection{Expansion of $W(\rho,S,\epsilon_1,\epsilon_2)$}\label{App:ExpansionW}
In \cite{Hohenegger:2015btj,Ahmed:2017hfr} the (quasi-)Jacobi form 
\begin{align}
&W(\rho,S,\epsilon_{1,2})=\frac{\theta_1(\rho,S+\epsilon_+)\theta_1(\rho,S-\epsilon_+)-\theta_1(\rho,S+\epsilon_-)\theta_1(\rho,S-\epsilon_-)}{\theta_1(\rho,\epsilon_1)\theta_2(\rho,\epsilon_2)}\,,&&\text{with} &&\epsilon_\pm =\frac{\epsilon_1\pm \epsilon_2}{2}\,,\label{ExpansionFunctionW}
\end{align}
was introduced, which governs the BPS-counting of a single M5-brane with on M2-brane ending on it on either side. In the NS-limit, expanding the latter in powers of $\epsilon_{1}$, we define
\begin{align}
\buildW{1}(\rho,S,\epsilon_{1})=\lim_{\epsilon_2\to 0} W(\rho,S,\epsilon_{1,2})=\sum_{s=0}^\infty \epsilon_1^{2s}\,W_{(2s)}(\rho,S)\,, \label{BuildingFct}
\end{align}
where to low orders in $s$, we find
\begin{align}
W_{(0)}&=\frac{1}{24}\,(\phi_{0,1}+2 E_2\,\phi_{-2,1})\,,\nonumber\\
W_{(2)}&=-\frac{1}{576}\,(E_4-E_2^2)\phi_{-2,1}\,,\nonumber\\
W_{(4)}&=\frac{5(E_4-E_2^2)\phi_{0,1}+2(5 E_2^3+3 E_2 E_4-8E_6)\phi_{-2,1}}{552960}\,,\nonumber\\
W_{(6)}&=\frac{\phi_{-2,1} \left(35 E_2 ^4+168 E_2 ^2 E_4 +16 E_2  E_6 -219 E_4 ^2\right)-7 \phi_{0,1} \left(5
   E_2 ^3+3 E_2  E_4 -8 E_6 \right)}{278691840}\,.\label{ExpansionWfunction}
\end{align}
While not a function of $R$, following the free energy for $N=1$ discussed in the previous appendix~\ref{App:N1Expansion}, we can define an extension of $W_{(2s)}$ to higher orders through
\begin{align}
W_{(2s)}^{(r)}(\rho,S)=\mathcal{H}_r\left(W_{(2s)}(\rho,S)\right)\,,
\end{align}
along with the building block
\begin{align}
\buildW{r}(\rho,S,\epsilon_{1})=\sum_{s=0}^\infty \epsilon_1^{2s}\,W^{(r)}_{(2s)}(\rho,S)\,.\label{DefBuildW}
\end{align}
For convenience we can give explicit expressions for the first few instances of $W^{(r)}_{(2s)}$. To this end, we introduce the decomposition 
\begin{align}
W^{(r)}_{(2s)}(\rho,S)=\sum_{i=0}^r\mathfrak{l}^{(r)}_{i,(2s)}(\rho)\,(\phi_{-2,1}(\rho,S))^{i}(\phi_{0,1}(\rho,S))^{r-i}\,,
\end{align}
where $\mathfrak{l}^{(r)}_{i,(2s)}$ is a quasi-modular form of weight $2s+2i$, which can be written as a polynomial in the Eisenstein series. For $r=1$, the expression (\ref{ExpansionWfunction}) can be tabulated as
\begin{center}
\begin{tabular}{|c||c|c|}\hline
&&\\[-12pt]
$s$ & $\mathfrak{l}^{(r=1)}_{0,(2s)}$ & $\mathfrak{l}^{(r=1)}_{1,(2s)}$\\[4pt]\hline\hline
&&\\[-12pt]
$0$ & $\frac{1}{24}$ & $\frac{E_2 }{12}$\\[4pt]\hline
&&\\[-12pt]
$1$ & $0$ & $-\frac{E_4-E_2 ^2}{576} $\\[4pt]\hline
&&\\[-12pt]
$2$ & $\frac{E_4 -E_2 ^2}{110592}$ & $\frac{5 E_2 ^3+3 E_2  E_4 -8 E_6 }{276480}$\\[4pt]\hline
&&\\[-12pt]
$3$ & $\frac{-5 E_2 ^3-3 E_2  E_4 +8 E_6 }{39813120}$ & $\frac{35 E_2 ^4+168 E_2 ^2 E_4 +16 E_2  E_6 -219 E_4 ^2}{278691840}$\\[4pt]\hline
&&\\[-12pt]
$4$ & $\frac{-35 E_2 ^4-126 E_2 ^2 E_4 -16 E_2  E_6 +177 E_4 ^2}{35672555520}$ & $\frac{175 E_2 ^5+2030 E_2 ^3 E_4 +2000 E_2 ^2 E_6 +1203 E_2  E_4 ^2-5408 E_4  E_6 }{267544166400}$\\[4pt]\hline
\end{tabular}
\end{center}
for $r=2$ we obtain
\begin{center}
\begin{tabular}{|c||c|c|c|}\hline
&&&\\[-12pt]
$s$ & $\mathfrak{l}^{(r=2)}_{0,(2s)}$ & $\mathfrak{l}^{(r=2)}_{1,(2s)}$ & $\mathfrak{l}^{(r=2)}_{2,(2s)}$\\[4pt]\hline\hline
&&&\\[-12pt]
$0$ & $\frac{1}{384}$ & $\frac{E_2 }{96}$ & $\frac{E_4 }{96}$\\[4pt]\hline
&&&\\[-12pt]
$1$ & $0$ & $\frac{E_2 ^2-E_4 }{2304}$ & $\frac{E_6 -E_2  E_4 }{576} $\\[4pt]\hline
&&&\\[-12pt]
$2$ & $\frac{E_4 -E_2 ^2}{442368}$ & $\frac{5 E_2 ^3+17 E_2  E_4 -22 E_6 }{552960}$ & $\frac{-3 E_2 ^2 E_4 -4 E_2  E_6 +7 E_4 ^2}{36864}$\\[4pt]\hline
&&&\\[-12pt]
$3$ & $\frac{-5 E_2 ^3-12 E_2  E_4 +17 E_6 }{79626240}$ & $\frac{35 E_2 ^4+462 E_2 ^2 E_4 +604 E_2  E_6 -1101 E_4 ^2}{278691840}$ & $\frac{-7 E_2 ^3 E_4 -24 E_2 ^2 E_6 -48 E_2  E_4 ^2+79 E_4  E_6 }{3981312}$\\[4pt]\hline
\end{tabular}
\end{center}
and for $r=3$ we find
\begin{center}
\begin{tabular}{|c||c|c|c|c|}\hline
&&&&\\[-12pt]
$s$ & $\mathfrak{l}^{(r=3)}_{0,(2s)}$ & $\mathfrak{l}^{(r=3)}_{1,(2s)}$ & $\mathfrak{l}^{(r=3)}_{2,(2s)}$ & $\mathfrak{l}^{(r=3)}_{3,(2s)}$\\[4pt]\hline\hline
&&&&\\[-12pt]
$0$ & $\frac{1}{10368}$ & $\frac{E_2 }{1728}$ & $\frac{E_4 }{864}$ & $\frac{9 E_2  E_4 -8 E_6 }{1296}$\\[4pt]\hline
&&&&\\[-12pt]
$1$ & $0$ & $\frac{E_2 ^2-E_4 }{27648}$ & $\frac{E_6 -E_2  E_4 }{1728}$ & $\frac{3 E_2 ^2 E_4 +8 E_2  E_6 -11 E_4 ^2}{6912}$\\[4pt]\hline
&&&&\\[-12pt]
$2$ & $\frac{E_4 -E_2 ^2}{5308416}$ & $\frac{15 E_2 ^3+121 E_2  E_4 -136 E_6 }{13271040}$ & $\frac{-51 E_2 ^2 E_4 -128 E_2  E_6 +179 E_4 ^2}{1327104}$ & $\frac{45 E_2 ^3 E_4 +280 E_2 ^2 E_6 +1003 E_2  E_4 ^2-1328 E_4  E_6 }{3317760}$\\[4pt]\hline
\end{tabular}
\end{center}
\section{The Paramodular Group $\Sigma_N^*$}\label{App:Paramodular}
Let $N\in\mathbb{N}$ with $N>1$. The degree two paramodular groups are subgroups of the symplectic group $Sp(4,\mathbb{Q})$ labelled by an integer $N$ and defined as \cite{HeimKrieg,RobertsSchmidt},  
\begin{align}
\Sigma_N=\left\{
\begin{pmatrix}
\star & N\,\star& \star &\star\\
\star &\star &\star &\star/N\\
\star & N\,\star& \star& \star\\
N\,\star&N\,\star &N\,\star& \star
\end{pmatrix}\in Sp(4,\mathbb{Q})\,,\bigg|\star\in \mathbb{Z}\right\}\,.
\end{align}
$\Sigma_N$ has the interesting property that $\Sigma_{N}\Gamma_{N}\subset \Gamma_{N}$ where $\Gamma_{N}$ is the lattice $\mathbb{Z}\oplus\mathbb{Z}\oplus \mathbb{Z}\oplus N\mathbb{Z}$ in $Sp(4,\mathbb{Q})$ and $\Sigma_N$ acts through simple matrix multiplication. A very useful review of the degree $n$ paramodular groups is given in \cite{zantout}.

In order to define the action of $\Sigma_N$ on the free energies discussed in the main body of this paper, we furthermore introduce the period matrix
\bea
\Omega=\left(\begin{array}{cc}
\rho& S\\
S& R
\end{array}\right)\in \mathbb{H}(2)
\eea
where $\mathbb{H}(2)$ is the space of $2\times 2$ matrices with positive imaginary part. We then define the action of $\Sigma_N$ by
\begin{align}
&g=\left(\begin{array}{cc} A & B \\ C & D\end{array}\right)\in\Sigma_N:&&\Omega\longmapsto\Omega'=g\circ\Omega=(A\cdot\Omega+B)\cdot (C\cdot \Omega+D)^{-1}\,,\label{ActionSp}
\end{align}
where $A$, $B$ ,$C$ and $D$ are $2\times 2$ matrices. 

Following \cite{RobertsSchmidt,Dern,Kappler} one can define an extension of $\Sigma_N$ to a subgroup of $Sp(4,\mathbb{R})$. To this end we introduce
\begin{align}
&h_N=\left(\begin{array}{cc}U_N & 0 \\ 0 & U_N^T\end{array}\right)\subset Sp(4,\mathbb{R})\,,&&\text{and} &&U_N=\frac{1}{\sqrt{N}}\left(\begin{array}{cc}0 & N \\ 1 & 0\end{array}\right)\,,\label{GeneratorSstar}
\end{align}
and define
\begin{align}
\Sigma_N^*=\Sigma_N\cup \Sigma_N h_N\subset Sp(4,\mathbb{R})\,.
\end{align}
Notice that $h_N$ in (\ref{GeneratorSstar}) acts as 
\begin{align}
h_N:\hspace{1cm}\Omega\longmapsto \Omega'=h_N\circ \Omega=\left(\begin{array}{cc}NR & S \\ S & \tfrac{\rho}{N}\end{array}\right)\,,
\end{align}
which implies the symmetry $f(R,\rho,S)=f(\tfrac{\rho}{N},NR,S)$ for paramodular forms.


\begin{thebibliography}{999}


\bibitem{Witten:1995zh}
E.~Witten,
{\it Some comments on string dynamics,}
[arXiv:hep-th/9507121 [hep-th]].


\bibitem{Aspinwall:1996vc}
P.~S.~Aspinwall,
{\it Point - like instantons and the spin (32) / Z(2) heterotic string,}
Nucl. Phys. B \textbf{496}, 149-176 (1997)
doi:10.1016/S0550-3213(97)00232-0
[arXiv:hep-th/9612108 [hep-th]].


\bibitem{Aspinwall:1997ye}
P.~S.~Aspinwall and D.~R.~Morrison,
{\it Point - like instantons on K3 orbifolds,}
Nucl. Phys. B \textbf{503}, 533-564 (1997)
doi:10.1016/S0550-3213(97)00516-6
[arXiv:hep-th/9705104 [hep-th]].


\bibitem{Seiberg:1997zk}
N.~Seiberg,
{\it New theories in six-dimensions and matrix description of M theory on T**5 and T**5 / Z(2),}
Phys. Lett. B \textbf{408}, 98-104 (1997)
doi:10.1016/S0370-2693(97)00805-8
[arXiv:hep-th/9705221 [hep-th]].


\bibitem{Intriligator:1997dh}
K.~A.~Intriligator,
{\it New string theories in six-dimensions via branes at orbifold singularities,}
Adv. Theor. Math. Phys. \textbf{1}, 271-282 (1998)
doi:10.4310/ATMP.1997.v1.n2.a5
[arXiv:hep-th/9708117 [hep-th]].


\bibitem{Hanany:1997gh}
A.~Hanany and A.~Zaffaroni,
{\it Branes and six-dimensional supersymmetric theories,}
Nucl. Phys. B \textbf{529}, 180-206 (1998)
doi:10.1016/S0550-3213(98)00355-1
[arXiv:hep-th/9712145 [hep-th]].


\bibitem{Brunner:1997gf}
I.~Brunner and A.~Karch,
{\it Branes at orbifolds versus Hanany Witten in six-dimensions,}
JHEP \textbf{03}, 003 (1998)
doi:10.1088/1126-6708/1998/03/003
[arXiv:hep-th/9712143 [hep-th]].


\bibitem{Heckman:2013pva}
J.~J.~Heckman, D.~R.~Morrison and C.~Vafa,
{\it On the Classification of 6D SCFTs and Generalized ADE Orbifolds,}
JHEP \textbf{05}, 028 (2014)
doi:10.1007/JHEP05(2014)028
[arXiv:1312.5746 [hep-th]].


\bibitem{Heckman:2015bfa}
J.~J.~Heckman, D.~R.~Morrison, T.~Rudelius and C.~Vafa,
{\it Atomic Classification of 6D SCFTs,}
Fortsch. Phys. \textbf{63}, 468-530 (2015)
doi:10.1002/prop.201500024
[arXiv:1502.05405 [hep-th]].

\bibitem{Xie:2015rpa}
D.~Xie and S.~T.~Yau,
{\it 4d N=2 SCFT and singularity theory Part I: Classification,}
[arXiv:1510.01324 [hep-th]].

\bibitem{Jefferson:2017ahm}
P.~Jefferson, H.~C.~Kim, C.~Vafa and G.~Zafrir,
{\it Towards Classification of 5d SCFTs: Single Gauge Node,}
[arXiv:1705.05836 [hep-th]].

\bibitem{Jefferson:2018irk}
P.~Jefferson, S.~Katz, H.~C.~Kim and C.~Vafa,
{\it On Geometric Classification of 5d SCFTs,}
JHEP \textbf{04}, 103 (2018)
doi:10.1007/JHEP04(2018)103
[arXiv:1801.04036 [hep-th]].

\bibitem{Caorsi:2018zsq}
M.~Caorsi and S.~Cecotti,
{\it Geometric classification of 4d $\mathcal{N}=2$ SCFTs,}
JHEP \textbf{07}, 138 (2018)
doi:10.1007/JHEP07(2018)138
[arXiv:1801.04542 [hep-th]].

\bibitem{Bhardwaj:2018vuu}
L.~Bhardwaj and P.~Jefferson,
{\it Classifying 5d SCFTs via 6d SCFTs: Arbitrary rank,}
JHEP \textbf{10}, 282 (2019)
doi:10.1007/JHEP10(2019)282
[arXiv:1811.10616 [hep-th]].

\bibitem{Bhardwaj:2019hhd}
L.~Bhardwaj,
{\it Revisiting the classifications of 6d SCFTs and LSTs,}
JHEP \textbf{03}, 171 (2020)
doi:10.1007/JHEP03(2020)171
[arXiv:1903.10503 [hep-th]].

\bibitem{Apruzzi:2019opn}
F.~Apruzzi, C.~Lawrie, L.~Lin, S.~Schäfer-Nameki and Y.~N.~Wang,
{\it Fibers add Flavor, Part I: Classification of 5d SCFTs, Flavor Symmetries and BPS States,}
JHEP \textbf{11}, 068 (2019)
doi:10.1007/JHEP11(2019)068
[arXiv:1907.05404 [hep-th]].

\bibitem{Bhardwaj:2019jtr}
L.~Bhardwaj,
{\it On the classification of $5d$ SCFTs,}
[arXiv:1909.09635 [hep-th]].

\bibitem{Martone:2020nsy}
M.~Martone,
{\it Towards the classification of rank-$r$ $\mathcal{N}=2$ SCFTs. Part I: twisted partition function and central charge formulae,}
[arXiv:2006.16255 [hep-th]].

\bibitem{Argyres:2020wmq}
P.~C.~Argyres and M.~Martone, {\it Towards a classification of rank $r$ $\mathcal{N}=2$ SCFTs Part II: special Kahler stratification of the Coulomb branch,}
[arXiv:2007.00012 [hep-th]].


\bibitem{Bhardwaj:2015oru}
L.~Bhardwaj, M.~Del Zotto, J.~J.~Heckman, D.~R.~Morrison, T.~Rudelius and C.~Vafa,
{\it F-theory and the Classification of Little Strings,}
Phys. Rev. D \textbf{93}, no.8, 086002 (2016)
doi:10.1103/PhysRevD.93.086002
[arXiv:1511.05565 [hep-th]].

\bibitem{Haghighat:2013gba}
B.~Haghighat, A.~Iqbal, C.~Koz\c{c}az, G.~Lockhart and C.~Vafa, {\it M-Strings,}
Commun. Math. Phys. \textbf{334}, no.2, 779-842 (2015)
doi:10.1007/s00220-014-2139-1
[arXiv:1305.6322 [hep-th]].

\bibitem{Hohenegger:2015btj}
S.~Hohenegger, A.~Iqbal and S.~J.~Rey, {\it Instanton-monopole correspondence from M-branes on $\mathbb S^1$ and little string theory,}
Phys. Rev. D \textbf{93}, no.6, 066016 (2016)
doi:10.1103/PhysRevD.93.066016
[arXiv:1511.02787 [hep-th]].

\bibitem{Hohenegger:2015cba}
S.~Hohenegger, A.~Iqbal and S.~Rey, {\it M-strings, monopole strings, and modular forms,}
Phys. Rev. D \textbf{92} (2015) no.6, 066005
doi:10.1103/PhysRevD.92.066005
[arXiv:1503.06983 [hep-th]]. 
  
\bibitem{Hohenegger:2016eqy}
S.~Hohenegger, A.~Iqbal and S.~Rey, {\it Self-Duality and Self-Similarity of Little String Orbifolds,}
Phys. Rev. D \textbf{94} (2016) no.4, 046006
doi:10.1103/PhysRevD.94.046006
arXiv:1605.02591 [hep-th].







\bibitem{Haghighat:2013tka}
B.~Haghighat, C.~Koz\c{c}az, G.~Lockhart and C.~Vafa, {\it Orbifolds of M-strings,}
Phys. Rev. D \textbf{89}, no.4, 046003 (2014)
doi:10.1103/PhysRevD.89.046003
[arXiv:1310.1185 [hep-th]].

\bibitem{Hohenegger:2013ala}
S.~Hohenegger and A.~Iqbal, {\it M-strings, elliptic genera and $\mathcal{N} = 4$ string amplitudes,}
Fortsch. Phys. \textbf{62}, 155-206 (2014)
doi:10.1002/prop.201300035
[arXiv:1310.1325 [hep-th]].


\bibitem{Kanazawa:2016tnt}
A.~Kanazawa and S.~C.~Lau, {\it Local Calabi-Yau manifolds of type $\tilde A$ via SYZ mirror symmetry,}
J. Geom. Phys. \textbf{139} (2019), 103-138
doi:10.1016/j.geomphys.2018.12.015
[arXiv:1605.00342 [math.AG]].



\bibitem{Aganagic:2003db}
M.~Aganagic, A.~Klemm, M.~Marino and C.~Vafa,
{\it The Topological vertex,}
Commun. Math. Phys. \textbf{254}, 425-478 (2005)
doi:10.1007/s00220-004-1162-z
[arXiv:hep-th/0305132 [hep-th]].

\bibitem{Iqbal:2007ii}
A.~Iqbal, C.~Kozcaz and C.~Vafa,
{\it The Refined Topological Vertex,}
JHEP \textbf{10}, 069 (2009)
doi:10.1088/1126-6708/2009/10/069
[arXiv:hep-th/0701156 [hep-th]].



\bibitem{GV1}
R.~Gopakumar and C.~Vafa,
{\it M theory and topological strings. 1.,}
[arXiv:hep-th/9809187 [hep-th]].

\bibitem{GV2}
R.~Gopakumar and C.~Vafa,
{\it M theory and topological strings. 2.,}
[arXiv:hep-th/9812127 [hep-th]].

\bibitem{Hollowood:2003cv}
T.~J.~Hollowood, A.~Iqbal and C.~Vafa,
{\it Matrix models, geometric engineering and elliptic genera,}
JHEP \textbf{03}, 069 (2008)
doi:10.1088/1126-6708/2008/03/069
[arXiv:hep-th/0310272 [hep-th]].

\bibitem{Antoniadis:2010iq}
I.~Antoniadis, S.~Hohenegger, K.~S.~Narain and T.~R.~Taylor, {\it Deformed Topological Partition Function and Nekrasov Backgrounds,}
Nucl. Phys. B \textbf{838} (2010), 253-265
doi:10.1016/j.nuclphysb.2010.04.021
[arXiv:1003.2832 [hep-th]].

\bibitem{Antoniadis:2013bja}
I.~Antoniadis, I.~Florakis, S.~Hohenegger, K.~S.~Narain and A.~Zein Assi, {\it Worldsheet Realization of the Refined Topological String,}
Nucl. Phys. B \textbf{875} (2013), 101-133
doi:10.1016/j.nuclphysb.2013.07.004
[arXiv:1302.6993 [hep-th]].

\bibitem{Antoniadis:2013mna}
I.~Antoniadis, I.~Florakis, S.~Hohenegger, K.~S.~Narain and A.~Zein Assi, {\it Non-Perturbative Nekrasov Partition Function from String Theory,}
Nucl. Phys. B \textbf{880} (2014), 87-108
doi:10.1016/j.nuclphysb.2014.01.006
[arXiv:1309.6688 [hep-th]].

\bibitem{Antoniadis:2015spa}
I.~Antoniadis, I.~Florakis, S.~Hohenegger, K.~S.~Narain and A.~Zein Assi, {\it Probing the moduli dependence of refined topological amplitudes,}
Nucl. Phys. B \textbf{901} (2015), 252-281
doi:10.1016/j.nuclphysb.2015.10.016
[arXiv:1508.01477 [hep-th]].



\bibitem{Hohenegger:2016yuv}
S.~Hohenegger, A.~Iqbal and S.~Rey, {\it Dual Little Strings from F-Theory and Flop Transitions,}
JHEP \textbf{07} (2017), 112
doi:10.1007/JHEP07(2017)112
[arXiv:1610.07916 [hep-th]]. 

\bibitem{Bastian:2017ing} B.~Bastian, S.~Hohenegger, A.~Iqbal and S.~J.~Rey, {\it Dual little strings and their partition functions,}
Phys. Rev. D \textbf{97} (2018) no.10, 106004
doi:10.1103/PhysRevD.97.106004
[arXiv:1710.02455 [hep-th]].

\bibitem{Bastian:2018dfu}
B.~Bastian, S.~Hohenegger, A.~Iqbal and S.~Rey, {\it Beyond Triality: Dual Quiver Gauge Theories and Little String Theories,}
JHEP \textbf{11} (2018), 016
doi:10.1007/JHEP11(2018)016
[arXiv:1807.00186 [hep-th]]. 


\bibitem{Haghighat:2018gqf}
B.~Haghighat and R.~Sun, {\it M5 branes and Theta Functions,}
JHEP \textbf{10} (2019), 192
doi:10.1007/JHEP10(2019)192
[arXiv:1811.04938 [hep-th]].

  
\bibitem{Bastian:2018jlf}
B.~Bastian and S.~Hohenegger,
{\it Dihedral Symmetries of Gauge Theories from Dual Calabi-Yau Threefolds,}
Phys. Rev. D \textbf{99} (2019) no.6, 066013
doi:10.1103/PhysRevD.99.066013
[arXiv:1811.03387 [hep-th]].

\bibitem{Bastian:2019hpx}
B.~Bastian and S.~Hohenegger, {\it Symmetries in A-type little string theories. Part I. Reduced free energy and paramodular groups,}
JHEP \textbf{03} (2020), 062
doi:10.1007/JHEP03(2020)062
[arXiv:1911.07276 [hep-th]].



\bibitem{Nekrasov:2009rc}
N.~A.~Nekrasov and S.~L.~Shatashvili, {\it Quantization of Integrable Systems and Four Dimensional Gauge Theories,}
doi:10.1142/978981430463$4_{-}$0015
[arXiv:0908.4052 [hep-th]].

\bibitem{Mironov:2009uv}
A.~Mironov and A.~Morozov, {\it Nekrasov Functions and Exact Bohr-Zommerfeld Integrals,}
JHEP \textbf{04} (2010), 040
doi:10.1007/JHEP04(2010)040
[arXiv:0910.5670 [hep-th]].


\bibitem{Ahmed:2017hfr}
  A.~Ahmed, S.~Hohenegger, A.~Iqbal and S.~J.~Rey, {\it Bound states of little strings and symmetric orbifold conformal field theories,}
  Phys.\ Rev.\ D {\bf 96} (2017) no.8,  081901
  doi:10.1103/PhysRevD.96.081901
  [arXiv:1706.04425 [hep-th]].





\bibitem{Hohenegger:2019tii}
S.~Hohenegger, {\it From Little String Free Energies Towards Modular Graph Functions,}
JHEP \textbf{03} (2020), 077
doi:10.1007/JHEP03(2020)077
[arXiv:1911.08172 [hep-th]].




\bibitem{Broedel:2014vla}
J.~Broedel, C.~R.~Mafra, N.~Matthes and O.~Schlotterer,
{\it Elliptic multiple zeta values and one-loop superstring amplitudes,}
JHEP \textbf{07}, 112 (2015)
doi:10.1007/JHEP07(2015)112
[arXiv:1412.5535 [hep-th]].


\bibitem{DHoker:2015wxz}
E.~D'Hoker, M.~B.~Green, \"O.~G\"urdogan and P.~Vanhove,
{\it Modular Graph Functions,}
Commun. Num. Theor. Phys. \textbf{11}, 165-218 (2017)
doi:10.4310/CNTP.2017.v11.n1.a4
[arXiv:1512.06779 [hep-th]].


\bibitem{DHoker:2016mwo}
E.~D'Hoker and M.~B.~Green,
{\it Identities between Modular Graph Forms,}
J. Number Theor. \textbf{189}, 25-88 (2018)
[arXiv:1603.00839 [hep-th]].


\bibitem{DHoker:2017pvk}
E.~D'Hoker, M.~B.~Green and B.~Pioline,
{\it Higher genus modular graph functions, string invariants, and their exact asymptotics,}
Commun. Math. Phys. \textbf{366}, no.3, 927-979 (2019)
doi:10.1007/s00220-018-3244-3
[arXiv:1712.06135 [hep-th]].

\bibitem{Zerbini:2018sox}
F.~Zerbini,
{\it Elliptic multiple zeta values, modular graph functions and genus 1 superstring scattering amplitudes,}
[arXiv:1804.07989 [math-ph]].

\bibitem{Zerbini:2018hgs}
F.~Zerbini,
{\it Modular and Holomorphic Graph Functions from Superstring Amplitudes,}
doi:10.1007/978-3-030-04480-$0_-$18
[arXiv:1807.04506 [math-ph]].


\bibitem{Gerken:2018jrq}
J.~E.~Gerken, A.~Kleinschmidt and O.~Schlotterer,
{\it Heterotic-string amplitudes at one loop: modular graph forms and relations to open strings,}
JHEP \textbf{01}, 052 (2019)
doi:10.1007/JHEP01(2019)052
[arXiv:1811.02548 [hep-th]].


\bibitem{Mafra:2019ddf}
C.~R.~Mafra and O.~Schlotterer,
{\it All Order $\alpha$ Expansion of One-Loop Open-String Integrals,}
Phys. Rev. Lett. \textbf{124}, no.10, 101603 (2020)
doi:10.1103/PhysRevLett.124.101603
[arXiv:1908.09848 [hep-th]].


\bibitem{Mafra:2019xms}
C.~R.~Mafra and O.~Schlotterer,
{\it One-loop open-string integrals from differential equations: all-order $\alpha$'-expansions at $n$ points,}
JHEP \textbf{03}, 007 (2020)
doi:10.1007/JHEP03(2020)007
[arXiv:1908.10830 [hep-th]].


\bibitem{Gerken:2019cxz}
J.~E.~Gerken, A.~Kleinschmidt and O.~Schlotterer,
{\it All-order differential equations for one-loop closed-string integrals and modular graph forms,}
JHEP \textbf{01}, 064 (2020)
doi:10.1007/JHEP01(2020)064
[arXiv:1911.03476 [hep-th]].


\bibitem{Bastian:2019wpx}
B.~Bastian and S.~Hohenegger, {\it Symmetries in A-type little string theories. Part II. Eisenstein series and generating functions of multiple divisor sums,}
JHEP \textbf{03} (2020), 016 doi:10.1007/JHEP03(2020)016 [arXiv:1911.07280 [hep-th]].  




\bibitem{Bershadsky:1993cx}
M.~Bershadsky, S.~Cecotti, H.~Ooguri and C.~Vafa,
{\it Kodaira-Spencer theory of gravity and exact results for quantum string amplitudes,}
Commun. Math. Phys. \textbf{165}, 311-428 (1994)
doi:10.1007/BF02099774
[arXiv:hep-th/9309140 [hep-th]].


\bibitem{EG3}
A.~Gadde and S.~Gukov,
{\it 2d Index and Surface operators,}
JHEP \textbf{03}, 080 (2014)
doi:10.1007/JHEP03(2014)080
[arXiv:1305.0266 [hep-th]].

\bibitem{EG1}
F.~Benini, R.~Eager, K.~Hori and Y.~Tachikawa,
{\it Elliptic genera of two-dimensional N=2 gauge theories with rank-one gauge groups,}
Lett. Math. Phys. \textbf{104}, 465-493 (2014)
doi:10.1007/s11005-013-0673-y
[arXiv:1305.0533 [hep-th]].



\bibitem{EG2}
F.~Benini, R.~Eager, K.~Hori and Y.~Tachikawa,
{\it Elliptic Genera of 2d ${\mathcal{N}}$ = 2 Gauge Theories,}
Commun. Math. Phys. \textbf{333}, no.3, 1241-1286 (2015)
doi:10.1007/s00220-014-2210-y
[arXiv:1308.4896 [hep-th]].



\bibitem{HNBook}
H. Nakajima, {\it Lectures on Hilbert Schemes of Points on Surfaces}, University Lecture Series, American Mathematical Society, 1999.

\bibitem{Nakajima:2003pg}
H.~Nakajima and K.~Yoshioka,
{\it Instanton counting on blowup. 1.,}
Invent. Math. \textbf{162}, 313-355 (2005)
doi:10.1007/s00222-005-0444-1
[arXiv:math/0306198 [math.AG]].








\bibitem{Bastian:2018fba}
B.~Bastian, S.~Hohenegger, A.~Iqbal and S.~J.~Rey, {\it Five-Dimensional Gauge Theories from Shifted Web Diagrams,}
Phys. Rev. D \textbf{99} (2019) no.4, 046012
doi:10.1103/PhysRevD.99.046012
[arXiv:1810.05109 [hep-th]].
  
\bibitem{Bastian:2017jje}
B.~Bastian and S.~Hohenegger, {\it Five-Brane Webs and Highest Weight Representations,}
JHEP \textbf{12} (2017), 020
doi:10.1007/JHEP12(2017)020
[arXiv:1706.08750 [hep-th]]. 


\bibitem{Bastian:2017ary}
B.~Bastian, S.~Hohenegger, A.~Iqbal and S.~Rey, {\it Triality in Little String Theories,}
Phys. Rev. D \textbf{97} (2018) no.4, 046004
doi:10.1103/PhysRevD.97.046004
[arXiv:1711.07921 [hep-th]].

\bibitem{krieg}
A. Krieg, {\it Jacobi Forms of Many variables and the Maa\ss\, Space,} Journal of Number Theory, {\bf 56}, 242-255 (1996).

\bibitem{Khoze:1998gy}
V.~V.~Khoze, M.~P.~Mattis and M.~J.~Slater,
{\it The Instanton Hunter's guide to supersymmetric SU(N) gauge theories,}
Nucl. Phys. B \textbf{536} (1998), 69-109
doi:10.1016/S0550-3213(98)00579-3
[arXiv:hep-th/9804009 [hep-th]].

\bibitem{Dorey:2002ik}
N.~Dorey, T.~J.~Hollowood, V.~V.~Khoze and M.~P.~Mattis,
{\it The Calculus of many instantons,}
Phys. Rept. \textbf{371}, 231-459 (2002)
doi:10.1016/S0370-1573(02)00301-0
[arXiv:hep-th/0206063 [hep-th]].
 





























































  


\bibitem{Dijkgraaf:1996xw}
R.~Dijkgraaf, G.~W.~Moore, E.~P.~Verlinde and H.~L.~Verlinde,
{\it Elliptic genera of symmetric products and second quantized strings,}
Commun. Math. Phys. \textbf{185}, 197-209 (1997)
doi:10.1007/s002200050087
[arXiv:hep-th/9608096 [hep-th]].


\bibitem{Belin:2018oza}
A.~Belin, A.~Castro, J.~Gomes and C.~A.~Keller,
{\it Siegel paramodular forms and sparseness in AdS$_{3}$/CFT$_{2}$,}
JHEP \textbf{11}, 037 (2018)
doi:10.1007/JHEP11(2018)037
[arXiv:1805.09336 [hep-th]].
  
\bibitem{P1}
K. Hulek, G. K. Sankaran, {\it The geometry of Siegel modular varieties}, 	arXiv:math/9810153 [math.AG].

\bibitem{P2}
V. Gritsenko, K. Hulek, {\it Minimal Siegel modular threefolds}, 	arXiv:alg-geom/9506017.





\bibitem{Dabholkar:2012nd}
A.~Dabholkar, S.~Murthy and D.~Zagier,
{\it Quantum Black Holes, Wall Crossing, and Mock Modular Forms,}
[arXiv:1208.4074 [hep-th]].  
 
 

 
\bibitem{AHK}
O. Aharony, A. Hanany, and B. Kol, {\it Webs of (p,q) five-branes, five-dimensional
field theories and grid diagrams,} JHEP 01 (1998) 002, arXiv:hep-th/9710116
[hep-th].


\bibitem{Ferlito:2017xdq}
G.~Ferlito, A.~Hanany, N.~Mekareeya and G.~Zafrir,
{\it 3d Coulomb branch and 5d Higgs branch at infinite coupling,}
JHEP \textbf{07}, 061 (2018)
doi:10.1007/JHEP07(2018)061
[arXiv:1712.06604 [hep-th]].

  


\bibitem{EichlerZagier} M.~Eichler and D.~Zagier, {\it The Theory of Jacobi Forms,} Springer Verlag, 1985    

\bibitem{Lang} S. Lang, {\it Introduction to Modular Forms,} Grundlehren der Mathematischen Wissenschaften 222, Springer Verlag, Berlin (1995).

\bibitem{Stein} W. Stein, {\it Modular Forms, a Computational Approach}, Graduate Studies in Mathematics 79, American Mathematical Society, Providence, RI (2007).
  

\bibitem{ambreenthesis}
A.~Ahmed, {\it Equivariant Integrals on $\mbox{Hilb}^{\bullet}[\mathbb{C}^2]$, Hecke Operators and Modular Forms}, PhD Thesis 2020.


\bibitem{HeimKrieg} B.~Heim and A.~Krieg, {\it The Maa\ss\, Space for Paramodular Groups,} [arXiv:1711.06619 [math.NT]].

\bibitem{RobertsSchmidt} B.~Roberts and R.~Schmidt, {\it On modular forms for the paramodular groups,} in "Automorphic forms and zeta functions", pp. 334-364, World Sci. Publ., Hackensack, NJ, 2006.

\bibitem{zantout}
D. M. Zantout, {\it On the Cuspidality of Maass-Gritsenko and Mixed Level Lifts}, PhD Thesis (2013).

\bibitem{Dern} T. Dern, {\it Paramodular Forms of Degree 2 and Level 3}, Commentarii Mathematici Universitatis Sancti Pauli, Vol. 51, No.2, 157-194, 2002.  

\bibitem{Kappler} F.O.~Kappler {\it \"Uber die Charaktere Siegelscher Stufengruppen}, Ph.D. -thesis, Freiburg 1977.

\end{thebibliography}
\end{document}